\pdfoutput=1
\documentclass[12pt, a4paper]{article}
\usepackage{jheppub}
\makeatletter\gdef\@fpheader{}\makeatother
\compress
\renewcommand\afterLogoSpace{}
\renewcommand\afterProceedingsSpace{}
\usepackage[usenames, dvipsnames]{xcolor}
\usepackage{graphicx}
\usepackage{microtype}
\usepackage{amssymb}
\usepackage{amsmath}
\usepackage{mathtools}
\usepackage{amsfonts}
\usepackage{bm}
\usepackage{dsfont}
\usepackage{framed}
\usepackage{subcaption}
\usepackage{enumitem}
\usepackage[hang, flushmargin]{footmisc}
\usepackage{hyperref}
\hypersetup{colorlinks=true,linkcolor=purple,citecolor=purple,urlcolor=purple}

\newcommand{\dd}{\mathrm{d}}
\newcommand{\ii}{\mathrm{i}}
\newcommand{\ee}{\mathrm{e}}
\newcommand{\Ga}{\Gamma}
\newcommand{\pic}{\pi_c}
\newcommand{\rhohat}{\hat{\rho}}
\newcommand{\mhat}{\hat{m}}
\newcommand{\mueff}{\mu_{\rm eff}}
\newcommand{\nueff}{\nu_{\rm eff}}
\newcommand{\chg}{\chi}
\newcommand{\ash}{\mathrm{arcsinh}(\rhohat/2x)}
\newcommand{\ucan}{u}
\newcommand{\vcan}{v}
\newcommand{\Rp}{Q_{\brh}}
\newcommand{\Rm}{Q_{\brl}}
\newcommand{\Rpm}{Q_{\brh,\brl}}
\newcommand{\brh}{{\mathrm{h}}}
\newcommand{\brl}{{\ell}}
\newcommand{\omh}{\omega_\brh}
\newcommand{\oml}{\omega_\brl}
\newcommand{\homh}{\hat\omega_\brh}
\newcommand{\homl}{\hat\omega_\brl}

\title{Inflation at strong mixing as a connection problem: effective field theory, Stokes phenomenon and the cosmological collider}

\author[a,b]{Sebastian Garcia-Saenz,}
\author[c,d]{Yizhou Lu,}
\author[e]{S\'ebastien Renaux-Petel}
\affiliation[a]{Department of Physics, Southern University of Science and Technology, Shenzhen 518055, China}
\affiliation[b]{Guangdong Provincial Key Laboratory of Topological Matter, Southern University of Science and Technology, Shenzhen 518055, China}
\affiliation[c]{Shanghai Institute for Mathematics and Interdisciplinary Sciences, Shanghai 200433, China}
\affiliation[d]{Research Institute of Intelligent Complex Systems, Fudan University, Shanghai 200433, China}
\affiliation[e]{Institut d'Astrophysique de Paris, CNRS, Sorbonne Universit\'e, 98 bis bd Arago, 75014 Paris, France}

\abstract{We study a massive field strongly mixed with the curvature perturbation during inflation, in the regime where the latter is described at low energies by a single-field effective field theory (EFT) with a quadratic dispersion relation. What this EFT captures, and what lies beyond it, are encoded in a connection problem, which relates the solutions that define the vacuum at early times to those that describe observables at late times. We solve it with a general method that does not rely on an exact solution. An expansion in the inverse mixing, keeping both fields, yields the EFT as its leading order, with corrections under systematic control. It selects the vacuum of the EFT, fixes a phase that the EFT cannot determine on its own, and describes the heavy mode at all times. Particle production lies beyond all orders of this asymptotic expansion. We show that it arises from a Stokes phenomenon at the complex branch points of the dispersion relation and that, in a rotating basis of the two fields, it reduces to the familiar production of particles with half the heavy mass. A reflection symmetry of the equations of motion, combined with the canonical commutation relations, then determines all exponentially small effects exactly, and with them the solution of the connection problem, which combines particle production with a conversion between light and heavy quanta. Together with the expansion, these exact results give a compact analytical formula for the cosmological collider signal, whose exponential suppression, half of the usual Boltzmann one, directly reflects the halved mass of the produced particles. The expansion only requires a separation of scales and the adiabaticity of the heavy mode, while the reflection symmetry, which already gives the particle production of a free field without solving its equation of motion, holds for many systems in de Sitter space. We therefore expect this approach to bring physical understanding and simple formulas well beyond the present case.}

\newenvironment{techinsert}[1]{\begin{framed}\small\noindent\textit{#1.}---}{\end{framed}}

\begin{document}
\enlargethispage{3\baselineskip}
\maketitle
\flushbottom

\section{Introduction}
\label{sec:intro}
Effective field theories (EFTs) are the language in which the physics of inflationary fluctuations is formulated, be it for the Goldstone boson of broken time translations alone~\cite{Creminelli:2006xe,Cheung:2007st} or in the presence of additional fields~\cite{Senatore:2010wk,Noumi:2012vr,Pinol:2023oux,Pinol:2024arz}. They describe the degrees of freedom that are kept, and encode the effects of the modes that have been integrated out in the coefficients of their operators. In cosmology the background depends on time and energy is not conserved, so that what such a description retains of the heavy fields, and what it leaves out, is a non-trivial question that has been examined from several angles. Heavy fields coupled to the inflaton can reduce the sound speed of the curvature perturbation, or modify its dispersion relation, and the conditions under which a single-field description then holds have been studied in detail~\cite{Tolley:2009fg,Cremonini:2010ua,Achucarro:2010jv,Baumann:2011su,Shiu:2011qw,Cespedes:2012hu,Achucarro:2012sm,Avgoustidis:2012yc,Achucarro:2012yr,Burgess:2012dz,Gwyn:2012mw,Castillo:2013sfa,Gwyn:2014doa,Garcia-Saenz:2019njm}. The production of heavy particles by the expansion of the universe is at the origin of the cosmological collider signal~\cite{Chen:2009zp,Noumi:2012vr,Arkani-Hamed:2015bza}, which is now searched for in cosmological data~\cite{Sohn:2024xzd,Cabass:2024wob,Philcox:2025bbo,Suman:2025vuf,Kumar:2026ogn,Philcox:2026tjj}. Since this production is exponentially small in the mass of the particles, it is absent at any finite order of an EFT. The expansion of cosmological EFTs, local or not, is indeed asymptotic~\cite{Jazayeri:2023xcj,Jazayeri:2023kji,DuasoPueyo:2025lmq}, and the possibility of recovering what it omits from its resummation, supplemented by the choice of vacuum, has recently been demonstrated~\cite{Li:2026xhw,Dorigoni:2026mbh}.

These questions take a concrete form for a massive field strongly mixed with the curvature perturbation. In the EFT of inflationary fluctuations with an additional massive field~\cite{Noumi:2012vr,Pinol:2023oux,Pinol:2024arz}, the two are coupled at quadratic order by a mixing $\rho$, which in two-field models of inflation is set by the rate at which the trajectory turns in field space~\cite{Gordon:2000hv,GrootNibbelink:2001qt}. When the mixing is larger than the mass $m$ of the additional field, it cannot be treated as a perturbation. We are interested here in a deeper regime, in which the mixing is also much larger than the Hubble rate $H$ and than $m^2/H$. The mixing then sets the separation of scales. The system contains a heavy mode, with a mass of the order of the mixing, and a light mode, described at low energies by a single-field effective theory with a quadratic dispersion relation~\cite{Baumann:2011su}, which we refer to as the EFT in what follows. This regime enhances the power spectrum of the curvature perturbation and leads to large non-Gaussianities. It has been studied with EFT and numerical methods~\cite{Cremonini:2010ua,An:2017hlx,Tong:2017iat,Garcia-Saenz:2018ifx,Werth:2023pfl,Pinol:2023oux,Jazayeri:2023xcj}, and more recently through exact solutions of the linear theory~\cite{Huenupi:2026abj,Belrhali:2026uxn} and their consequences for the bispectrum~\cite{Huenupi:2026aqc,Wang:2026lff,Pinol:2026xnl,Belrhali:2026uxn,Pinol:2026hey}. In particular, exact integral representations of the bispectra, evaluated numerically, were obtained in~\cite{Huenupi:2026aqc,Pinol:2026xnl,Pinol:2026hey}, and analytical expressions in the form of series were obtained in~\cite{Belrhali:2026uxn}, where the cosmological collider signal was found to carry only half of the usual Boltzmann suppression in this regime. Convergent series at any mass and mixing were also obtained in~\cite{Chen:2026khu}, by a different method based on the Mellin-Barnes representation of the mixed propagator.

At the level of the linear theory, these questions can be phrased as a single problem for the mode functions. The vacuum is defined at early times, deep inside the horizon, where the solutions that represent it oscillate with a single positive frequency. Observables are evaluated at late times, where the natural solutions are instead those that behave as a definite power of conformal time. These are a constant curvature perturbation and decaying oscillations of the massive field, of positive or negative frequency. Expressing the former solutions in terms of the latter is known as a connection problem (figure~\ref{fig:connection}), and its coefficients contain the physical information, namely the late-time amplitude of the fluctuations and the Bogoliubov coefficients that measure the production of particles by the cosmic expansion. If the evolution were adiabatic at all times, a solution that starts with a single positive frequency would keep it, and it would end on a single late-time solution. The problem is non-trivial because adiabaticity fails, at real or at complex times. For a free field in de Sitter space, the problem is solved by the classical formula that relates a Hankel function to Bessel functions. Our aim is to solve it for the strongly mixed system in a way that does not rely on the knowledge of an exact solution, and that identifies the origin of each coefficient.

We proceed in three steps. We first solve the linear theory as an expansion in the inverse mixing, keeping the two fields throughout, and show that the EFT emerges as its leading order. This determines some of the coefficients, as series in the inverse mixing. The others vanish at all orders of this expansion, since they are exponentially small in the mixing. In a second step, we trace them to the complex branch points of the dispersion relation, where a Stokes phenomenon, the appearance of an exponentially small wave across a line in the complex time plane, produces particles, and we obtain them exactly, in terms of the coefficients given by the expansion, from a reflection symmetry of the equations and the canonical commutation relations. Finally, we combine the two to derive the cosmological collider signal in the squeezed limit of the bispectrum in closed form.

The first step rests on the local dispersion relation of the mixed system, which has two branches. These two branches have a simple interpretation. A time-dependent rotation of the two fields removes the mixing, the same that leads, in two-field models of inflation, from the fluctuations along and across the background trajectory to those of the original fields. In this basis, which we call the field basis, the system reduces at high frequencies to two free fields of mass $\rho/2$, and the two branches are their two circular polarisations. The light mode turns in the same sense as the rotation, and the heavy mode in the opposite sense. The heavy mode has a frequency that remains larger than the Hubble rate at all times, and it evolves adiabatically. The light mode evolves adiabatically at early times, crosses its own horizon earlier than a massless field would, and then freezes to become the late-time curvature perturbation. In the adiabatic regimes, the solutions are given by the WKB approximation, which we work out to second order of the WKB expansion. Across the horizon crossing of the light mode, the equations of motion reduce to those of the EFT, whose solutions are Hankel functions with an argument quadratic in conformal time. The two descriptions of the light mode overlap, and matching them, in the manner of matched asymptotic expansions~\cite{Bender:1999box}, gives the solution at all times.

The effective theory thus appears as the leading order of a systematic expansion of the two-field system. The matching determines what the EFT would otherwise take as an input, namely its vacuum, which is selected by the Bunch-Davies vacuum of the two-field system, and a phase accumulated before the EFT becomes valid. This phase is observable for excited initial states and, as we show below, it also contributes to the phase of the cosmological collider signal. We also obtain the heavy mode, which the EFT does not contain, in closed form and at all times. At this stage, the light mode ends as a constant curvature perturbation, which sets the power spectrum, and the heavy mode as a positive-frequency oscillation of the massive field, with no admixture of the other late-time solutions. The expansion thus determines the amplitudes of these two late-time solutions. Together, these results provide approximations of the mode functions in terms of elementary and Hankel functions, valid uniformly in time, which we compare with the exact solution.

Since the two fields are kept throughout, one can systematically identify the effects that the EFT does not capture. We distinguish three kinds of effects: (i) corrections to the dynamics of the light mode, in powers of the inverse mixing, which the expansion organises order by order; (ii) the heavy mode, whose contributions are not exponentially suppressed when the massive field propagates on internal lines, and to which the closed form of the heavy mode gives access; (iii) effects that are exponentially small in the mixing, among which the production of particles at the origin of the cosmological collider signal. The expansion controls the first two. The third lies beyond all its orders, and the expansion is indeed asymptotic, with coefficients that grow factorially.

The effects of the third kind are the subject of the second step. We show that the production of particles arises from a Stokes phenomenon at the complex branch points of the dispersion relation, which lie at the scale of the heavy mass. In the field basis, this phenomenon reduces to the familiar production of particles of a free field of mass $\rho/2$ by the cosmic expansion. A reflection symmetry of the equations, combined with the canonical commutation relations, then expresses all the exponentially small coefficients exactly, in terms of the two coefficients given by the expansion. The argument is general. For a free field in de Sitter space, including in the presence of a chemical potential, it gives the Bogoliubov coefficients of particle production without solving the equation of motion. For the mixed system, it expresses the late-time curvature perturbation of the heavy mode in terms of the light mode, and the late-time oscillations of the massive field in the light mode in terms of the heavy one, at any mass and mixing. In particular, it readily yields the exponential dependence of the cosmological collider signal on the mass and the mixing found in~\cite{Pinol:2026hey}, which interpolates between the usual Boltzmann suppression at small mixing and half of it at large mixing~\cite{Belrhali:2026uxn}.

We thereby obtain the solution of the connection problem exactly, in terms of the mass, the mixing and the phase of the late-time curvature perturbation of the light mode. The early-time and the late-time solutions are related by a Bogoliubov transformation, which contains two physical effects. The first is particle production, which mixes positive and negative frequencies as for a free field, and which is the Stokes phenomenon just described. It creates pairs of one light and one heavy quantum. The second is a conversion of the light mode into the heavy one, and conversely, which changes the species of the quanta without changing their number. It has no analogue for a single field. Both effects take place at the scale of the heavy mass, well before the light mode crosses its horizon. The symmetry gives them exactly, while the expansion provides the mode functions at all times and the phase accumulated by the light mode, which appears in the conversion.

The third step combines the exact relations of the second step with the adiabatic solutions of the first. In the squeezed limit of the bispectrum, the cosmological collider signal comes from the exponentially small oscillating part of the long mode, given by the reflection symmetry, and from the adiabatic evolution of the two short ones. The resonance between this oscillation and the two short quanta, which generates the signal, takes place at the energy scale of the heavy mass, where the EFT does not apply. We obtain its amplitude and its phase in closed form for the cubic interactions of the EFT of inflationary fluctuations. This provides, at large mixing, a compact expression for the cosmological collider signal contained in the bispectra of~\cite{Huenupi:2026aqc,Pinol:2026xnl,Belrhali:2026uxn,Pinol:2026hey,Chen:2026khu}. We recover in particular that the signal grows with the number of time derivatives of the curvature perturbation at the vertex, as found and explained in~\cite{Belrhali:2026uxn}.

The paper is organised as follows. In section~\ref{sec:setup} we present the theory, its equations of motion, and the so-called field basis in which the mixing disappears. In section~\ref{sec:connection} we define the solutions adapted to early and to late times, formulate the problem of relating them, and recall the exact solution of the system. In section~\ref{sec:regimes} we describe the two branches of the dispersion relation, their interpretation in the field basis, and the regimes in which each of them evolves adiabatically. Section~\ref{sec:wkb} is devoted to the WKB solutions, and section~\ref{sec:eft} to the effective theory of the light branch. We match the two in section~\ref{sec:matching}, where we derive the vacuum of the EFT and the power spectrum, give approximations of the mode functions that are valid at all times, and extend the matching to excited initial states. Section~\ref{sec:beyond} is devoted to the effects that lie beyond all orders of the expansion, which we trace to a Stokes phenomenon at the complex branch points of the dispersion relation. In section~\ref{sec:reflection} we show how a reflection symmetry of the equations gives the particle production of a free field, derive from it these effects exactly, deduce the solution of the connection problem, and interpret it in terms of particle production and conversion. In section~\ref{sec:collider} we use these results to compute the cosmological collider signal in the squeezed limit of the bispectrum. We conclude in section~\ref{sec:conclusions}, while six appendices collect technical details.

\section{Strongly mixed system}
\label{sec:setup}

\subsection{Theory and equations of motion}
\label{sec:theory}

We work in quasi-de Sitter space, in the decoupling limit of the effective field theory of inflationary fluctuations~\cite{Creminelli:2006xe,Cheung:2007st}, extended by a massive scalar field $\sigma$~\cite{Noumi:2012vr,Pinol:2023oux,Pinol:2024arz}. The curvature sector is described by the canonically normalised Goldstone boson $\pic$, related to the curvature perturbation by $\zeta=-H\pic/f_\pi^2$, with $f_\pi^4=2M_{\rm Pl}^2|\dot H|$ the symmetry breaking scale. At quadratic order, at unit sound speed and at leading order in derivatives, the Lagrangian reads
\begin{equation}
\label{eq:lagrangian}
\mathcal L/a^{3}=\frac12\dot{\pi}_c^{2}-\frac{(\partial_i\pic)^{2}}{2a^{2}}
+\frac12\dot\sigma^{2}-\frac{(\partial_i\sigma)^{2}}{2a^{2}}-\frac12 m^{2}\sigma^{2}
+\rho\,\dot{\pi}_c\,\sigma\,,
\end{equation}
where $m$ is the mass of $\sigma$ and $\rho$ the linear mixing between the two fields, both taken constant. We use the same conventions as in~\cite{Belrhali:2026uxn}, and denote by $\rhohat\equiv\rho/H$ and $\mhat\equiv m/H$ the two parameters of the theory in Hubble units. The Lagrangian~\eqref{eq:lagrangian} arises in particular, at leading order in the slow-roll approximation, from a two-field model of inflation whose trajectory turns in field space at a constant rate~\cite{Gordon:2000hv,GrootNibbelink:2001qt}, in which case $\rho$ is twice the turn rate. Whatever its origin, $\pic$ and $\sigma$ are commonly referred to as the adiabatic and the entropic fluctuation, respectively.

This paper is concerned with the regime
\begin{equation}
\label{eq:regime}
\rhohat\gg1\,,\qquad |\mhat^{2}|\ll\rhohat\,,
\end{equation}
that is, $\rho\gg H$ and $\rho\gg m^2/H$. The second condition ensures that the mass is a small correction to the dynamics set by the mixing. Indeed, it enters the results below only at relative order $\mhat^2/\rhohat$ (sections~\ref{sec:wkb-second} and~\ref{sec:eft-constraint}). In this regime, the light mode has the quadratic dispersion relation identified in~\cite{Baumann:2011su}. We assume nothing about the sign of $m^2$ beyond the stability condition $m^2+\rho^2>0$, which is automatic in this regime. When instead $|\mhat^2|\gtrsim\rhohat$, the mass term dominates at the horizon crossing of the light mode, which then propagates with the reduced sound speed $c_s^2=m^2/(m^2+\rho^2)$~\cite{Achucarro:2010jv,Baumann:2011su}, imaginary when $m^2<0$~\cite{Garcia-Saenz:2018ifx,Garcia-Saenz:2018vqf,Fumagalli:2019noh,Ferreira:2020qkf,Garcia-Saenz:2025jis}. The regime~\eqref{eq:regime} is the complementary one. The mixing cannot, however, be arbitrarily large. Perturbative control requires $\rhohat\lesssim\kappa^{1/2}/\Delta_\zeta\simeq5\times10^3$, with $\kappa=2\Ga(5/4)^2/\pi^3$ and $\Delta_\zeta$ the amplitude of the curvature power spectrum. This bound was obtained in~\cite{Pinol:2023oux} by requiring that the cubic interaction $(\partial_i\pic)^2\sigma$, which accompanies the mixing in the EFT of inflationary fluctuations, be weakly coupled at the Hubble scale, given the amplified power spectrum~\eqref{eq:R-leading}; see also~\cite{Baumann:2011su}. The regime~\eqref{eq:regime} covers most of this range.

We now turn to the equations of motion. The Fourier modes of the two fields, of comoving wavenumber $k$, evolve independently of one another, and for each of them we use the time variable
\begin{equation}
\label{eq:xdef}
x\equiv-k\tau=\frac{k}{aH}>0\,,
\end{equation}
with $\tau$ the conformal time, so that early times correspond to $x\to\infty$ and late times to $x\to0$.\footnote{The variable $x$ is minus the variable $z$ used in Ref.~\cite{Belrhali:2026uxn}.} In terms of it, the equations of motion that follow from~\eqref{eq:lagrangian} read
\begin{align}
\label{eq:system}
\pic''-\frac{2}{x}\pic'+\pic&=\frac{\rhohat}{x}\,\sigma'-\frac{3\rhohat}{x^{2}}\,\sigma\,,\notag\\
\sigma''-\frac{2}{x}\sigma'+\Big[1+\frac{\mhat^{2}}{x^{2}}\Big]\sigma&=-\frac{\rhohat}{x}\,\pic'\,,
\end{align}
where primes denote $\dd/\dd x$. The asymptotic analysis we will perform is best carried out on the canonically normalised variables
\begin{equation}
\label{eq:canonical}
\ucan\equiv\frac{\pic}{x}\,,\qquad \vcan\equiv\frac{\sigma}{x}\,,
\end{equation}
which are, up to the constant factor $k/H$, the fields multiplied by the scale factor, i.e.~the Sasaki-Mukhanov variables $a\pic$ and $a\sigma$. In terms of them, the system~\eqref{eq:system} becomes
\begin{align}
\label{eq:canonical-system}
\ucan''+\Big(1-\frac{2}{x^{2}}\Big)\ucan-\frac{\rhohat}{x}\Big(\vcan'-\frac{2}{x}\vcan\Big)&=0\,,\notag\\
\vcan''+\Big(1+\frac{\mhat^{2}-2}{x^{2}}\Big)\vcan+\frac{\rhohat}{x}\Big(\ucan'+\frac{1}{x}\ucan\Big)&=0\,.
\end{align}
Without mixing, these are the familiar equations of a massless and of a massive field in de Sitter space, with the term $-2/x^2$ coming from the cosmic expansion. The mixing enters through the terms proportional to $\rhohat/x$ and $\rhohat/x^2$. Comparing them with the frequency term, of order one, shows that the mixing is a small correction to the frequency at very early times, $x\gg\rhohat$, and dominant at late times, $x\ll\rhohat$. The difficulty of the strongly mixed regime lies in the transition between these two limits.

Let us also mention that the two equations can be decoupled at the price of raising their order. Each field then obeys a fourth-order equation, and $\sigma$ is determined algebraically by $\pic$ and its derivatives. These equations, which are used in appendix~\ref{app:series}, are given in appendix~\ref{app:fourth-order}. An equivalent decoupling was used in~\cite{Huenupi:2026abj}.

\subsection{Field basis}
\label{sec:field-basis}

The mixing $\rho\,\dot{\pi}_c\,\sigma$ of~\eqref{eq:lagrangian} is a kinetic one, with a single time derivative, and such a coupling can always be removed by a time-dependent rotation of the fields. This change of basis only relies on the form of the Lagrangian, and it will help interpret several results of this paper.

We define two fields $\varphi_1$ and $\varphi_2$ by
\begin{equation}
\label{eq:rotation}
\pic+\ii\,\sigma=\ee^{\ii\rho t/2}\,\big(\varphi_1+\ii\,\varphi_2\big)\,,
\end{equation}
with $t$ the cosmic time, so that at each time the two pairs of fields are related by a rotation of angle $\rho t/2$. In terms of the new fields, the Lagrangian~\eqref{eq:lagrangian} reads, up to a total derivative,
\begin{equation}
\label{eq:lagrangian-fixed}
\mathcal L/a^{3}=\sum_{A=1,2}\bigg[\frac12\dot\varphi_A^{2}-\frac{(\partial_i\varphi_A)^{2}}{2a^{2}}-\frac12\Big(\frac{\rho}{2}\Big)^{2}\varphi_A^{2}\bigg]-\frac32\,H\rho\,\pic\,\sigma-\frac12 m^{2}\sigma^{2}\,.
\end{equation}
The kinetic mixing has disappeared, and the sum describes two decoupled free fields with the same mass $\rho/2$. This sum is invariant under rotations in the plane of the two fields, a property that relies on their common sound speed, here equal to one. By contrast, the last two terms, which we have kept in terms of $\pic$ and $\sigma$, are not invariant, and they still couple the two fields.

This rotation has a familiar meaning in the best-known realisation of the Lagrangian~\eqref{eq:lagrangian}, a two-field model of inflation whose trajectory turns in field space (section~\ref{sec:theory}). There, the adiabatic fluctuation $\pic$ and the entropic fluctuation $\sigma$ are the fluctuations along the background trajectory and orthogonal to it. The basis that they define turns with the trajectory, at the angular velocity $\rho/2$, and the kinetic mixing is the trace that this rotation leaves in the Lagrangian. The fields $\varphi_1$ and $\varphi_2$ are then the fluctuations of the original fields, along directions that do not turn (parallel-transported along the trajectory if the field space is curved). Borrowing the language of this case, we call $(\varphi_1,\varphi_2)$ the \emph{field basis} and $(\pic,\sigma)$ the rotating basis, independently of the origin of the Lagrangian.

The value of the common mass has a simple origin. The Lagrangian~\eqref{eq:lagrangian} depends on $\pic$ through its derivatives only, so that a constant and uniform $\pic$, with $\sigma=0$, is a solution, corresponding to a constant curvature perturbation on super-horizon scales. In the field basis, this configuration is represented by a vector of constant length that rotates at the angular velocity $\rho/2$, $\varphi_1+\ii\,\varphi_2\propto\ee^{-\ii\rho t/2}$, and the common mass of $\varphi_1$ and $\varphi_2$ is precisely the one that allows this motion. The last two terms of~\eqref{eq:lagrangian-fixed} do not alter this conclusion. Indeed, the mass term exerts no force on a configuration with $\sigma=0$, and the term proportional to $H\rho$ exactly compensates the Hubble friction on the rotating vector. Equivalently, going to the rotating basis subtracts $\rho^2/4$ from the two diagonal mass terms, the centrifugal contribution familiar from the entropic mass in multifield inflation~\cite{GrootNibbelink:2001qt}, and leaves none for $\pic$.

For a Fourier mode, $\rho t/2=-(\rhohat/2)\log x$ up to a constant, which can be absorbed in a constant rotation of $\varphi_1$ and $\varphi_2$, and~\eqref{eq:rotation} reads
\begin{equation}
\label{eq:rotation-x}
\pic\pm\ii\,\sigma=x^{\mp\ii\rhohat/2}\,\big(\varphi_1\pm\ii\,\varphi_2\big)\,.
\end{equation}
When the last two terms of~\eqref{eq:lagrangian-fixed} are left aside, each of the two fields obeys the equation of a free field of mass $\rho/2$ in de Sitter space,
\begin{equation}
\label{eq:field-eom}
\Big(\frac{\varphi_A}{x}\Big)''+\Big(1+\frac{\rhohat^{2}/4-2}{x^{2}}\Big)\frac{\varphi_A}{x}=0\,,\qquad A=1,2\,.
\end{equation}
We come back to the role of these two terms in section~\ref{sec:rotating}. In the following we keep working with $\pic$ and $\sigma$, the fields in terms of which the curvature perturbation and the interactions are defined, and we use the field basis to interpret several results below.

\section{From early to late times: the connection problem}
\label{sec:connection}

Among the solutions of the equations of motion~\eqref{eq:system}, two sets play a special role. The first is adapted to early times, $x\to\infty$, where the vacuum is defined, and consists of solutions that reduce there to a single positive-frequency oscillation. The second is adapted to late times, $x\to0$, where the correlation functions are evaluated, and consists of solutions that behave there as a single power of $x$. A solution of the first set is in general a superposition of solutions of the second. In this section we define the two sets of solutions of the mixed system, and we formulate the question that the rest of the paper answers, namely how the former decompose in terms of the latter.

\subsection{Early times: channels and the Bunch-Davies vacuum}
\label{sec:channels}

The way to quantise the system is well known~\cite{GrootNibbelink:2001qt}, and it is transparent in the field basis~\cite{Cremonini:2010ua,Assassi:2013gxa,An:2017hlx,Belrhali:2026uxn}. Deep inside the horizon, $x\gg\rhohat$, all the mass terms of~\eqref{eq:lagrangian-fixed} are negligible, and $\varphi_1$ and $\varphi_2$ are two free massless fields, whose positive-frequency mode functions are $x\,\ee^{\ii x}$ up to corrections in $1/x$. Two such fields are quantised with two independent quanta, one for each of two orthogonal polarisations in the plane $(\varphi_1,\varphi_2)$. The two circular polarisations, $\varphi_2=-\ii a\,\varphi_1$ with $a=\pm1$, are singled out by the rotation~\eqref{eq:rotation-x}, which only multiplies them by a phase. For these polarisations, $\pic-\ii a\,\sigma=0$ and
\begin{equation}
\label{eq:earlytime}
\pic+\ii a\,\sigma\;\propto\;x\,\ee^{\ii x}\,x^{-\ii a\rhohat/2}\qquad (x\to\infty)\,.
\end{equation}
The mixing shifts the frequency $k/a(t)$ of these oscillations by the constant $-a\rho/2$. Integrated in time, this shift produces the logarithmic phase $x^{-\ii a\rhohat/2}$ of~\eqref{eq:earlytime}, of opposite signs for the two polarisations, which is precisely the phase introduced by the rotation~\eqref{eq:rotation-x}. Following~\cite{Belrhali:2026uxn}, we call these two degrees of freedom the \emph{channels} of the mixed system, and expand both fields on the same pair of quanta,
\begin{equation}
\label{eq:quanta}
X_{\bm k}=\sum_{a=\pm1}\Big[\,X_a(x)\,\hat b_{a,\bm k}+X^{*}_a(x)\,\hat b^{\dagger}_{a,-\bm k}\Big]\,,\qquad X=\pic,\sigma\,,
\end{equation}
with $[\hat b_{a,\bm k},\hat b^{\dagger}_{b,\bm k'}]=(2\pi)^{3}\delta_{ab}\,\delta^{(3)}(\bm k-\bm k')$, the other commutators vanishing, $[\hat b_{a,\bm k},\hat b_{b,\bm k'}]=[\hat b^{\dagger}_{a,\bm k},\hat b^{\dagger}_{b,\bm k'}]=0$. The mode functions $(\pi_{c,a},\sigma_a)$ form the channel-$a$ solution of~\eqref{eq:system}, the solution that reduces to the positive-frequency wave~\eqref{eq:earlytime} at early times with the other combination vanishing,
\begin{equation}
\label{eq:BD}
\pi_{c,a}=N_a\,x^{1-\ii a\rhohat/2}\,\ee^{\ii x}\big[1+\mathcal O(1/x)\big]\,,\qquad
\sigma_a=-\ii a\,\pi_{c,a}\big[1+\mathcal O(1/x)\big]\,,\qquad x\to\infty\,.
\end{equation}
These conditions define the Bunch-Davies vacuum of the mixed system. Two-point correlation functions are expressed as sums over the channels, $\langle X_{\bm k}(x_1)Y_{\bm k'}(x_2)\rangle=(2\pi)^{3}\delta^{(3)}(\bm k+\bm k')\sum_{a}X_a(x_1)Y^{*}_a(x_2)$.

The normalisation $N_a$ follows from that of the free fields $\varphi_1$ and $\varphi_2$. At early times each of them has the mode functions of a free massless field, of squared modulus $H^2x^2/(2k^3)$. A quantum of circular polarisation, of unit polarisation vector $(1,-\ii a)/\sqrt2$, is shared equally between the two fields, so that $|\varphi_{1,a}|^2=H^2x^2/(4k^3)$. Since $\pi_{c,a}=x^{-\ii a\rhohat/2}\varphi_{1,a}$ (see~\eqref{eq:rotation-x}), one obtains $|N_a|^2=H^2/(4k^3)$. The two channels are thus equally excited in the vacuum. The phases of the $N_a$ are conventional, and we take
\begin{equation}
\label{eq:Na}
N_+=N_-=\frac{H}{2k^{3/2}}\,.
\end{equation}

\paragraph{Commutation relations.} The canonical commutation relations hold at all times, and they constrain the mode functions in a way that we use throughout. For two solutions $f$ and $g$ of~\eqref{eq:system}, with canonical variables $(u_f,v_f)$ and $(u_g,v_g)$ defined as in~\eqref{eq:canonical}, the Wronskian
\begin{equation}
\label{eq:wronskian}
W[f,g]\equiv u_f\,u_g'-u_f'\,u_g+v_f\,v_g'-v_f'\,v_g-\frac{\rhohat}{x}\,\big(u_f\,v_g-v_f\,u_g\big)
\end{equation}
is conserved by the system~\eqref{eq:canonical-system}. In terms of it, the commutation relations between the two fields and their conjugate momenta are equivalent to the following relations between the channels $f_a=(\pi_{c,a},\sigma_a)$ and their complex conjugates,
\begin{equation}
\label{eq:exact-wronskian}
W[f_a,f_b^{*}]=-4\ii\,|N_a|^{2}\,\delta_{ab}\,,\qquad W[f_a,f_b]=0\,.
\end{equation}
They hold at early times, as one checks with~\eqref{eq:BD}, and therefore at all times. For $a=b$, the first relation fixes $|N_a|^2$, and gives back the value found above. The others relate the two channels, and they constrain their exponentially small coefficients in sections~\ref{sec:stokes} and~\ref{sec:reflection-stokes}. We will also use one of the commutation relations in its original form (section~\ref{sec:reflection-mixed}), namely that the two fields commute with each other at equal times, which reads
\begin{equation}
\label{eq:ccr}
\sum_{a=\pm}\Big[\pi_{c,a}(x)\,\sigma_a^{*}(x)-\pi_{c,a}^{*}(x)\,\sigma_a(x)\Big]=0
\end{equation}
at all times.

\subsection{Late times: light and heavy waves}
\label{sec:waves}

\paragraph{Late-time solutions.} Near $x=0$, we look for solutions of~\eqref{eq:system} that behave as a single power, $(\pic,\sigma)\simeq(p,q)\,x^{s}$ with two constants $p$ and $q$. Inserting this form in~\eqref{eq:system} gives four possible exponents,
\begin{equation}
\label{eq:exponents}
s=0\,,\qquad s=3\,,\qquad s=\tfrac32+\ii\mueff\,,\qquad s=\tfrac32-\ii\mueff\,,
\end{equation}
where
\begin{equation}
\label{eq:mueff}
\mueff\equiv\sqrt{\rhohat^{2}+\mhat^{2}-\tfrac94}=\rhohat+\frac{\mhat^2-9/4}{2\rhohat}+\mathcal O(\rhohat^{-3})
\end{equation}
is real in the regime~\eqref{eq:regime}. Since the equations are invariant under $x\to-x$, each of these powers is multiplied by a series in powers of $x^{2}$, and the equations fix all the coefficients of this series, as no two exponents differ by an even integer. Each exponent thus defines a single solution, up to its normalisation. For $s=0$ the equations give $q=0$, and the solution is constant, with $\pic\to1$ while $\sigma$ vanishes as $x^{2}$. The exponent $s=3$ gives a decaying solution, and the last two exponents give two oscillating solutions, complex conjugates of each other. These four solutions form a basis, on which any solution of~\eqref{eq:system} can be decomposed with constant coefficients.

The constant solution is a curvature perturbation frozen on super-horizon scales, as for a massless field. By contrast, the two oscillating solutions decay as $a^{-3/2}$ while oscillating at the frequency $\mueff H$, as a free field of mass $m_{\rm eff}$ would, with $m_{\rm eff}^2=m^2+\rho^2$. This effective mass has long been understood to govern the late-time dynamics of the system~\cite{Castillo:2013sfa,An:2017hlx,Iyer:2017qzw,Werth:2023pfl,Pinol:2023oux}. Since $x^{\ii\mueff}\propto\ee^{-\ii\mueff Ht}$, the solution of exponent $3/2+\ii\mueff$ has a positive frequency, and its complex conjugate a negative one.

\paragraph{Late-time coefficients.} For $x\to0$, we write the decomposition of the channel $f_a=(\pi_{c,a},\sigma_a)$ in terms of these four solutions as
\begin{equation}
\label{eq:late-coefficients}
\begin{aligned}
\sigma_a(x)&\to T_a\,x^{3/2+\ii\mueff}+\bar T_a\,x^{3/2-\ii\mueff}\,,\\[4pt]
\pi_{c,a}(x)&\to\pi_{c,a}(0)+\rhohat\,\bigg[\frac{T_a\,x^{3/2+\ii\mueff}}{3/2+\ii\mueff}+\frac{\bar T_a\,x^{3/2-\ii\mueff}}{3/2-\ii\mueff}\bigg]\,,
\end{aligned}
\end{equation}
up to non-oscillating terms that decay as $x^2$ or faster, among which the decaying solution, whose coefficient we will not need. The oscillating tails of $\pic$ follow from those of $\sigma$ by the second equation of~\eqref{eq:system}. Each channel is thus characterised at late times by three coefficients, namely its curvature perturbation $\pi_{c,a}(0)$, and the amplitudes $T_a$ and $\bar T_a$ of its positive-frequency and negative-frequency oscillations, where $\bar T_a$ is an independent coefficient, not the complex conjugate of $T_a$.

\paragraph{Light and heavy waves.} As we will show in sections~\ref{sec:regimes} to~\ref{sec:matching}, the two channels behave very differently at late times, at all orders in $1/\rhohat$. The channel $a=+1$ contains neither of the two oscillating solutions. It is a combination of the constant and decaying solutions, and it freezes to a constant curvature perturbation. By contrast, the channel $a=-1$ is the oscillating solution of exponent $3/2+\ii\mueff$ alone, a positive-frequency oscillation of the massive field. We therefore call them the \emph{light channel} and the \emph{heavy channel}, and we use them to define two exact solutions by their behaviour at late times. The \emph{heavy wave}, denoted by $f_\brh=(\pi_{c,\brh},\sigma_\brh)$, is the oscillating solution of exponent $3/2+\ii\mueff$, with the normalisation that it has in the heavy channel. The \emph{light wave}, denoted by $f_\brl=(\pi_{c,\brl},\sigma_\brl)$, is the light channel without its two oscillating tails, which are exponentially small (sections~\ref{sec:beyond} and~\ref{sec:reflection}), that is, its part along the constant and decaying solutions. At late times,
\begin{equation}
\label{eq:waves-late}
\sigma_\brh(x)\to T_-\,x^{3/2+\ii\mueff}\,,\qquad \pi_{c,\brl}(x)\to\pi_{c,+}(0)\,.
\end{equation}
The complex conjugate $f_\brh^{*}$ of the heavy wave is the oscillating solution of exponent $3/2-\ii\mueff$, and $f_\brl^{*}$ is another combination of the constant and decaying solutions. The four solutions $f_\brl$, $f_\brh$, $f_\brl^{*}$ and $f_\brh^{*}$ thus form a basis adapted to late times.

\subsection{Statement of the problem and strategy}
\label{sec:problem}

\begin{figure}[tp]
\centering
\includegraphics[width=0.92\textwidth]{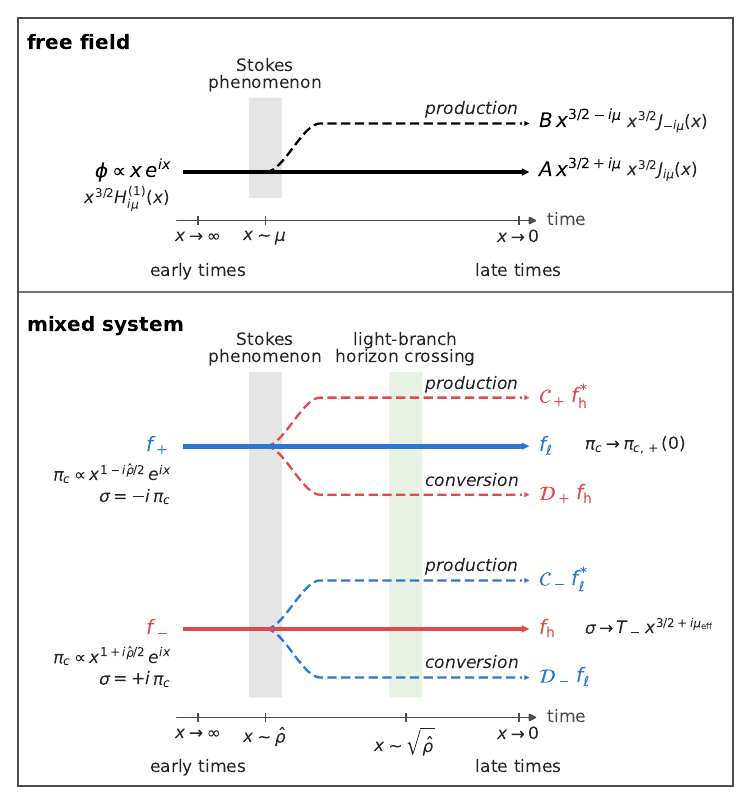}
\caption{The connection problem, for a free field (\emph{top}) and for the mixed system (\emph{bottom}). Time runs from left to right, as $x=-k\tau$ decreases. A solution defined at early times, where it is a single positive-frequency wave, is decomposed on the solutions defined at late times by their power of $x$. \emph{Top}: the Hankel function of the Bunch-Davies vacuum becomes two powers, the leading terms of two Bessel functions (section~\ref{sec:reflection-argument}). The dashed one, exponentially small, measures particle production. \emph{Bottom}: the decomposition~\eqref{eq:connection-problem} of the two channels (light in blue, heavy in red). Solid lines are given by the expansion in powers of $1/\rhohat$, with $\rhohat=\rho/H$, and dashed lines are exponentially small in $\rhohat$. The production ($\mathcal C_\pm$) is acquired across the Stokes line of the branch points (section~\ref{sec:stokes}), and the conversion ($\mathcal D_\pm$) in the same range, without a branch point of its own (section~\ref{sec:production}). Both arise around $x\sim\rhohat$, well before the horizon crossing of the light branch. The constant $\mathcal E$ of~\eqref{eq:connection-problem}, even smaller, is not shown.}
\label{fig:connection}
\end{figure}

\paragraph{Channels and waves.} Channels and waves are two sets of exact solutions of~\eqref{eq:system}, defined at early and at late times respectively, which should not be confused. In the language of particle production, the channels are the in-modes of the problem and the waves its out-modes. In the late-time basis of section~\ref{sec:waves}, the two channels read
\begin{equation}
\label{eq:connection-problem}
\begin{aligned}
f_+&=f_\brl+\mathcal C_+\,f_\brh^{*}+\mathcal D_+\,f_\brh\,,\\[4pt]
f_-&=f_\brh+\mathcal C_-\,f_\brl^{*}+\mathcal D_-\,f_\brl+\mathcal E\,f_\brh^{*}\,,
\end{aligned}
\end{equation}
with five constants. From the late-time behaviour~\eqref{eq:late-coefficients} and the normalisation~\eqref{eq:waves-late} of the heavy wave, the oscillating tails of the two channels give $\mathcal C_+=\bar T_+/T_-^{*}$, $\mathcal D_+=T_+/T_-$ and $\mathcal E=\bar T_-/T_-^{*}$, and the constants $\mathcal C_-$ and $\mathcal D_-$ are the coefficients of the non-oscillating part of the heavy channel on $f_\brl^{*}$ and $f_\brl$. The connection problem for the mixed system consists in determining the two waves, in particular their late-time coefficients $\pi_{c,+}(0)$ and $T_-$, and the five constants of~\eqref{eq:connection-problem}, which measure by how much each channel differs from a single wave. This decomposition is represented in the bottom panel of figure~\ref{fig:connection}.

\paragraph{Free field.} The best-known connection problem is that of a free massive field in de Sitter space, to which we come back in section~\ref{sec:reflection-argument}. Its Bunch-Davies mode function is proportional to a Hankel function, $x^{3/2}H^{(1)}_{\ii\mu}(x)$ with $\mu=\sqrt{\mhat^{2}-9/4}$, which is a single positive-frequency wave at early times. Its late-time solutions are Bessel functions, $x^{3/2}J_{\pm\ii\mu}(x)$, each of which is a power $x^{3/2\pm\ii\mu}$ times a series in powers of $x^{2}$. The connection formula that expresses the Hankel function in terms of these two Bessel functions~\cite{NIST:DLMF} gives the coefficients of the positive-frequency and negative-frequency waves at late times, which are the Bogoliubov coefficients describing particle production (top panel of figure~\ref{fig:connection}). The problem is solved in this case because its solutions are known special functions. A closed-form solution is however not needed to pose the problem, nor to solve it.

\paragraph{Strategy.} For the mixed system, we proceed in three steps. We first solve the system as an expansion in powers of $1/\rhohat$ (sections~\ref{sec:regimes} to~\ref{sec:matching}). As we will show, each channel reduces to a single wave at all orders of this expansion, $f_+=f_\brl$ and $f_-=f_\brh$. In this description the five constants of~\eqref{eq:connection-problem} vanish, and the expansion determines the two waves, including $\pi_{c,+}(0)$ and $T_-$. The five constants are therefore effects beyond all orders, exponentially small in $\rhohat$ (dashed lines in figure~\ref{fig:connection}). In a second step, we explain where they come from (section~\ref{sec:beyond}), and we obtain them exactly from a symmetry of the equations of motion and the canonical commutation relations (section~\ref{sec:reflection}). These exact relations fix the five constants in terms of the mass, the mixing and the phase of $\pi_{c,+}(0)$, but they do not give the two waves themselves. The two steps are thus complementary. The expansion describes the dynamics. It gives the waves at all times, including the phases accumulated by the two branches, on which the constants and the observables depend, and it shows where and when the exponentially small effects arise. The symmetry, for its part, gives the exact relations between the channels and the waves. Finally, we combine the two steps to compute the cosmological collider signal (section~\ref{sec:collider}). This signal originates from the oscillating tails of the light channel, given by the symmetry, and from a resonance at the scale of the heavy mass, described by the expansion.

\subsection{Exact solution as a reference}
\label{sec:exact}

The mixed system also happens to have a closed-form solution, recently obtained for arbitrary $\rhohat$ and $\mhat$ in~\cite{Huenupi:2026abj} and, by a different method, in~\cite{Belrhali:2026uxn}. We shall use the exact solution as the reference against which the approximations of this paper are measured (appendix~\ref{app:numerics}), and we recall the form it takes in the Laplace-space representation, which is the most convenient for our purposes. This representation, introduced recently in~\cite{Belrhali:2026ygh,Belrhali:2026jqe} and applied to the present system in~\cite{Belrhali:2026uxn}, writes a mode function as a continuous superposition of flat-space plane waves, everything specific to the theory being carried by the spectral density of that superposition. In this representation, the channel-$a$ mode function of the massive field and its spectral density read
\begin{equation}
\label{eq:sigmaray}
\sigma_a(x)=-a\,N_a\,\kappa_a\;x^{2}\!\int_1^{\infty}\!\dd\lambda\;B_a(\lambda)\,\ee^{\ii\lambda x}\,,
\qquad
B_a(\lambda)=\Ga(1-2s)\,P^{\,2s}_{\nueff-\frac12}(\lambda)\,,
\end{equation}
with $P^\mu_\nu$ the associated Legendre function, and
\begin{equation}
\label{eq:kernel-params}
s=-\frac{a\,\ii\rhohat}{2}\,,\qquad \nueff^{2}=\frac94-(\mhat^{2}+\rhohat^{2})\,,\qquad
\kappa_a=\frac{\ee^{a\pi\rhohat/4}\,2^{-s}}{\Ga(1-s)}\,.
\end{equation}
The equations of motion then determine $\pic$ algebraically from $\sigma$, and one finds
\begin{equation}
\label{eq:piray}
\pi_{c,a}(x)=\frac{a\,N_a\,\kappa_a}{\rhohat}\int_1^{\infty}\!\dd\lambda\;B_a(\lambda)\,\ee^{\ii\lambda x}\Big[\ii\lambda(\lambda^{2}-1)\,x^{3}+(3\lambda^{2}-1)\,x^{2}+\big(\mhat^{2}+\rhohat^{2}-2\big)(1-\ii\lambda x)\Big]\,.
\end{equation}
In the regime~\eqref{eq:regime}, $\nueff$ is imaginary, $\nueff=\ii\mueff$ with $\mueff$ given in~\eqref{eq:mueff}. The late-time coefficients~\eqref{eq:late-coefficients} of the channels follow from this representation in closed form~\cite{Belrhali:2026uxn}. The constant $\pi_{c,a}(0)$ follows from the integral of the spectral density over $\lambda$, and the two oscillating tails from its behaviour at large $\lambda$ (appendix~\ref{app:numerics}). With the normalisation~\eqref{eq:BD} and~\eqref{eq:Na} of the channels, the late-time curvature perturbation reads
\begin{equation}
\label{eq:pi-exact}
\pi_{c,a}(0)=\ii\,N_a\,\ee^{a\pi\rhohat/4}\,2^{s}\,\frac{\Ga(1-2s)}{\Ga(1-s)}\;
\frac{\Ga\big(\tfrac34-\tfrac{\nueff}{2}\big)\,\Ga\big(\tfrac34+\tfrac{\nueff}{2}\big)}
{\Ga\big(\tfrac34-\tfrac{\nueff}{2}-s\big)\,\Ga\big(\tfrac34+\tfrac{\nueff}{2}-s\big)}\,.
\end{equation}

Two consequences of~\eqref{eq:pi-exact} will serve as references in the following, one for each kind of quantity in the decomposition~\eqref{eq:connection-problem}. Since $|\Ga(1-2s)/\Ga(1-s)|^2=1/\cosh(\pi\rhohat/2)$, and since the last factor of~\eqref{eq:pi-exact} for one channel is the complex conjugate of that for the other, the two late-time values differ in modulus by the factor $\ee^{a\pi\rhohat/4}$ only. The first consequence is therefore an exponentially small ratio between the late-time values of the curvature mode functions of the two channels,
\begin{equation}
\label{eq:asymmetry}
\frac{|\pi_{c,-}(0)|}{|\pi_{c,+}(0)|}=\ee^{-\pi\rhohat/2}\,,
\end{equation}
for any mass. The light channel thus carries essentially all of the late-time curvature perturbation, and the heavy channel contributes at the level $\ee^{-\pi\rhohat/2}$ only, through the constants $\mathcal C_-$ and $\mathcal D_-$ of~\eqref{eq:connection-problem}. No expansion in powers of $1/\rhohat$ can produce such a factor, which we explain in section~\ref{sec:beyond} and derive in section~\ref{sec:reflection}. The second consequence is the power spectrum of the curvature perturbation, a sum over the two channels, which is amplified with respect to its value in the absence of mixing, $\Delta^2_{\zeta,0}=(4\pi^2)^{-1}(H/f_\pi)^4$,
\begin{equation}
\label{eq:R-def}
\Delta^2_\zeta=\frac{k^3}{2\pi^2}\,\frac{H^2}{f_\pi^4}\sum_a|\pi_{c,a}(0)|^2\equiv R\,\Delta^2_{\zeta,0}\,,
\end{equation}
where, using~\eqref{eq:pi-exact} and~\eqref{eq:Na}, the amplification factor is the squared modulus of the last factor of~\eqref{eq:pi-exact},
\begin{equation}
\label{eq:R}
R=\left|\frac{\Ga\big(\tfrac34-\tfrac{\ii\mueff}{2}\big)\Ga\big(\tfrac34+\tfrac{\ii\mueff}{2}\big)}
{\Ga\big(\tfrac34-\tfrac{\ii\mueff}{2}+\tfrac{\ii\rhohat}{2}\big)\Ga\big(\tfrac34+\tfrac{\ii\mueff}{2}+\tfrac{\ii\rhohat}{2}\big)}\right|^{2}\,,
\end{equation}
in agreement with~\cite{Huenupi:2026abj,Pinol:2026xnl}. Unlike~\eqref{eq:asymmetry}, $R$ has an expansion in powers of $1/\rhohat$ in the regime~\eqref{eq:regime}, against which the results of section~\ref{sec:matching} will be compared.

\section{Two branches and three regimes}
\label{sec:regimes}

We now turn to the first step of the strategy of section~\ref{sec:problem}, the solution of the system as an expansion in powers of $1/\rhohat$, which occupies this section and the next three. We start from the local dispersion relation, which has two branches, and from the three regimes that it defines in the evolution of a mode.

\subsection{Local dispersion relation}
\label{sec:dispersion}

We first look for solutions that oscillate rapidly compared with the cosmic expansion, that is, whose physical frequency $\omega$ is much larger than $H$. Writing $(\ucan,\vcan)=(U,V)\ee^{\ii S(x)}$ with slowly varying amplitudes, the physical frequency is $\omega=xHS'$, and we denote $\hat\omega\equiv\omega/H=xS'$. Keeping in the equations of motion~\eqref{eq:canonical-system} only the terms in $S'^2$, $1$ and $\rhohat S'/x$, the others being suppressed by powers of $1/\hat\omega$ in the regime~\eqref{eq:regime}, one finds the algebraic system
\begin{equation}
\label{eq:dispersion-system}
(x^{2}-\hat\omega^{2})\,U-\ii\rhohat\,\hat\omega\,V=0\,,\qquad
(x^{2}-\hat\omega^{2})\,V+\ii\rhohat\,\hat\omega\,U=0\,,
\end{equation}
whose determinant vanishes for $\hat\omega^2-x^2=\pm\rhohat\,\hat\omega$, that is $\omega^2-k^2/a^2=\pm\rho\,\omega$. The local dispersion relation of the mixed system therefore has two positive branches, labelled $\brh$ and $\brl$ for reasons given below,
\begin{equation}
\label{eq:omega-pm}
\boxed{\;\hat\omega_{\brh,\brl}(x)=\sqrt{x^{2}+\frac{\rhohat^{2}}{4}}\;\pm\;\frac{\rhohat}{2}\,,\qquad\text{i.e.}\qquad \omega_{\brh,\brl}=\sqrt{\frac{k^{2}}{a^{2}}+\frac{\rho^{2}}{4}}\;\pm\;\frac{\rho}{2}\;}
\end{equation}
with the upper sign for $\brh$ and the lower one for $\brl$, satisfying $\omh\oml=k^2/a^2$. On the branch $\omh$, the system~\eqref{eq:dispersion-system} imposes $\sigma=+\ii\pic$, and on the branch $\oml$, $\sigma=-\ii\pic$.

The two limits of~\eqref{eq:omega-pm} are transparent. At early times, $x\gg\rhohat$, $\omega_{\brh,\brl}\simeq k/a\pm\rho/2$. Both branches are then relativistic, and the mixing only shifts their frequencies by $\pm\rho/2$. By contrast, at $x\ll\rhohat$, the two branches separate widely,
\begin{equation}
\label{eq:omega-limits}
\homh\simeq\rhohat\,,\qquad \homl\simeq\frac{x^{2}}{\rhohat}\,,\qquad\text{i.e.}\qquad \omh\simeq\rho\,,\qquad \oml\simeq\frac{k^2}{a^2\rho}\,.
\end{equation}
The branch $\omh$ describes a heavy mode, of mass $\rho$, which is the mode of effective mass $m_{\rm eff}\simeq\rho$ that governs the late-time oscillations~\eqref{eq:mueff}. The branch $\oml$ describes a light mode with the quadratic dispersion relation $\omega\propto k^2$ familiar from the effective single-field theory of a curvature perturbation strongly mixed with a heavy field~\cite{Baumann:2011su,Pinol:2023oux}. We call them the \emph{heavy} and the \emph{light} branch, hence their labels.

A branch should not be confused with the channels and the waves of section~\ref{sec:connection}, which are exact solutions of~\eqref{eq:system}, defined once and for all by their behaviour at early or at late times. A branch, by contrast, is a local notion, which describes how a solution oscillates at a given time, within the adiabatic approximation. By extension, we also call light and heavy branch the solutions that follow these two roots, including after the light one stops oscillating. A priori, a channel could therefore be a superposition of the two branches, with weights that change in time. Here, however, this does not happen at any order in $1/\rhohat$, and each channel follows a single branch. At early times, where the two branches differ only by the frequency shift $\pm\rho/2$, whose integrated phase is $\int(\hat\omega_{\brh,\brl}-x)\,\dd x/x\simeq\pm(\rhohat/2)\log x$, the comparison with the phase of the channels in~\eqref{eq:earlytime} gives
\begin{equation}
\label{eq:channel-branch}
\text{channel }a=-1\;\longleftrightarrow\;\text{heavy branch}\,,\qquad
\text{channel }a=+1\;\longleftrightarrow\;\text{light branch}\,,
\end{equation}
and the relation between the two fields in each channel, $\sigma_a=-\ii a\,\pi_{c,a}$ in~\eqref{eq:BD}, is indeed that of the corresponding branch. The identification then persists at all later times, since the adiabatic solutions of the two branches, constructed in section~\ref{sec:wkb}, are not coupled at any order in $1/\rhohat$. They are coupled only through effects exponentially small in $\rhohat$ (section~\ref{sec:beyond}). The four notions used in this paper are summarised in table~\ref{tab:notions}.

\begin{table}[t]
\centering
\small\renewcommand{\arraystretch}{1.3}
\begin{tabular}{lp{6.0cm}ll}
\hline
Object & Definition & Status & Labels\\\hline
Channel & single positive-frequency wave at early times,~\eqref{eq:BD} & exact solution & $a=\pm1$\\
Wave & single oscillating power of $x$ at late times, or none (section~\ref{sec:waves}) & exact solution & $\brh$, $\brl$\\
Branch & root of the local dispersion relation~\eqref{eq:omega-pm} & adiabatic notion & $\brh$, $\brl$\\
Circular polarisation & sense of rotation in the field basis,~\eqref{eq:branches-rotation} & adiabatic notion & $a=\pm1$\\\hline
\end{tabular}
\caption{The objects of the connection problem. The channel $a=+1$, the light wave, the light branch and the polarisation that turns with the rotating basis coincide at all orders in $1/\rhohat$, and similarly for $a=-1$ and the heavy objects. We use the terms light and heavy modes when this distinction does not matter.}
\label{tab:notions}
\end{table}

\subsection{Two branches in the field basis}
\label{sec:rotating}

\begin{figure}[t]
\centering
\includegraphics[width=0.80\textwidth]{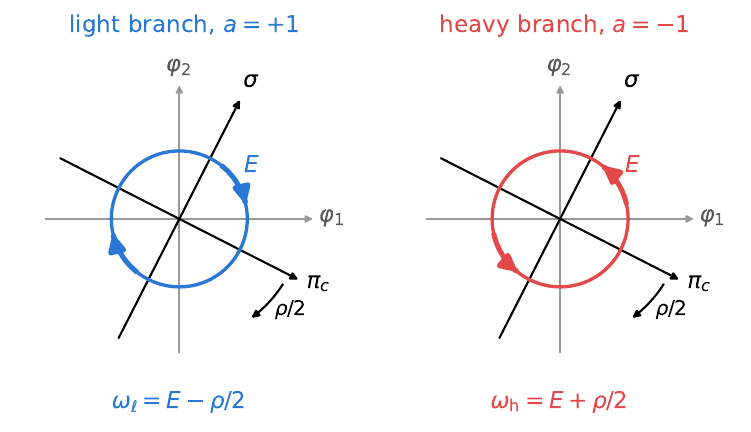}
\caption{The plane of the two fields. The rotating basis $(\pic,\sigma)$ turns with respect to the field basis $(\varphi_1,\varphi_2)$ at the angular velocity $\rho/2$, clockwise with the convention~\eqref{eq:rotation}, for which a constant $\pic$ is the vector $\varphi_1+\ii\,\varphi_2\propto\ee^{-\ii\rho t/2}$. In the adiabatic regime, $\varphi_1$ and $\varphi_2$ oscillate at the same frequency $E$ of~\eqref{eq:E-field}, and a circular polarisation describes a circle in this plane, in the same sense as the rotating basis for $a=+1$ (\emph{left}) and in the opposite sense for $a=-1$ (\emph{right}). Seen from the rotating basis, the two angular velocities are $E-\rho/2$ and $E+\rho/2$, the frequencies~\eqref{eq:omega-pm} of the light and of the heavy branch. At $k/a\ll\rho$, where $E\simeq\rho/2$, the first polarisation is almost at rest in the rotating basis, as is a constant $\pic$.}
\label{fig:bases}
\end{figure}

The dispersion relation~\eqref{eq:omega-pm} is readily interpreted in the field basis of section~\ref{sec:field-basis}. As long as the last two terms of~\eqref{eq:lagrangian-fixed} can be neglected, $\varphi_1$ and $\varphi_2$ are two decoupled fields of mass $\rho/2$. Their rapid oscillations, $\varphi_A/x\propto\ee^{\ii S(x)}$, have the frequency that follows from~\eqref{eq:field-eom},
\begin{equation}
\label{eq:E-field}
\hat{E}(x)\equiv xS'=\sqrt{x^{2}+\frac{\rhohat^{2}}{4}}\,,\qquad\text{i.e.}\qquad E=\sqrt{\frac{k^{2}}{a^{2}}+\frac{\rho^{2}}{4}}\,,
\end{equation}
that of a relativistic particle of mass $\rho/2$. Since this frequency is the same for the two fields, the two oscillations can be combined into a circular motion in the plane $(\varphi_1,\varphi_2)$, in one sense or in the other, which gives the circular polarisations $\varphi_2=-\ii a\,\varphi_1$ of section~\ref{sec:channels}. Through the rotation~\eqref{eq:rotation-x}, they correspond to $\pic=x^{-\ii a\rhohat/2}\varphi_1$ and $\sigma=-\ii a\,\pic$, and the factor $x^{-\ii a\rhohat/2}$ shifts the frequency $xS'$ by $-a\rhohat/2$,
\begin{equation}
\label{eq:branches-rotation}
\varphi_2=-\ii a\,\varphi_1\,,\quad\hat\omega=\hat{E}\qquad\Longleftrightarrow\qquad\sigma=-\ii a\,\pic\,,\quad\hat\omega=\hat{E}-\frac{a\rhohat}{2}\,.
\end{equation}
One recognises the two branches of~\eqref{eq:omega-pm}, with the relation between the two fields found there. The light branch, $a=+1$, is the circular polarisation that turns in the same sense as the rotating basis, and the heavy branch, $a=-1$, the one that turns in the opposite sense (figure~\ref{fig:bases}). The identification~\eqref{eq:channel-branch} of the channels with the branches is then immediate. Indeed, a channel is a circular polarisation at early times, it remains one as long as the Lagrangian is invariant under rotations, and a circular polarisation is a single branch.

The limits of the two branches discussed in section~\ref{sec:dispersion} are read in the same way. At early times, $k/a\gg\rho$, the mass $\rho/2$ of the two fields is negligible in~\eqref{eq:E-field}, and the rotation only shifts the frequencies $E\simeq k/a$ by $\mp\rho/2$. At $k/a\ll\rho$, on the other hand, the two fields oscillate at the frequency $\rho/2$, which is the angular velocity of the rotating basis. The polarisation that turns with the basis is then almost at rest in it, corresponding to the solution with constant $\pic$ described in section~\ref{sec:field-basis}, and the frequency of the light branch vanishes at small $k/a$. Its frequency in~\eqref{eq:omega-limits}, $\oml\simeq(k/a)^2/\rho$, is the kinetic energy $p^2/2M$ of a non-relativistic particle of momentum $p=k/a$ and mass $M=\rho/2$, the rotation having removed its rest energy. The polarisation that turns in the opposite sense has instead twice the frequency $\rho/2$, which is the mass $\rho$ of the heavy mode.

The same description can be phrased in two other languages. In mechanical terms, the mixing $\rho\,\dot{\pi}_c\,\sigma$ is the Coriolis coupling of the rotating basis. In terms of the complex field $\varphi_1+\ii\,\varphi_2$, the two circular polarisations are its particles and antiparticles, which carry opposite charges under the rotations, and going to the rotating basis shifts their energies by $\mp\rho/2$, as a chemical potential does.

Let us finally come back to the last two terms of~\eqref{eq:lagrangian-fixed}, which are not invariant under rotations and limit this description. The term proportional to $H\rho$ is due to the cosmic expansion, and it is required by the shift symmetry of $\pic$. In an expanding universe this symmetry is indeed incompatible with the invariance under rotations. If the Lagrangian were invariant under them, that is, reduced to the sum in~\eqref{eq:lagrangian-fixed}, uniform fields would obey $\ddot\varphi_A+3H\dot\varphi_A+(\rho^2/4)\,\varphi_A=0$, and the Hubble friction would make them decay, as $a^{-3/2}$. A constant and uniform $\pic$, the rotating vector of section~\ref{sec:field-basis}, would not be a solution, and the term in $H\rho$ is what restores it. It is negligible as long as the frequencies are much larger than the Hubble rate, which is the condition under which~\eqref{eq:omega-pm} was derived, and it becomes important at the horizon crossing of the light branch, where it allows the curvature perturbation to become constant.

The mass term, which only involves $\sigma$, is of relative size $\mhat^2/\rhohat$ compared with the previous one. The regime~\eqref{eq:regime} is thus the one in which the mass breaks the invariance under rotations less than the cosmic expansion does, and it is then a small correction at all times. Compared with the common mass term, it is small when $m\ll\rho$ and it dominates when $m\gg\rho$. In the first case the natural modes of the system are the two circular polarisations, that is, the channels. In the second they are $\pic$ and $\sigma$ themselves, weakly coupled by the mixing. This gives a transparent physical picture of the distinction between strong and weak mixing.

\subsection{Adiabaticity and the three regimes}
\label{sec:adiabaticity}

A branch can be followed adiabatically as long as its frequency varies slowly over a period. For the canonical variables~\eqref{eq:canonical}, on which the WKB expansion of section~\ref{sec:wkb} is built, the adiabaticity parameter is $|S''/S'^2|=|H/\omega+\dot\omega/\omega^2|$, with a dot the derivative with respect to cosmic time. Besides the variation of the frequency, this parameter includes the effect of the cosmic expansion on the canonical variables. Recalling that $\hat\omega=xS'$ and using~\eqref{eq:omega-pm}, one finds on each branch, with $\hat\omega=\homh$ or $\homl$,
\begin{equation}
\label{eq:adiabaticity}
\left|\frac{S''}{S'^{2}}\right|=\frac{\rhohat}{2\hat\omega\sqrt{x^{2}+\rhohat^{2}/4}}\;\le\;\frac{1}{\hat\omega}\,,\qquad \left|\frac{S''}{S'^{2}}\right|\simeq\begin{cases}\dfrac{\rhohat}{2x^{2}}\,, & x\gg\rhohat\,,\\[6pt] \dfrac{1}{\hat\omega}\,, & x\ll\rhohat\,.\end{cases}
\end{equation}
A branch is therefore adiabatic as long as its physical frequency is much larger than the Hubble rate, as assumed in section~\ref{sec:dispersion}. The two branches nonetheless behave very differently. The heavy branch is adiabatic at all times, since $\omh\ge\rho\gg H$, and its adiabaticity parameter is bounded by $1/\rhohat$, a bound saturated at late times. The light branch is adiabatic at early times, but its adiabaticity parameter grows as $\rhohat/x^2$ and becomes of order one when $\oml\sim H$, at
\begin{equation}
\label{eq:xstar}
x\sim\sqrt{\rhohat}\,,\qquad\text{i.e.}\qquad \frac{k^2}{a^2\rho}\sim H\,,
\end{equation}
which is the horizon crossing of the light branch. Below this point the light branch freezes, and the adiabatic description of it must be replaced by an exact solution of its equation of motion.

\begin{figure}[t]
\centering
\includegraphics[width=0.72\textwidth]{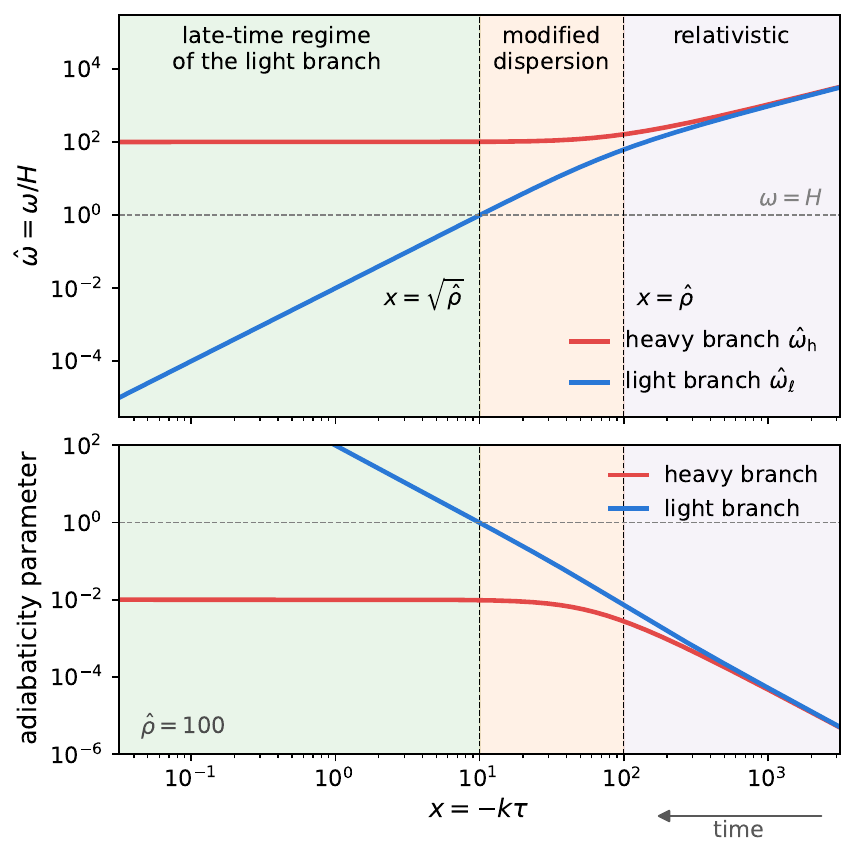}
\caption{\emph{Top}: the two branches $\hat\omega_{\brh,\brl}=\omega_{\brh,\brl}/H$ of the local dispersion relation~\eqref{eq:omega-pm}, as functions of $x=-k\tau$ for $\rhohat=100$. Time runs from right to left. The three regimes of section~\ref{sec:adiabaticity} are shaded, separated by $x=\rhohat$ and $x=\sqrt{\rhohat}$. The light branch crosses the Hubble rate, $\oml=H$ (dashed), at $x=\sqrt{1+\rhohat}$. \emph{Bottom}: the adiabaticity parameters~\eqref{eq:adiabaticity}. That of the heavy branch is bounded by $1/\rhohat$ at all times, while that of the light branch grows as $\rhohat/x^2$ and reaches unity at the horizon crossing of the light branch.}
\label{fig:regimes}
\end{figure}

The evolution of a mode of momentum $k$ therefore goes through three regimes, sketched in figure~\ref{fig:regimes}:
\begin{enumerate}[label=(\roman*),leftmargin=*]
\item the \emph{relativistic regime}, $x\gg\rhohat$, where the two branches differ only by the constant $\rho$, small compared with $k/a$, and the Bunch-Davies vacuum~\eqref{eq:BD} is set;
\item the \emph{modified dispersion regime}, $\sqrt{\rhohat}\ll x\ll\rhohat$, where the two branches have separated, the heavy one oscillating at frequency $\rho$ and the light one following the dispersion relation $\omega=k^2/(a^2\rho)$, both adiabatically;
\item the \emph{late-time regime of the light branch}, $x\lesssim\sqrt{\rhohat}$, which begins at its horizon crossing, where the light branch stops oscillating and freezes to the constant that becomes the late-time curvature perturbation, while the heavy branch keeps oscillating and decays as $x^{3/2}$.
\end{enumerate}
The transitions between these regimes take place at $x\sim\rhohat$ and $x\sim\sqrt{\rhohat}$, parametrically separated when $\rhohat\gg1$. This separation is what makes the problem solvable by matched asymptotic expansions. In their language, the adiabatic solutions form the outer expansion, at fixed $x/\rhohat$, and the effective theory of the light branch the inner one, at fixed $x^2/\rhohat$. In the adiabatic regimes, the WKB approximation applies to each branch, and we construct it in section~\ref{sec:wkb}, to second order, for both branches. Across its horizon crossing, the equation obeyed by the light branch simplifies enough to be solved exactly, in terms of a Hankel function. This equation is that of the single-field EFT, which we derive from the full two-field system in section~\ref{sec:eft}. The two descriptions of the light branch overlap in the modified dispersion regime, where they are matched in section~\ref{sec:matching}. Since the heavy branch is adiabatic throughout, its WKB solution needs no matching at all.

\section{Adiabatic regime: WKB solutions of the two branches}
\label{sec:wkb}

\subsection{Expansion and leading order}
\label{sec:wkb-leading}

We now construct the adiabatic solutions of~\eqref{eq:canonical-system} on each branch, as a systematic expansion in $1/\rhohat$. The expansion is organised in the variable
\begin{equation}
\label{eq:xi}
\xi\equiv\frac{2x}{\rhohat}\,,
\end{equation}
which is of order one in the region where the two branches separate, and in the small parameter $\varepsilon\equiv2/\rhohat$. We write
\begin{equation}
\label{eq:wkb-ansatz}
\ucan=\exp\!\Big[\frac{\ii}{\varepsilon}\,\Psi(\xi)\Big]\,,\qquad \vcan=Z(\xi)\,\ucan\,,\qquad
\begin{aligned}
\Psi'&=p_0+\varepsilon p_1+\varepsilon^2p_2+\dots\,,\\
Z&=Z_0+\varepsilon Z_1+\varepsilon^2Z_2+\dots\,,
\end{aligned}
\end{equation}
where primes now denote $\dd/\dd\xi$, so that $\Psi=\Psi_0+\varepsilon\Psi_1+\varepsilon^2\Psi_2+\dots$ with $\Psi_n'=p_n$, and we substitute into~\eqref{eq:canonical-system}, in which $\dd/\dd x=\varepsilon\,\dd/\dd\xi$ and the mixing terms $\rhohat/x$ and $\rhohat/x^2$ become $2/\xi$ and $2\varepsilon/\xi^2$. At each order in $\varepsilon$ one obtains two linear equations for the two unknowns $p_n$ and $Z_n$, with the same non-degenerate matrix at every order beyond the leading one, so that the expansion is algebraic and can be pushed as far as desired. The details are given in appendix~\ref{app:wkb}. The imaginary part of $\Psi$ gives the amplitude of the solution, and its real part gives the phase. In the variable $\xi$, the two frequencies~\eqref{eq:omega-pm} read
\begin{equation}
\label{eq:omega-xi}
\hat\omega_{\brh,\brl}=\frac{\rhohat}{2}\,(r\pm1)\,,\qquad r\equiv\sqrt{1+\xi^{2}}\,,
\end{equation}
and we also use $q\equiv r/\xi=\sqrt{1+1/\xi^{2}}$. The function
\begin{equation}
\label{eq:arcsinh}
\mathrm{arcsinh}\,\frac{\rhohat}{2x}=\log\frac{1+r}{\xi}\;\simeq\;\begin{cases}\rhohat/2x\,, & x\gg\rhohat\,,\\ \log(\rhohat/x)\,, & x\ll\rhohat\,,\end{cases}
\end{equation}
will appear repeatedly in the phases. It vanishes at early times and tends to $\log(\rhohat/x)$ at late times.

At leading order, $p_0=(r\pm1)/\xi$ and $Z_0=\pm\ii$. Since $\dd\Psi/\dd\xi=\dd S/\dd x$, we recover the dispersion relation~\eqref{eq:omega-pm}, $\dd S/\dd x=\hat\omega_{\brh,\brl}/x$, together with the relation between the two fields on each branch. At first order, one finds for both branches
\begin{equation}
\label{eq:wkb-order1}
p_1=\frac{\ii}{2}\frac{q'}{q}\,,\qquad Z_1=\pm\frac{3}{2(r\pm1)}\,,
\end{equation}
where the upper signs refer to the heavy branch and the lower ones to the light branch. Since $p_1$ is purely imaginary, the first-order term only contributes to the amplitude, $\exp(\ii\Psi_1)=q^{-1/2}$, and the phase receives no correction of order $\rhohat^{0}$. The first correction to the phase, of order $1/\rhohat$, thus comes from $p_2$, and is computed in section~\ref{sec:wkb-second}. At this order, the relation between the two fields becomes $Z=\pm\ii\pm3/[\rhohat(r\pm1)]$, with a real correction of relative size $3/[\rhohat(r\pm1)]$. On the heavy branch this correction is bounded by $3/2\rhohat$ at all times. On the light branch, by contrast, where $r-1\simeq2x^2/\rhohat^2$ at late times, it grows as $3\rhohat/2x^2$ and becomes of order one at $x\sim\sqrt{\rhohat}$, which is a first sign, within the expansion, that the adiabatic description of the light branch fails at its horizon crossing~\eqref{eq:xstar}. Integrating the phase, $\int p_0\,\dd\xi=r\pm\log\xi-\mathrm{arcsinh}(1/\xi)$ up to a constant, we define the leading phase of each channel,
\begin{equation}
\label{eq:Sa}
S_a(x)\equiv \sqrt{x^{2}+\frac{\rhohat^{2}}{4}}\;-\;a\,\frac{\rhohat}{2}\log x\;-\;\frac{\rhohat}{2}\,\ash\,,
\end{equation}
and we obtain the leading WKB mode functions of the two branches, which we write directly for the channels~\eqref{eq:channel-branch} with the normalisation~\eqref{eq:BD},
\begin{equation}
\label{eq:wkb-leading}
\pi_{c,a}(x)=N_a\;\frac{x\;\ee^{\ii S_a(x)}}{\left(1+\rhohat^{2}/4x^{2}\right)^{1/4}}\;\big[1+\mathcal O(\rhohat^{-1})\big]\,,\qquad
\sigma_a(x)=-\ii a\,\pi_{c,a}(x)\,\big[1+\mathcal O(\rhohat^{-1})\big]\,.
\end{equation}
At early times, the denominator tends to one, $\sqrt{x^2+\rhohat^2/4}\to x$ and $\ash\to0$, so that~\eqref{eq:wkb-leading} reduces to the Bunch-Davies behaviour~\eqref{eq:BD}. The corrections, given below, are of relative order $1/\hat\omega$, that is $1/\rhohat$ on the heavy branch and $\max(1/\rhohat,\rhohat/x^2)$ on the light one. Appendix~\ref{app:series} gives an independent derivation of~\eqref{eq:wkb-leading}, by resumming the expansion of the fourth-order equations of appendix~\ref{app:fourth-order} in powers of $\rhohat/x$. The present derivation explains why this resummation exists and why it does not depend on the mass. Indeed, the resummed series is the adiabatic solution of the two branches, and the mass does not enter the dispersion relation~\eqref{eq:omega-pm}.

It is instructive to follow~\eqref{eq:wkb-leading} on each branch into the modified dispersion regime, $x\ll\rhohat$, where $\sqrt{x^2+\rhohat^2/4}\simeq\rhohat/2+x^2/\rhohat$, $\ash\simeq\log(\rhohat/x)+x^2/\rhohat^2$ and the prefactor becomes $x^{3/2}(2/\rhohat)^{1/2}$. On the heavy branch, $a=-1$, the two logarithms add up and
\begin{equation}
\label{eq:wkb-heavy-late}
\pi_{c,-}(x)\;\simeq\;N_-\,\Big(\frac{2}{\rhohat}\Big)^{1/2}\,\ee^{\ii\frac{\rhohat}{2}(1-\log\rhohat)}\;x^{3/2+\ii\rhohat}\,\ee^{\ii x^{2}/2\rhohat}\,.
\end{equation}
On the light branch, $a=+1$, the two logarithms instead cancel and
\begin{equation}
\label{eq:wkb-light-mdr}
\pi_{c,+}(x)\;\simeq\;N_+\,\Big(\frac{2}{\rhohat}\Big)^{1/2}\,\ee^{\ii\frac{\rhohat}{2}(1-\log\rhohat)}\;x^{3/2}\,\ee^{\ii x^{2}/2\rhohat}\,,
\end{equation}
in which the only time-dependent phase is $x^2/2\rhohat=\int\homl\,\dd x/x$, accumulated at the frequency $\oml\simeq k^2/(a^2\rho)$ of the light branch. Both branches share the constant phase $\ee^{\ii\gamma_0}$, with $\gamma_0\equiv(\rhohat/2)(1-\log\rhohat)$. This phase records the evolution from the Bunch-Davies vacuum down to the modified dispersion regime, and the matching of section~\ref{sec:matching} will transfer it to the EFT. They also share the quadratic phase $x^2/2\rhohat$. Indeed, since $\homh-\homl=\rhohat$ exactly, the phase of the heavy branch is that of the light branch plus $\rhohat\log x$. The quadratic phase is negligible below the horizon crossing of the light branch, $x\ll\sqrt{\rhohat}$, where~\eqref{eq:wkb-heavy-late} reduces to the late-time behaviour of a heavy field of mass $\rho$.

\subsection{Second order and the limits of the adiabatic description}
\label{sec:wkb-second}

At second order, $p_2$ is real for both branches. The WKB amplitude $q^{-1/2}$ thus receives no correction at order $1/\rhohat$, and the second-order effect only modifies the phase, by $2\Psi_2/\rhohat$. Integrating $p_2$, we find (appendix~\ref{app:wkb})
\begin{equation}
\label{eq:Psi2}
\Psi_2^{(\brh,\brl)}=-\frac{4\mhat^{2}-9}{16}\,\ash+\Rpm(r)\,,\qquad
\Rpm(r)=\frac{57r^{3}\mp6r^{2}+10r\pm10}{48\,r^{3}(r\pm1)}\,,
\end{equation}
again with the upper signs for $\brh$ and the lower ones for $\brl$. The second-order phase has two parts. The first, proportional to $\ash$, is where the mass enters, and only through the combination that shifts $\rhohat$ to $\mueff$ in~\eqref{eq:mueff}. Indeed, since $2\Psi_2/\rhohat=-(\mueff-\rhohat)\ash+2\Rpm/\rhohat+\mathcal O(\rhohat^{-3})$, the coefficient $-\rhohat/2$ of $\ash$ in the leading phase~\eqref{eq:Sa} becomes $-(\mueff-\rhohat/2)$. This term vanishes at early times, leaving the Bunch-Davies behaviour untouched. At late times, where $\ash\simeq\log(\rhohat/x)$, it adds $(\mueff-\rhohat)\log x$ to the phase, that is, it shifts the frequency of both branches by the same amount, $(\mueff-\rhohat)H\simeq(m^2-9H^2/4)/2\rho$. On the heavy branch, this turns $\rho$ into $\mueff H$, and the exponent $x^{3/2+\ii\rhohat}$ of~\eqref{eq:wkb-heavy-late} into the exact one, $x^{3/2+\ii\mueff}$. On the light branch, it is a small correction to $\oml\simeq k^2/(a^2\rho)$, of relative size $(\mhat^2-9/4)/2x^2$, hence of order $\mhat^2/\rhohat$ at horizon crossing, where the phase it accumulates, $(\mueff-\rhohat)\log x$, is of order $\rhohat^{-1}\log\rhohat$. The mass enters the light branch in no other way at this order. The second part, $\Rpm$, is independent of the mass, and it alone distinguishes the two branches.

On the heavy branch, $\Rp$ is bounded, decreasing from $71/96$ at late times to zero at early times, so that the second-order phase is uniformly small, of order $1/\rhohat$. The second-order result then gives a mode function valid at all times,
\begin{equation}
\label{eq:heavy-uniform}
\pi_{c,-}(x)=\frac{N_-\,x}{\left(1+\rhohat^{2}/4x^{2}\right)^{1/4}}\,\exp\!\left\{\ii\left[S_-(x)-(\mueff-\rhohat)\ash+\frac{2\Rp(r)}{\rhohat}\right]\right\},
\end{equation}
with an error of order $\rhohat^{-2}$ in amplitude and $\rhohat^{-3}$ in phase, uniformly in $x$. It approximates the heavy wave of section~\ref{sec:waves} uniformly. For the curvature perturbation of the heavy channel, it holds down to the times at which the exponentially small constant~\eqref{eq:asymmetry}, which no order of the expansion contains, takes over. The massive field follows as $\sigma_-=Z\pi_{c,-}$ with $Z=\ii+\frac{3}{\rhohat(r+1)}+\mathcal O(\rhohat^{-2})$. With elementary functions only,~\eqref{eq:heavy-uniform} interpolates between the Bunch-Davies wave $x^{1+\ii\rhohat/2}\ee^{\ii x}$ at early times and the heavy-field decay $x^{3/2+\ii\mueff}$ at late times.

On the light branch, by contrast, $\Rm$ diverges at late times,
\begin{equation}
\label{eq:Rminus-div}
\Rm\simeq\frac{21}{8\,\xi^{2}}\qquad(\xi\to0)\,,\qquad\text{that is}\qquad \frac{2\Rm}{\rhohat}\simeq\frac{21\,\rhohat}{16\,x^{2}}\,,
\end{equation}
and the second-order phase becomes of order one at $x\sim\sqrt{\rhohat}$, like the first-order correction to $Z$. The second-order solution of the light branch,~\eqref{eq:heavy-uniform} with $a=+1$ and $\Rm$ in place of $\Rp$, therefore holds for $x\gg\sqrt{\rhohat}$ only. Across its horizon crossing, the light branch requires another description, the EFT of section~\ref{sec:eft}, to which it is matched in section~\ref{sec:matching}.

\paragraph{Comparison with exact solution.}\label{sec:wkb-numerics} We have compared the WKB solutions with the exact solution, for $\rhohat=20$, $40$ and $80$ and $\mhat^2=1$ (appendix~\ref{app:numerics}). On the heavy branch, the errors are those stated after~\eqref{eq:heavy-uniform}, over the full range of $x$. On the light branch, they are the same for $x\gg\sqrt{\rhohat}$, and they grow when $x/\sqrt{\rhohat}$ becomes of order one, as the divergence~\eqref{eq:Rminus-div} anticipates.

\section{Effective theory of the light branch}
\label{sec:eft}
\subsection{Constrained dynamics of the light branch}
\label{sec:eft-constraint}

Below $x\sim\rhohat$ the light branch oscillates slowly compared with $k/a$, at the frequency $\oml\simeq k^2/(a^2\rho)$, and below $x\sim\sqrt{\rhohat}$, where this frequency falls below $H$, it does not oscillate at all. In both cases the massive field on this branch varies slowly compared to the frequency scale $k/a$ set by the term $+1$ in its equation of motion, the second line of~\eqref{eq:system}. More precisely, on the light branch, $\sigma''$ and $\sigma'/x$ are smaller than $\sigma$ by a factor of order $\max(x^2/\rhohat^2,1/x^2)$, and the mass term $\mhat^2\sigma/x^2$ is smaller by $\mhat^2/x^2$. In the range
\begin{equation}
\label{eq:eft-range}
1\ll x\ll\rhohat\qquad(\text{and }x\gg\mhat)\,,
\end{equation}
which contains the horizon crossing $x\sim\sqrt{\rhohat}$ of the light branch, the equation of motion of $\sigma$ therefore reduces at leading order in $1/\rhohat$ to an algebraic constraint,
\begin{equation}
\label{eq:constraint}
\sigma=-\frac{\rhohat}{x}\,\pic'\,.
\end{equation}
On the light branch, the massive field thus stops being a dynamical degree of freedom and follows the curvature perturbation. As long as the light branch oscillates, $\pic'\simeq\ii(x/\rhohat)\pic$ and~\eqref{eq:constraint} reduces to the relation $\sigma=-\ii\pic$ of section~\ref{sec:dispersion}, which it extends across the horizon crossing, where the light branch stops oscillating. Inserting~\eqref{eq:constraint} into the first line of~\eqref{eq:system} and keeping the leading terms for $x\ll\rhohat$ gives a closed second-order equation for the curvature perturbation,
\begin{equation}
\label{eq:eft-eq}
\pic''-\frac{4}{x}\,\pic'+\frac{x^{2}}{\rhohat^{2}}\,\pic=0\,,
\end{equation}
in which the mixing has replaced the relativistic frequency term $\pic$ by $x^2\pic/\rhohat^2$, which gives the modified dispersion relation $\oml\simeq k^2/(a^2\rho)$ of~\eqref{eq:omega-limits}, now with the friction that governs horizon crossing. The mass has disappeared from this equation, since the term $\mhat^2\sigma/x^2$ was neglected in the constraint along with the time derivatives of $\sigma$. Keeping it gives $(1+\mhat^2/x^2)\,\sigma=-\rhohat\,\pic'/x$, so that the mass enters the light branch only at relative order $\mhat^2/\rhohat$ around horizon crossing, through the shift of the frequency $\oml$ noted in section~\ref{sec:wkb-second}. Equation~\eqref{eq:eft-eq} is the equation of motion of the effective theory with quadratic dispersion relation~\cite{Baumann:2011su,Pinol:2023oux}, obtained in~\cite{Baumann:2011su} by dropping the relativistic kinetic terms of both fields in the two-field action. Here, by contrast, it appears as the leading order of a systematic expansion in $1/\rhohat$ on the light branch of the full two-field system, which also controls its corrections and its range of validity (section~\ref{sec:eft-solution} and appendix~\ref{app:series}). The same counting also explains why the EFT describes the light branch only. On the heavy branch, which oscillates at the frequency $\rho$, $\sigma''\simeq-(\rhohat/x)^2\sigma$ exceeds $\sigma$ by a factor $\rhohat^2/x^2$. The time derivatives then dominate the equation of motion of $\sigma$ instead of being negligible, and no constraint arises.

\subsection{Solution and its two limits}
\label{sec:eft-solution}

Equation~\eqref{eq:eft-eq} is solved by Bessel functions of order $5/4$ and argument $x^2/2\rhohat$. Writing the general solution in terms of Hankel functions, and using the constraint~\eqref{eq:constraint} together with the identity $\dd[z^{\nu}H_\nu(z)]/\dd z=z^\nu H_{\nu-1}(z)$, one finds
\begin{equation}
\label{eq:eft-solution}
\begin{aligned}
\pic^{\rm eff}(x)&=x^{5/2}\Big[A_+\,H^{(1)}_{5/4}\Big(\frac{x^{2}}{2\rhohat}\Big)+A_-\,H^{(2)}_{5/4}\Big(\frac{x^{2}}{2\rhohat}\Big)\Big]\,,\\[4pt]
\sigma^{\rm eff}(x)&=-x^{5/2}\Big[A_+\,H^{(1)}_{1/4}\Big(\frac{x^{2}}{2\rhohat}\Big)+A_-\,H^{(2)}_{1/4}\Big(\frac{x^{2}}{2\rhohat}\Big)\Big]\,,
\end{aligned}
\end{equation}
with the same two constants $A_\pm$ in both fields. The EFT has two solutions where the full system has four. The two that are missing are the heavy branch, with its positive and negative frequencies, which the adiabatic solution~\eqref{eq:heavy-uniform} describes for all $x$. The mode functions of the EFT are elementary in their two limits. At early times within its range of validity, $\sqrt{\rhohat}\ll x\ll\rhohat$, the Hankel functions are in their asymptotic regime, $H^{(1,2)}_\nu(z)\simeq\sqrt{2/\pi z}\,\ee^{\pm\ii(z-\nu\pi/2-\pi/4)}$, and
\begin{equation}
\label{eq:eft-early}
\pic^{\rm eff}(x)\;\simeq\;2\sqrt{\frac{\rhohat}{\pi}}\;x^{3/2}\sum_{\pm}A_\pm\,\ee^{\pm\ii\theta(x)}\,,
\end{equation}
with $\theta(x)\equiv x^{2}/2\rhohat-7\pi/8$. These are the two WKB waves of the light branch in the modified dispersion regime, of phase $\pm x^2/2\rhohat$ and amplitude $x^{3/2}$. The positive-frequency one has the form of the adiabatic solution~\eqref{eq:wkb-light-mdr}. At late times, $x\ll\sqrt{\rhohat}$,\footnote{The constraint~\eqref{eq:constraint} was derived in the range~\eqref{eq:eft-range}, whereas~\eqref{eq:eft-solution} is used down to $x\to0$. Its use there is legitimate at leading order in $1/\rhohat$. Indeed, below the range~\eqref{eq:eft-range} the curvature perturbation is frozen and the massive field is small, $\sigma\simeq-x^2\pic/\rhohat$, so that the terms neglected in~\eqref{eq:constraint} modify the evolution only at relative order $\rhohat^{-2}$. The systematic expansion of appendix~\ref{app:series} confirms that the leading order is exactly~\eqref{eq:eft-solution} down to $x\to0$. Around horizon crossing, at fixed $x^2/\rhohat$, its corrections are of relative order $1/\rhohat$. Since the late-time constant~\eqref{eq:eft-late} is set at horizon crossing, it receives corrections of that order, and not $\rhohat^{-2}$. We will obtain the first of them from the exact solution in section~\ref{sec:matching-spectrum}.} using $H^{(1,2)}_\nu(z)\simeq\mp(\ii/\pi)\Ga(\nu)(z/2)^{-\nu}$,
\begin{equation}
\label{eq:eft-late}
\pic^{\rm eff}(x)\;\to\;-\frac{\ii}{\pi}\,\Ga\big(\tfrac54\big)\,(4\rhohat)^{5/4}\,\big(A_+-A_-\big)\,,\quad
\sigma^{\rm eff}(x)\;\to\;\frac{\ii}{\pi}\,\Ga\big(\tfrac14\big)\,(4\rhohat)^{1/4}\,\big(A_+-A_-\big)\,x^{2},
\end{equation}
the curvature perturbation freezes and the massive field decays as $x^2$, that is as $a^{-2}$. The constant in~\eqref{eq:eft-late} is the late-time curvature perturbation, fixed by the constants $A_\pm$. In the EFT one usually imposes a positive-frequency condition, $A_-=0$, and normalises the mode function by the Wronskian of the effective action~\cite{Baumann:2011su,Pinol:2023oux}. This fixes the modulus of $A_+$, but the EFT cannot tell whether this is the state inherited from the Bunch-Davies vacuum of the full system, set at $x\gg\rhohat$, nor determine the phase of $A_+$. Both follow from matching the effective solution to the adiabatic solution of the light branch in the modified dispersion regime, which is the object of the next section.

\section{Matching and uniform approximations}
\label{sec:matching}

In this section we match the descriptions of the two previous sections. The matching follows each channel from early to late times, gives the two late-time coefficients of the connection problem of section~\ref{sec:problem} that are not exponentially small, and yields an approximation of each channel that is valid at all times. We finally extend it to excited initial states.

\subsection{Matching the light branch: the vacuum of the effective theory}
\label{sec:matching-light}

In the modified dispersion regime, $\sqrt{\rhohat}\ll x\ll\rhohat$, the light branch is described both by the adiabatic solution~\eqref{eq:wkb-leading}, valid from early times, and by the effective solution~\eqref{eq:eft-solution}, valid down to late times. The two descriptions must agree there. Since errors on the frequency accumulate in the phase, the leading effective solution is in fact accurate for $x\ll\rhohat^{3/4}$ only, where its phase differs from that of the adiabatic solution by $x^4/4\rhohat^3\ll1$, so that the two leading-order descriptions overlap for $\sqrt{\rhohat}\ll x\ll\rhohat^{3/4}$. Comparing~\eqref{eq:wkb-light-mdr} with~\eqref{eq:eft-early} there determines the constants of the effective theory,
\begin{equation}
\label{eq:Apm}
A_+=N_+\,\frac{1}{\rhohat}\sqrt{\frac{\pi}{2}}\;\ee^{\ii\left(\gamma_0+7\pi/8\right)}\,,\qquad A_-=0\,,
\end{equation}
with $\gamma_0$ the constant phase of~\eqref{eq:wkb-light-mdr}.

\paragraph{Vacuum and normalisation.} Only the positive-frequency solution of the effective theory is excited, with no admixture of $H^{(2)}$ at any order in $1/\rhohat$. The positive-frequency condition that the EFT imposes as a definition of its vacuum is thus derived here, from the Bunch-Davies vacuum of the full two-field system, set deep in the relativistic regime. As a check beyond leading order, the first phase correction in the asymptotic expansion of $H^{(1)}_{5/4}(z)$ at large argument is $(4\nu^2-1)/8z$, with $\nu=5/4$ and $z=x^2/2\rhohat$, that is $21\rhohat/16x^2$, exactly the second-order phase~\eqref{eq:Rminus-div} of the light branch. The modulus of $A_+$ reproduces the normalisation obtained in the EFT from its own Wronskian~\cite{Baumann:2011su,Pinol:2023oux}.

\paragraph{Phase of the light branch.} The phase $\gamma_0$, on the other hand, is not accessible to the effective theory. This phase is accumulated by the light branch between the relativistic regime and the modified dispersion regime, as the time integral of $\oml-k/a$ across the transition at $x\sim\rhohat$, below which the EFT applies. Equivalently, $\gamma_0=S_+(0)$, the leading phase~\eqref{eq:Sa} of the light branch at the point $x=0$, where its frequency vanishes. This phase does not affect the power spectrum in the Bunch-Davies vacuum, but it enters any observable in which the light branch interferes with another wave, as we will see with the excited states of section~\ref{sec:excited}. The second-order phase of the adiabatic solution, through which the mass enters the light branch, adds to $\gamma_0$ a constant of order $\rhohat^{-1}\log\rhohat$, larger than the corrections to the EFT (see the insert below). We include this constant by replacing $\gamma_0$, in~\eqref{eq:Apm} and in what follows, with
\begin{equation}
\label{eq:gamma}
\gamma\equiv\gamma_0-\frac{\mueff-\rhohat}{2}\,\log\rhohat=\frac{\rhohat}{2}-\frac{\mueff}{2}\,\log\rhohat\,.
\end{equation}

\begin{techinsert}{Logarithmic phase}The matching above involves the leading phase of the adiabatic solution only. Its second-order phase~\eqref{eq:Psi2} also contains the phase through which the mass enters the light branch, $-(\mueff-\rhohat)\,\ash$ (section~\ref{sec:wkb-second}). The second-order phase has two parts. The first one, $2\Rm/\rhohat$, is reproduced by the Hankel function, as noted above. In the region of the matching, where $\ash\simeq\log(\rhohat/x)$, the second one reads
\begin{equation}
\label{eq:log-split}
-(\mueff-\rhohat)\,\log\frac{\rhohat}{x}=-\frac{\mueff-\rhohat}{2}\,\log\rhohat+\frac{\mueff-\rhohat}{2}\,\log\frac{x^{2}}{\rhohat}\,.
\end{equation}
The last term depends on time through $x^2/\rhohat$, the variable of the effective solution, and it is of order $1/\rhohat$ at a fixed value of this variable. It therefore belongs to the first corrections to the EFT (section~\ref{sec:eft-solution}), which we do not compute. The other term is a constant phase, which the matching transfers to $A_+$. Being of order $\rhohat^{-1}\log\rhohat$, it is larger than these corrections, and it gives~\eqref{eq:gamma}. The logarithm counts the e-folds during which the mass acts on the light branch. Indeed, the shift $(\mueff-\rhohat)H$ of its frequency (section~\ref{sec:wkb-second}) acts below $x\sim\rhohat$, until the light branch freezes at $x\sim\sqrt{\rhohat}$, that is, during $\log(\rhohat/\sqrt{\rhohat})=\tfrac12\log\rhohat$ e-folds. The heavy branch, by contrast, never freezes, and it accumulates this phase at all times, which turns its exponent into $\ii\mueff$ and leaves the constant $-(\mueff-\rhohat)\log\rhohat$ in~\eqref{eq:heavy-late} below.
\end{techinsert}

\subsection{Late-time behaviour of the two channels}
\label{sec:matching-spectrum}

\paragraph{Light channel and power spectrum.} With~\eqref{eq:Apm} and~\eqref{eq:gamma} in~\eqref{eq:eft-late}, the late-time value of the curvature mode function of the light channel is
\begin{equation}
\label{eq:pi-late}
\boxed{\;\pi_{c,+}(0)=N_+\;\frac{4\,\Ga(5/4)}{\sqrt\pi}\;\rhohat^{1/4}\;\ee^{\ii\left(\gamma+3\pi/8\right)}\;\big[1+\mathcal O(\rhohat^{-1})\big]\,.\;}
\end{equation}
Unlike the result for the heavy channel below, this is a leading-order formula, except for the logarithmic term of its phase. The heavy channel contributes nothing at any order of the expansion, since its adiabatic solution~\eqref{eq:heavy-uniform} decays as $x^{3/2}$. With~\eqref{eq:Na}, the amplification~\eqref{eq:R-def} of the power spectrum reads
\begin{equation}
\label{eq:R-leading}
R=\frac{8\,\Ga(5/4)^{2}}{\pi}\,\sqrt{\rhohat}=\frac{\pi}{\Ga(3/4)^{2}}\,\sqrt{\rhohat}\;\simeq\;2.09\,\sqrt{\rhohat}\,,
\end{equation}
the enhancement $\propto\sqrt{\rhohat}$ of the curvature power spectrum first found in~\cite{Cremonini:2010ua}, and whose coefficient follows from the mode functions of the EFT~\cite{Baumann:2011su}.\footnote{The two expressions of the coefficient are related by $\Ga(5/4)\,\Ga(3/4)=\pi\sqrt2/4$.} Equation~\eqref{eq:R-leading} is the limit of the exact result~\eqref{eq:R} at large $\rhohat$ and fixed $\mhat$, where the latter becomes\footnote{We use $|\Ga(a+\ii y)|^2=2\pi\,y^{2a-1}\ee^{-\pi y}[1+\mathcal O(y^{-2})]$ for the three Gamma functions of large argument in~\eqref{eq:R}, and $\Ga(\tfrac34+\tfrac{\ii}{2}(\rhohat-\mueff))\to\Ga(\tfrac34)$ since $\mueff-\rhohat\to0$.}
\begin{equation}
\label{eq:R-NLO}
R=\frac{\pi\sqrt{\rhohat}}{\Ga(3/4)^{2}}\;\ee^{\,\pi(\rhohat-\mueff)/2}\;\big[1+\mathcal O(\rhohat^{-2})\big]
=\frac{\pi\sqrt{\rhohat}}{\Ga(3/4)^{2}}\left[1+\frac{\pi\,(9-4\mhat^{2})}{16\,\rhohat}+\mathcal O(\rhohat^{-2})\right]\,.
\end{equation}
The exponential factor in the first expression is of order one. Indeed, at fixed mass, $\mueff-\rhohat=(\mhat^2-9/4)/2\rhohat+\mathcal O(\rhohat^{-3})$, so that this factor is a power series in $1/\rhohat$, whose first term is displayed in the second expression. This first correction to the EFT contains the mass, in the same combination $\mhat^2-9/4$ as the second-order phase~\eqref{eq:Psi2}. Within our expansion, it comes from the first corrections to the equation~\eqref{eq:eft-eq} of the EFT at horizon crossing, through which the mass enters the light branch (section~\ref{sec:eft-constraint} and appendix~\ref{app:series}). We have not computed them.

\paragraph{Heavy channel.}\label{sec:matching-heavy} The heavy channel needs no matching, and its late-time behaviour follows from~\eqref{eq:heavy-uniform} by letting $x\to0$, where $\ash\simeq\log(\rhohat/x)$,
\begin{equation}
\label{eq:heavy-late}
\begin{aligned}
\pi_{c,-}(x)\;&\to\;N_-\,\sqrt{\frac{2}{\rhohat}}\;x^{3/2+\ii\mueff}\;\exp\Big\{\ii\Big[\gamma_0-\big(\mueff-\rhohat\big)\log\rhohat+\frac{71}{48\,\rhohat}\Big]\Big\}\,,\\[4pt]
\sigma_{-}(x)\;&\to\;\Big(\ii+\frac{3}{2\rhohat}\Big)\,\pi_{c,-}(x)\,,
\end{aligned}
\end{equation}
up to corrections of relative order $\rhohat^{-2}$ in modulus and $\rhohat^{-3}$ in phase, as we have checked against the exact late-time coefficients (section~\ref{sec:exact}). The heavy channel is thus the positive-frequency oscillating solution of section~\ref{sec:waves}, with the envelope $|\sigma_-|^2\to(H^2/2k^3)\,x^3/\rhohat$.\footnote{The modulus of the oscillating part at $x\to0$ follows exactly from the Wronskian~\eqref{eq:exact-wronskian}. It gives $|\pi_{c,-}|^2\to\frac{H^2}{2k^3}\frac{\rhohat^2x^3}{\mueff(\rhohat^2+\mhat^2)}$ and $|\sigma_-|^2\to\frac{H^2}{2k^3}\frac{x^3}{\mueff}$, in agreement with~\eqref{eq:heavy-late} up to relative corrections of order $\rhohat^{-2}$ at fixed mass.} At late times, and at all orders in $1/\rhohat$, the massive field is therefore carried by the heavy channel, as a decaying oscillation, whereas in the light channel it is tied to the curvature perturbation by the constraint~\eqref{eq:constraint}, and decays as $x^2$ once the latter freezes. In the curvature perturbation, $|\pi_{c,-}(x)|^2/|\pi_{c,+}(0)|^2=\Ga(3/4)^2\,x^3/\pi\rhohat^{3/2}$. At fixed $x$, the heavy channel is thus smaller than the light one by a power of the mixing, not by an exponential, and the two are comparable at the horizon crossing of the light branch, $x\sim\sqrt{\rhohat}$, when the latter freezes. This matters when the massive field propagates on an internal line.

\paragraph{Coefficients given by the expansion.} We can now state what the expansion in powers of $1/\rhohat$ gives for the connection problem of section~\ref{sec:problem}. In the notation~\eqref{eq:late-coefficients}, the late-time behaviour~\eqref{eq:heavy-late} of the heavy channel is a single tail, of amplitude
\begin{equation}
\label{eq:T-minus}
T_-=N_-\,\Big(\ii+\frac{3}{2\rhohat}\Big)\sqrt{\frac{2}{\rhohat}}\;\exp\Big\{\ii\Big[\gamma_0-\big(\mueff-\rhohat\big)\log\rhohat+\frac{71}{48\,\rhohat}\Big]\Big\}\,,
\end{equation}
with the accuracy quoted there. With the late-time curvature perturbation~\eqref{eq:pi-late} of the light channel, this gives the two late-time coefficients that are not exponentially small, $\pi_{c,+}(0)$ and $T_-$. The other four, $\pi_{c,-}(0)$, $\bar T_+$, $T_+$ and $\bar T_-$, vanish at all orders in $1/\rhohat$, so that at all these orders each channel is a single wave in the decomposition~\eqref{eq:connection-problem}.

\subsection{Uniform approximations and their accuracy}
\label{sec:uniform}

We now assemble, for each channel, a single formula valid at all times. For the heavy channel there is nothing to assemble, since the second-order adiabatic solution~\eqref{eq:heavy-uniform} is uniformly valid. We restate it together with the massive field, $\sigma_-=Z\pi_{c,-}$ with $Z$ to first order,
\begin{equation}
\label{eq:heavy-uniform-bis}
\boxed{\;
\begin{aligned}
\pi_{c,-}^{\rm unif}(x)&=\frac{N_-\,x\;\ee^{\ii S_-(x)}}{\left(1+\rhohat^{2}/4x^{2}\right)^{1/4}}\,\exp\!\left\{\ii\left[-(\mueff-\rhohat)\ash+\frac{2\Rp(r)}{\rhohat}\right]\right\}\,,\\[4pt]
\sigma_-^{\rm unif}(x)&=\Big[\ii+\frac{3}{\rhohat\,(r+1)}\Big]\,\pi_{c,-}^{\rm unif}(x)\,.
\end{aligned}\;}
\end{equation}
For the light channel, by contrast, we have two descriptions with overlapping domains, the adiabatic solution~\eqref{eq:wkb-leading} for $x\gg\sqrt{\rhohat}$ and the effective solution~\eqref{eq:eft-solution} with the matched constants~\eqref{eq:Apm} for $x\ll\rhohat$, and they must be combined into one. The effective solution differs from its asymptotic form~\eqref{eq:eft-early} by the series of corrections $1+\ii\,21\rhohat/(16x^2)+\dots$ of the Hankel function, which matter around horizon crossing and become negligible for $x\gg\sqrt{\rhohat}$. The natural way to combine the two descriptions is to multiply by these corrections the adiabatic wave, which has the correct phase $S_+(x)$ at all times. We define the uniform approximation of the light channel as the adiabatic solution multiplied by the ratio of the effective solution to this asymptotic form, for each of the two fields,
\begin{equation}
\label{eq:light-uniform}
\boxed{\;
\begin{aligned}
\pi_{c,+}^{\rm unif}(x)&=\frac{N_+\,x\;\ee^{\ii S_+(x)}}{\left(1+\rhohat^{2}/4x^{2}\right)^{1/4}}\;h_{5/4}\Big(\frac{x^{2}}{2\rhohat}\Big)\,,\\[4pt]
\sigma_+^{\rm unif}(x)&=-\ii\,\frac{N_+\,x\;\ee^{\ii S_+(x)}}{\left(1+\rhohat^{2}/4x^{2}\right)^{1/4}}\;h_{1/4}\Big(\frac{x^{2}}{2\rhohat}\Big)\,,
\end{aligned}\;}
\end{equation}
with
\begin{equation}
\label{eq:hnu}
h_\nu(z)\equiv\sqrt{\frac{\pi z}{2}}\;\ee^{-\ii(z-\nu\pi/2-\pi/4)}\,H^{(1)}_{\nu}(z)\,,
\end{equation}
the Hankel function normalised to its asymptotic form. Since $h_\nu$ tends to one at large argument,~\eqref{eq:light-uniform} reduces to the adiabatic solution for $x\gg\sqrt{\rhohat}$. For $x\ll\rhohat$, the adiabatic prefactor reduces to the common form~\eqref{eq:wkb-light-mdr}, up to a phase $x^4/4\rhohat^3$ and an amplitude $x^2/\rhohat^2$ that are both small below $x\sim\sqrt{\rhohat}$, so that~\eqref{eq:light-uniform} reduces to the effective solution across horizon crossing and down to late times. Equations~\eqref{eq:heavy-uniform-bis} and~\eqref{eq:light-uniform} thus give each channel at all times, in terms of elementary functions and, for the light channel, one Hankel function.

\begin{figure}[t]
\centering
\includegraphics[width=0.5\textwidth]{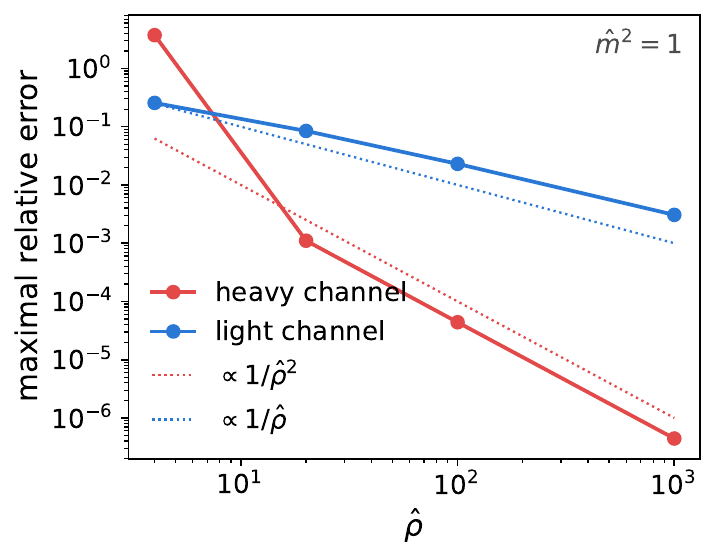}
\caption{Maximal relative error of the uniform approximations~\eqref{eq:heavy-uniform-bis} and~\eqref{eq:light-uniform} on the mode function of $\pic$, over time, as a function of the mixing, for $\mhat^2=1$. The dotted lines indicate the scalings $\rhohat^{-1}$ and $\rhohat^{-2}$. The point at $\rhohat=4$ for the heavy channel is dominated by the late-time constant of section~\ref{sec:beyond}, which the adiabatic solution misses entirely. The error on $\sigma_-$, to which this constant does not contribute, stays at the few percent level.}
\label{fig:uniform-rho}
\end{figure}

\paragraph{Accuracy.}\label{sec:uniform-accuracy} We have compared~\eqref{eq:heavy-uniform-bis} and~\eqref{eq:light-uniform} with the exact solution, channel by channel, on the mode function of $\pic$, for $\mhat^2=1$ and $\rhohat=4$, $20$, $100$ and $1000$. We summarise the results here, and leave the details, including the errors as functions of time, to appendix~\ref{app:numerics}. Figure~\ref{fig:uniform-rho} shows the maximal relative error over time as a function of the mixing.

For the heavy channel the error is of relative order $\rhohat^{-2}$, as anticipated. At $\rhohat=4$ it is dominated by late times, where the exact heavy channel freezes to the exponentially small constant~\eqref{eq:asymmetry}, discussed in section~\ref{sec:beyond}. For the light channel, the error is instead of relative order $\rhohat^{-1}$, and still $25\%$ at $\rhohat=4$, where the expansion parameter is no longer small. Finally, for the amplification of the power spectrum, the leading formula~\eqref{eq:R-leading} approaches the exact result~\eqref{eq:R} as $1/\rhohat$, with errors of $9\%$ at $\rhohat=10$ and $1\%$ at $\rhohat=100$, and the exponential form~\eqref{eq:R-NLO} as $1/\rhohat^2$, with an error of $0.3\%$ already at $\rhohat=10$ (see figure~\ref{fig:R} in appendix~\ref{app:numerics}).

\subsection{Excited initial states}
\label{sec:excited}

The phase $\gamma$ of~\eqref{eq:gamma} contributes to the phase of the cosmological collider signal (section~\ref{sec:collider}). It is already observable in the power spectrum when the initial state is excited, as we now show. A general homogeneous, pure Gaussian state of the two-field system has mode functions
\begin{equation}
\label{eq:excited-modes}
\tilde X_a=\sum_{b=\pm}\Big[\alpha_{ab}\,X_b+\beta_{ab}\,X_b^{*}\Big]\,,\qquad X=\pic,\sigma\,,
\end{equation}
where $X_b$ are the Bunch-Davies mode functions obtained in the previous sections and $\alpha$, $\beta$ are $2\times2$ complex matrices subject to the Bogoliubov conditions $\alpha\alpha^{\dagger}-\beta\beta^{\dagger}=\mathds{1}$ and $\alpha\beta^{T}=\beta\alpha^{T}$, which preserve the canonical commutation relations~\cite{GrootNibbelink:2001qt}. Since the heavy channel decays, only $\alpha_{a+}$ and $\beta_{a+}$ enter the late-time curvature perturbation, and the power spectrum is multiplied, with respect to its value in the vacuum, by (see appendix~\ref{app:excited})
\begin{equation}
\label{eq:excited-late}
1+2n_+ +2\,\mathrm{Re}\Big[\mathcal I\;\ee^{2\ii\theta_0}\Big]\,,\qquad
n_+\equiv\sum_a|\beta_{a+}|^{2}\,,\qquad
\mathcal I\equiv\sum_a\alpha_{a+}\,\beta^{*}_{a+}\,,
\end{equation}
with $\theta_0\equiv\gamma+3\pi/8$ the phase of~\eqref{eq:pi-late}. The first two terms are the enhancement by the occupation number $n_+$ of the light channel, familiar from a single field. The last one is the interference between the positive- and negative-frequency waves of the light branch, and its phase contains $\gamma$. It vanishes for any state invariant under rotations of $(\pic,\sigma)$ in field space, a symmetry of the dynamics deep inside the horizon (sections~\ref{sec:field-basis} and~\ref{sec:channels}), such as the state produced by a feature of $a''/a$ at $x\gg\rhohat$, which acts on the two fields identically. By contrast, this term is present as soon as the excitation distinguishes the two fields. For a single-field Bogoliubov transformation of coefficients $\alpha$ and $\beta$ acting on $\pic$ alone at a time $x_f\gg\rhohat$, for instance, $n_+=|\beta|^2/2$ and $|\mathcal I|=|\alpha\beta^*|/2$, half of the single-field values, since only the part of the excitation carried by the light channel survives (see appendix~\ref{app:excited}), and $\gamma$ is observable.

\section{Beyond all orders: Stokes phenomenon}
\label{sec:beyond}

In the expansion in powers of $1/\rhohat$ of the previous sections, each channel is a single wave, and the constants of the decomposition~\eqref{eq:connection-problem} are effects beyond all orders, exponentially small in $\rhohat$. In this section we explain where they come from in the two-field theory. In section~\ref{sec:reflection} we compute them exactly, and in section~\ref{sec:collider} we use them to obtain the cosmological collider signal in the squeezed limit of the bispectrum.

\subsection{Why the expansion is asymptotic}
\label{sec:beyond-why}

The expansion is in fact asymptotic. Expanding the amplification~\eqref{eq:R} to high order in powers of $1/\rhohat$, we find that its coefficients grow factorially, as $(n-1)!/\pi^n$ at order $n$. The terms of the series therefore decrease until $n\simeq\pi\rhohat$ and grow afterwards, and stopping at the smallest term leaves an error of order $\ee^{-\pi\rhohat}$, the best precision the series can reach.\footnote{In the language of Borel summation, the Borel transform of the series has a singularity at a distance $\pi$ from the origin, which is why the series determines $R$ only up to terms of order $\ee^{-\pi\rhohat}$.} This error has the size of the contribution of the heavy channel to the power spectrum, which is the square of the factor $\ee^{-\pi\rhohat/2}$ of~\eqref{eq:asymmetry}. The situation is familiar from a free field of mass $m\gg H$, for which the adiabatic expansion in $H/m$ gives the mode function to any order, but not the Bogoliubov coefficient $\ee^{-\pi m/H}$ associated with particle production.

In both cases the origin is the same. An adiabatic wave is attached to each root of the local dispersion relation, and these waves evolve independently of one another, at all orders of the adiabatic expansion, as long as the roots are distinct. The adiabatic description can thus fail only where two roots coincide. For the free field, the two roots $\pm\sqrt{x^2+\mhat^2}$ never coincide at real times. They do, however, at $x=\pm\ii\mhat$, in the complex plane, at a distance $\mhat$ from the real axis. The positive-frequency and negative-frequency waves are thus coupled only by an effect that is exponentially small in this distance, that is, the Bogoliubov coefficient, as section~\ref{sec:stokes} makes precise.

In the two-field theory, by contrast, the roots are the four frequencies $\pm\homh$ and $\pm\homl$ of~\eqref{eq:omega-pm}. Since $\homh-\homl=\rhohat$ and $\homh\homl=x^2$, two of them can coincide in two ways only.\footnote{One may wonder what becomes of the points $x=\pm\ii\mhat$ of the free massive field once the mixing is present. They do not disappear but change branch. Keeping the mass of $\sigma$ in the dispersion relation, the two frequencies obey $\homh^{2}\,\homl^{2}=x^{2}(x^{2}+\mhat^{2})$ and $\homh^{2}+\homl^{2}=2x^{2}+\rhohat^{2}+\mhat^{2}$, which reduce to $\homh\homl=x^2$ and $\homh-\homl=\rhohat$ when the mass is dropped. One of the two frequencies therefore vanishes at $x=0$ and at $x=\pm\ii\mhat$. Following the two roots from $x=0$ along the imaginary axis, one finds that, as soon as $\rho^2>m^2$, the frequency that vanishes at the latter points is $\homl$. In other words, the mixing has transferred them from the massive field to the light branch. In the regime~\eqref{eq:regime}, $|\mhat|\ll\sqrt{\rhohat}$, so that they lie well inside the region of the horizon crossing of the light branch (figure~\ref{fig:stokes}), where this branch is described by the effective theory, in which the mass contributes only at relative order $\mhat^2/\rhohat$ (section~\ref{sec:eft-constraint}). They therefore produce no separate exponentially small effect.} The first is $\homl=0$, at $x=0$, where the positive-frequency and negative-frequency waves of the light branch coalesce. This point lies on the real axis, at late times, and its effect is not small, since it corresponds to the horizon crossing of the light branch, where the adiabatic description fails for $x\lesssim\sqrt{\rhohat}$, and which we have treated with the effective theory and the matching of sections~\ref{sec:eft} and~\ref{sec:matching}. The second is $\homh+\homl=\sqrt{\rhohat^2+4x^2}=0$, at
\begin{equation}
\label{eq:branch-points}
x=\pm\,\frac{\ii\rhohat}{2}\,,
\end{equation}
where $\homh=-\homl$. At these points, the positive-frequency wave of the heavy branch and the negative-frequency wave of the light branch coalesce, and so do their complex conjugates (figure~\ref{fig:stokes}). Unlike $x=0$, they lie at a distance $\rhohat/2$ from the real axis, on which the two branches never cross, and their effect is therefore exponentially small in this distance, as for the free field. The adiabatic description of the system thus fails in two places only, namely the point $x=0$, which the matching of section~\ref{sec:matching} has dealt with, and the branch points~\eqref{eq:branch-points}, which are the subject of the present section. We will see in section~\ref{sec:reflection-stokes} that one more exponentially small effect, the conversion, involves both at once.

The adiabatic expansion itself signals these branch points, as its second-order phases~\eqref{eq:Psi2} are singular at $r=0$, where $r=\sqrt{1+4x^2/\rhohat^2}$ vanishes, that is, precisely at~\eqref{eq:branch-points}. The branch points lie at $|x|=\rhohat/2$, at the scale of the heavy mass. The dispersion relation of the effective theory, on the other hand, $\omega\propto k^2/(a^2\rho)$, which is the limit $x\ll\rhohat$ of the light branch, has no such point. The EFT therefore cannot produce these effects at any order, and they show up in the expansion in powers of $1/\rhohat$ only through the factorial growth of its coefficients.

\subsection{Stokes phenomenon of the adiabatic solutions}
\label{sec:stokes}

We now determine how much the branch points mix the two branches, and when. The question is that of particle production in a time-dependent background, in which a positive-frequency wave acquires a negative-frequency component with an exponentially small amplitude. For a single field, the answer is known~\cite{Dumlu:2010ua}. The component is acquired when the solution crosses a Stokes line, and its amplitude is determined by an integral of the frequency up to the complex branch point. In cosmology, this description has been applied to particle production~\cite{Hashiba:2021npn,Sou:2021juh} and to the computation of correlators~\cite{Qin:2025xct}. We apply the same type of analysis here, to two waves that belong to the two branches of the mixed system rather than to the positive and negative frequencies of a single field.

As far as the branch points are concerned, the field basis of section~\ref{sec:field-basis} shows that this difference is one of basis only. In this basis, once the two terms of~\eqref{eq:lagrangian-fixed} that are not invariant under rotations are neglected, as they are in the dispersion relation~\eqref{eq:omega-pm}, the two fields are free, with the mass $\rho/2$, and the two circular polarisations evolve independently. The points~\eqref{eq:branch-points} are those where their frequency $\hat{E}$ of~\eqref{eq:E-field} vanishes, as $x=\pm\ii\mhat$ for a free field of mass $m$. In this approximation, the problem is therefore that of particle production for a free field of mass $\rho/2$, in which a positive-frequency wave of one polarisation, of frequency $\hat{E}$, acquires the negative-frequency wave of the same polarisation, of frequency $-\hat{E}$. In terms of $\pic$ and $\sigma$, the two have the frequencies $\hat{E}-a\rhohat/2$ and $-(\hat{E}+a\rhohat/2)$ (see~\eqref{eq:branches-rotation}). The first is thus a wave of one branch, and the second the complex conjugate of a wave of the other. We do not pursue the analysis in the field basis, however. The reason is that the connection problem is posed on the channels and the waves, and the acquired wave must be followed down to late times, where the two terms that break the invariance under rotations are no longer negligible and the two polarisations no longer evolve independently (see section~\ref{sec:rotating}). We therefore translate the standard results into the language of the two branches.

\begin{figure}[t]
\centering
\includegraphics[width=0.62\textwidth]{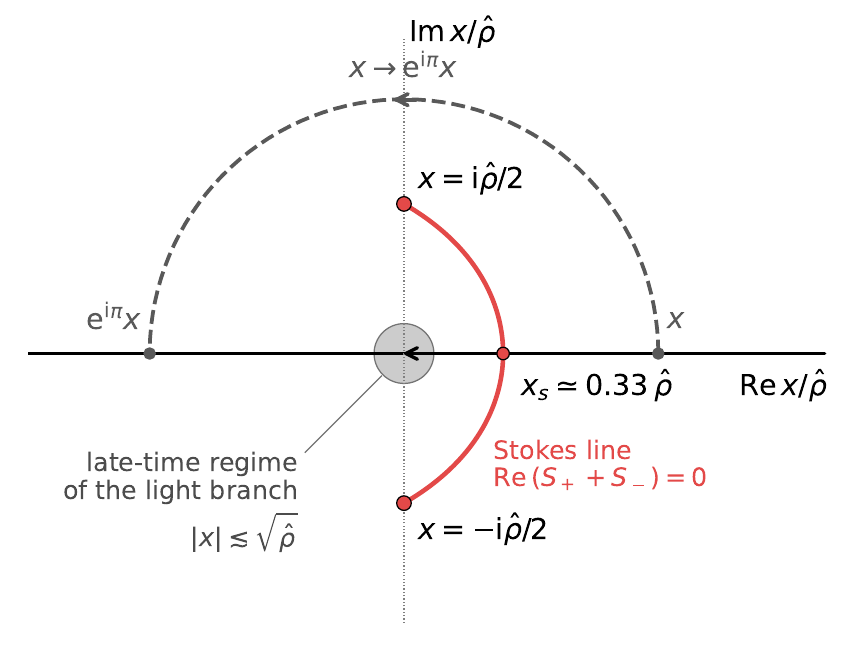}
\caption{The complex $x$ plane, in units of $\rhohat$. The shaded disc is the region $|x|\lesssim\sqrt{\rhohat}$, below the horizon crossing of the light branch, drawn for $\rhohat=100$, where the adiabatic description fails and the effective theory applies. The branch points~\eqref{eq:branch-points} of the dispersion relation (dots) are joined by the Stokes line $\mathrm{Re}\,(S_++S_-)=0$ (red), which crosses the real axis at $x_s\simeq0.33\,\rhohat$, in the modified dispersion regime. The imaginary part of $S_++S_-$ grows along the Stokes line, from zero at $x_s$ to $\pm\pi\rhohat/2$ at the branch points. The other Stokes lines that start at the branch points are not shown. The arrow on the real axis is the direction of time. The dashed half-turn $x\to\ee^{\ii\pi}x$ (grey) is the continuation used in section~\ref{sec:reflection}, where its radius is taken large.}
\label{fig:stokes}
\end{figure}

Let us consider the two waves that coalesce at~\eqref{eq:branch-points}, $\ee^{\ii S_-}$ for the heavy branch and $\ee^{-\ii S_+}$ for the negative frequency of the light branch. Their phases~\eqref{eq:Sa} are the logarithmic integrals of the frequencies, $x\,\dd S_-/\dd x=\homh$ and $x\,\dd S_+/\dd x=\homl$. The quantity that governs the exchange between the two waves is their relative phase measured from the branch point $x_c=\ii\rhohat/2$,
\begin{equation}
\label{eq:Stokes-phase}
F(x)\equiv\ii\int_{x_c}^{x}\big(\homh+\homl\big)\,\frac{\dd x'}{x'}=\frac{\pi\rhohat}{2}+\ii\,\bigg[2\sqrt{x^2+\frac{\rhohat^2}{4}}-\rhohat\,\ash\bigg]\,,
\end{equation}
where the bracket is $S_++S_-$. The Stokes line of the pair is the curve that starts at the branch point and on which $F$ is real, so that the ratio of the two waves is a real exponential and one of them is maximally dominant over the other. A solution that follows one of the two waves acquires a component along the other precisely across this line~\cite{Berry:1972,Berry:1990}. Here a single Stokes line, on which $S_++S_-$ is purely imaginary, joins the two branch points (figure~\ref{fig:stokes}). It crosses the real axis at $x_s\simeq0.33\,\rhohat$, the root of $2\sqrt{x^2+\rhohat^2/4}=\rhohat\,\ash$, in the modified dispersion regime. On the real axis, where $S_++S_-$ is real, one has
\begin{equation}
\label{eq:Dykhne}
\mathrm{Re}\,F=\frac{\pi\rhohat}{2}\,,\qquad \mathrm{Im}\,F=S_++S_-\,,
\end{equation}
and $\mathrm{Im}\,F$ vanishes at $x_s$. The amplitude of the acquired component is $\ee^{-\mathrm{Re}\,F}=\ee^{-\pi\rhohat/2}$, times a constant of modulus one at leading order in the adiabatic expansion, a result known as the Dykhne formula~\cite{Dykhne:1962,Davis:1976}. As expected, this amplitude coincides with the factor $\ee^{-\pi\mu}$ of particle production for a free field of mass $\rho/2$ in de Sitter space, for which $\mu\simeq\rhohat/2$. The switching is smooth, as shown by Berry~\cite{Berry:1990}. More precisely, along the real axis, the acquired component grows as
\begin{equation}
\label{eq:Stokes-multiplier}
\mathcal S(x)=\frac12\bigg[1+\mathrm{erf}\bigg(\frac{-\mathrm{Im}\,F}{\sqrt{2\,\mathrm{Re}\,F}}\bigg)\bigg]=\frac12\bigg[1+\mathrm{erf}\bigg(-\frac{S_++S_-}{\sqrt{\pi\rhohat}}\bigg)\bigg]\,,
\end{equation}
from zero at early times to one at late times. Since $S_++S_-\simeq3.6\,(x-x_s)$ near the crossing, the switching takes place over $|x-x_s|\lesssim0.5\sqrt{\rhohat}$, where both branches are adiabatic, and it is complete well before the light branch crosses its horizon at $x\sim\sqrt{\rhohat}$. This profile depends on the precise definition of the adiabatic waves, but the connection problem involves only the amplitude finally acquired.

Let us now translate this result into the language of the two channels. Across the Stokes line, each channel acquires the complex conjugate of the wave of the other branch. With the adiabatic waves described by the uniform approximations~\eqref{eq:heavy-uniform-bis} and~\eqref{eq:light-uniform}, these acquired terms read
\begin{equation}
\label{eq:connection}
\delta\pi_{c,-}(x)=\mathcal C_-\,\mathcal S(x)\,\big[\pi^{\rm unif}_{c,+}(x)\big]^{*}\,,\qquad
\delta\pi_{c,+}(x)=\mathcal C_+\,\mathcal S(x)\,\big[\pi^{\rm unif}_{c,-}(x)\big]^{*}\qquad(x\gg\sqrt{\rhohat})\,,
\end{equation}
and similarly for $\sigma$, with two constants of modulus $\ee^{-\pi\rhohat/2}$. We denote them by the same symbols as in~\eqref{eq:connection-problem} because they are the same constants. Indeed, in the adiabatic regime, the light and heavy waves of section~\ref{sec:waves} are the adiabatic waves of the two branches, as we will see in section~\ref{sec:reflection-stokes}, so that~\eqref{eq:connection} describes the terms $\mathcal C_+\,f_\brh^{*}$ and $\mathcal C_-\,f_\brl^{*}$ of~\eqref{eq:connection-problem}, and the time at which they appear. At this order, the mass does not appear in the exponent, since the branch points are those of the leading dispersion relation, from which it is absent. It does appear, however, in the exact constant~\eqref{eq:bogoliubov-waves}, through $\mueff-\rhohat$, which is of order $1/\rhohat$.

The two constants are in fact equal, as the commutation relations~\eqref{eq:exact-wronskian} require. To see this, note that the adiabatic waves of the two channels, which we denote here by $f^{\rm ad}_\pm$, satisfy these relations by themselves, at each order in $1/\rhohat$, since the acquired terms~\eqref{eq:connection} are smaller than any power of $1/\rhohat$. After the switching, each channel is its adiabatic wave plus the acquired term, $f_\pm=f^{\rm ad}_\pm+\mathcal C_\pm\,\big(f^{\rm ad}_\mp\big)^{*}$, and the relation $W[f_+,f_-]=0$ reads, at first order in the acquired terms,
\begin{equation}
\label{eq:connection-ccr}
W[f_+,f_-]=\mathcal C_-\,W\big[f^{\rm ad}_+,\big(f^{\rm ad}_+\big)^{*}\big]+\mathcal C_+\,W\big[\big(f^{\rm ad}_-\big)^{*},f^{\rm ad}_-\big]=-4\ii\,|N_+|^{2}\,\big(\mathcal C_--\mathcal C_+\big)=0\,,
\end{equation}
where we used that the Wronskian is antisymmetric and that $N_+=N_-$. Hence
\begin{equation}
\label{eq:connection-equal}
\mathcal C_+=\mathcal C_-\equiv\mathcal C\,.
\end{equation}
We come back to this equality in section~\ref{sec:reflection-stokes}, where the same relations, applied to the light and heavy waves, give the constants of the heavy channel exactly.

Once acquired, each wave evolves as a solution of~\eqref{eq:system} in its own right. In the light channel, the acquired wave belongs to the heavy branch, so that it keeps oscillating and becomes at late times a tail $x^{3/2-\ii\mueff}$ of the massive field. In the heavy channel, the acquired wave belongs instead to the light branch. It therefore freezes at horizon crossing, like the light channel itself, whereas the heavy wave decays as $x^{3/2}$. The heavy channel is thus left with a late-time curvature perturbation of order $\ee^{-\pi\rhohat/2}\,\pi_{c,+}^{*}(0)$, as in the exact result~\eqref{eq:asymmetry}.

The leading-order analysis determines the modulus of $\mathcal C$ only. Its phase, and its corrections in powers of $1/\rhohat$, would require a local solution at the branch point, where the adiabatic waves fail, matched to them on both sides in the manner of section~\ref{sec:matching}. However, a symmetry of the equations spares us this work. In the next section we use it to obtain the late-time behaviour of the two channels exactly, which will determine the constant $\mathcal C$. It will also show that the terms~\eqref{eq:connection} are not the only ones that each channel acquires. More precisely, the constants $\mathcal D_\pm$ of~\eqref{eq:connection-problem}, which multiply the wave of the other branch itself rather than its complex conjugate, are of the same order. Since no branch point relates these two waves, the analysis of this section does not capture them. These constants describe the conversion, which has no analogue for a single field (see section~\ref{sec:production}).

\section{Exact results from reflection symmetry}
\label{sec:reflection}

In this section we obtain exactly, from a symmetry of the equations of motion and the canonical commutation relations, what the analysis of section~\ref{sec:stokes} leaves open, namely the phase of the constant $\mathcal C$ of~\eqref{eq:connection} and~\eqref{eq:connection-equal}, the constants $\mathcal D_\pm$ and $\mathcal E$ of~\eqref{eq:connection-problem}, and the exponentially small late-time coefficients~\eqref{eq:late-coefficients} of the two channels. Since the argument is not specific to the mixed system, we first show how it gives the particle production of a free field, then state it in general and apply it to a chemical potential (section~\ref{sec:reflection-argument}). We then apply it to the mixed system, where it gives the exponentially small late-time coefficients of the two channels (section~\ref{sec:reflection-mixed}). Finally, we deduce the constants of the decomposition~\eqref{eq:connection-problem} on the light and heavy waves, which solves the connection problem of section~\ref{sec:connection} (section~\ref{sec:reflection-stokes}), and we give the physical reading of this solution, in terms of particle production and conversion, and its relation to the Stokes phenomenon (section~\ref{sec:production}).

\subsection{Particle production of free fields}
\label{sec:reflection-argument}

\paragraph{Free field.} Let us first consider a free field $\phi$ of mass $m$ in de Sitter space. We take $\mu\equiv\sqrt{\mhat^{2}-9/4}$ real, the case in which the field oscillates at late times. Its mode function obeys the equation of $\sigma$ in~\eqref{eq:system}, without the mixing,
\begin{equation}
\label{eq:free-eom}
\phi''-\frac{2}{x}\,\phi'+\Big[1+\frac{\mhat^{2}}{x^{2}}\Big]\phi=0\,.
\end{equation}
In the Bunch-Davies vacuum, the mode function starts as the positive-frequency wave $\phi\propto x\,\ee^{\ii x}$ at early times, and it becomes a sum of two powers at late times,
\begin{equation}
\label{eq:free-late}
\phi(x)\to A\,x^{3/2+\ii\mu}+B\,x^{3/2-\ii\mu}\,,
\end{equation}
the second of which is the negative-frequency wave that the cosmic expansion has produced. The canonical commutation relation between the field and its conjugate momentum holds at all times. In terms of the mode function, it states that the Wronskian $(\phi\,\phi^{*\prime}-\phi^{*}\phi')/x^{2}$ is conserved, as also follows from~\eqref{eq:free-eom}. Comparing its values at early and at late times gives $\mu\,(|A|^{2}-|B|^{2})=1$ for the mode function normalised to $\phi\to x\,\ee^{\ii x}$. The quantities $\alpha\equiv\sqrt{\mu}\,A$ and $\beta\equiv\sqrt{\mu}\,B$ are therefore the Bogoliubov coefficients that relate the Bunch-Davies vacuum to the late-time waves, with $|\alpha|^{2}-|\beta|^{2}=1$, and $|\beta|^{2}$ is the usual measure of particle production. The coefficients $A$ and $B$ are usually read from the expansion of a Hankel function at small argument. We now show that their ratio follows instead from a symmetry of the equation, without solving it.

Indeed, equation~\eqref{eq:free-eom} is invariant under $x\to-x$ and its coefficients are real. If $\phi(x)$ is a solution, so are $\phi(-x)$ and $\phi^{*}(x)$. Continuing the mode function along a half-turn in the upper half-plane, $x\to\ee^{\ii\pi}x$, thus gives another solution, which we identify from its early-time behaviour. In the upper half-plane, the early-time wave $x\,\ee^{\ii x}$ decays and the other early-time solution, $x\,\ee^{-\ii x}$, grows. Since a Stokes phenomenon can only add to a solution a component that is exponentially smaller than itself, the early-time wave cannot acquire a component along the growing one there. Its expansion therefore remains valid along a half-turn of large radius, on which it becomes $\ee^{\ii\pi}x\,\ee^{-\ii x}=-\big[x\,\ee^{\ii x}\big]^{*}$. The continued solution and $-\phi^{*}$ have the same early-time behaviour, which determines a solution uniquely, and hence, one obtains that, for all $x>0$,
\begin{equation}
\label{eq:free-reflection}
\phi\big(\ee^{\ii\pi}x\big)=-\,\phi^{*}(x)\,.
\end{equation}
For the free field, whose mode function is a Hankel function (section~\ref{sec:problem}), this relation is the continuation formula $H^{(1)}_{\nu}(\ee^{\ii\pi}z)=-\ee^{-\ii\nu\pi}H^{(2)}_{\nu}(z)$~\cite{NIST:DLMF}. Our derivation, however, uses only the symmetry of the equation and the early-time behaviour of the solution. It therefore extends, as we show in the rest of this section, to mode functions that are not Hankel functions. Let us now read this relation at late times. Under the half-turn, each of the two powers of~\eqref{eq:free-late} is multiplied by a constant,
\begin{equation}
\label{eq:half-turn-powers}
\big(\ee^{\ii\pi}x\big)^{3/2\pm\ii\mu}=\ee^{\ii\pi(3/2\pm\ii\mu)}\,x^{3/2\pm\ii\mu}=-\,\ii\,\ee^{\mp\pi\mu}\,x^{3/2\pm\ii\mu}\,,
\end{equation}
and equating the coefficients of $x^{3/2-\ii\mu}$ on the two sides of~\eqref{eq:free-reflection} gives
\begin{equation}
\label{eq:free-bogoliubov}
B=-\,\ii\,\ee^{-\pi\mu}\,A^{*}\,.
\end{equation}
One recognises the factor $\ee^{-\pi\mu}$ of particle production in de Sitter space, obtained here exactly and with its phase. Together with the commutation relation $|\alpha|^{2}-|\beta|^{2}=1$, it gives $|\beta|^{2}=1/(\ee^{2\pi\mu}-1)$.

\paragraph{General statement.} The argument uses three properties only. (i) The equations of motion are linear, with real coefficients, and they are invariant under $x\to-x$, possibly combined with an exchange of the modes of the system. (ii) Each mode starts in the Bunch-Davies vacuum, as a wave
\begin{equation}
\label{eq:general-early}
f_{\chg}(x)\propto x^{1-\ii\chg}\,\ee^{\ii x}\qquad(x\to\infty)\,,
\end{equation}
where the real number $\chg$ allows for a logarithmic phase, which lowers the frequency of the wave by $\chg H$. (iii) The modes behave as powers of $x$ at late times, as they generically do in de Sitter space. Under the half-turn, the wave~\eqref{eq:general-early} becomes
\begin{equation}
\label{eq:half-turn-wave}
\big(\ee^{\ii\pi}x\big)^{1-\ii\chg}\,\ee^{\ii\,(\ee^{\ii\pi}x)}=-\,\ee^{\pi\chg}\;x^{1-\ii\chg}\,\ee^{-\ii x}=-\,\ee^{\pi\chg}\,\Big[x^{1+\ii\chg}\,\ee^{\ii x}\Big]^{*}\,,
\end{equation}
which is $-\ee^{\pi\chg}$ times the complex conjugate of the wave with the opposite logarithmic phase. We call the mode that starts as the latter the conjugate mode, and denote it by $f_{-\chg}$. With the same normalisation for the two modes, the continued solution is therefore
\begin{equation}
\label{eq:general-reflection}
f_{\chg}\big(\ee^{\ii\pi}x\big)=-\,\ee^{\pi\chg}\,f_{-\chg}^{*}(x)\,.
\end{equation}
For a pair of oscillating powers at late times, $f_{\chg}\to A_{\chg}\,x^{3/2+\ii\mu}+B_{\chg}\,x^{3/2-\ii\mu}$, equations~\eqref{eq:half-turn-powers} and~\eqref{eq:general-reflection} give
\begin{equation}
\label{eq:reflection-rule}
B_{\chg}=-\,\ii\,\ee^{-\pi(\mu-\chg)}\,A_{-\chg}^{*}\,.
\end{equation}
The negative-frequency wave of a mode is thus given by the positive-frequency wave of its conjugate mode, and the logarithmic phase lowers the exponent from $\pi\mu$ to $\pi(\mu-\chg)$.

\paragraph{Chemical potential.} A chemical potential produces precisely such a logarithmic phase~\cite{Sou:2021juh,Qin:2025xct}. For the standard example of a massive vector field with a chemical potential, generated by a Chern-Simons coupling to a rolling scalar field, the mode function $v=f/x$ of each transverse mode obeys
\begin{equation}
\label{eq:chemical-eom}
v''+\Big[1-\frac{2\chg}{x}+\frac{\mu^{2}+1/4}{x^{2}}\Big]v=0\,,
\end{equation}
where $\chg$ is the chemical potential in units of $H$, with a sign that depends on the helicity of the mode. The Bunch-Davies solution is of the form~\eqref{eq:general-early}. Since the term in $\chg/x$ is odd in $x$, the reflection $x\to-x$ exchanges the modes with opposite signs of $\chg$, and~\eqref{eq:reflection-rule} relates them. The Wronskian is conserved as for the free field above, so that the Bogoliubov coefficients $\alpha_{\chg}\equiv\sqrt{\mu}\,A_{\chg}$ and $\beta_{\chg}\equiv\sqrt{\mu}\,B_{\chg}$ obey $|\alpha_{\chg}|^{2}-|\beta_{\chg}|^{2}=1$ for each sign of $\chg$. Since $|\beta_{\chg}|=\ee^{-\pi(\mu-\chg)}\,|\alpha_{-\chg}|$ (see~\eqref{eq:reflection-rule}), these relations are two linear equations for $|\alpha_{\chg}|^{2}$ and $|\alpha_{-\chg}|^{2}$, whose solution gives
\begin{equation}
\label{eq:chemical-production}
|\beta_{\chg}|^{2}=\frac{\ee^{2\pi\chg}+\ee^{-2\pi\mu}}{2\sinh(2\pi\mu)}\,,
\end{equation}
in agreement with the result obtained in~\cite{Sou:2021juh} from the Whittaker functions that solve~\eqref{eq:chemical-eom}. For $\chg=0$, it reduces to the result for the free field.

\paragraph{Hermitian analyticity.} For a single field without logarithmic phase, the relation~\eqref{eq:free-reflection} is known, up to the normalisation of the mode function, as the Hermitian analyticity of the mode functions of the Bunch-Davies vacuum~\cite{Goodhew:2020hob,Goodhew:2021oqg}. In these works, it is formulated as a property under $k\to-k^{*}$ at fixed time, and it constrains the wavefunction of the universe through the analytic continuation of the propagators in $k$. We read the same property here on the late-time powers of the mode functions instead, where it gives the coefficients of particle production.

\subsection{Late-time coefficients of the mixed system}
\label{sec:reflection-mixed}

The expansion in powers of $1/\rhohat$ gives two of the six late-time coefficients~\eqref{eq:late-coefficients} of the channels, $\pi_{c,+}(0)$ and $T_-$ (section~\ref{sec:matching-spectrum}). We now show that the other four, $\pi_{c,-}(0)$, $\bar T_+$, $T_+$ and $\bar T_-$, which vanish at all orders of this expansion, are determined by these two.

\paragraph{Reflection relation.} The system~\eqref{eq:system} has the three properties of section~\ref{sec:reflection-argument}. In particular, its equations are invariant under $x\to-x$, and their coefficients are real and analytic except at $x=0$. The channels can therefore be continued in the complex plane, where the branch points~\eqref{eq:branch-points} are singular for the adiabatic waves only, not for the solutions themselves. As in section~\ref{sec:reflection-argument}, we consider a half-turn of large radius (figure~\ref{fig:stokes}), on which the channels keep their early-time form. The two channels share the same exponential $\ee^{\ii x}$ and differ only by a power of $x$, so that neither is exponentially small compared with the other, and no Stokes phenomenon can add one to the other along the half-turn. On this half-turn, the channel $a$ is the wave~\eqref{eq:BD}, which is of the form~\eqref{eq:general-early} with $\chg=a\rhohat/2$. The two fields are related in the same way on both sides of~\eqref{eq:half-turn-wave}, with $\sigma_a=-\ii a\,\pi_{c,a}$ on the continued solution, and $\sigma_{-a}^{*}=(\ii a\,\pi_{c,-a})^{*}=-\ii a\,\pi_{c,-a}^{*}$ for the complex conjugate of the other channel. The conjugate mode of a channel is thus the other channel and, since $N_+=N_-$ (see~\eqref{eq:Na}), equation~\eqref{eq:general-reflection} reads, for $f=(\pic,\sigma)$ and all $x>0$,
\begin{equation}
\label{eq:reflection}
\boxed{\;f_a\big(\ee^{\ii\pi}x\big)=-\,\ee^{a\pi\rhohat/2}\,f_{-a}^{*}(x)\,.\;}
\end{equation}
The exponential comes from the logarithmic phase of~\eqref{eq:BD}, which the mixing gives to the two channels with opposite signs. Although established at large radius, this relation holds for all $x$, since the solutions are analytic away from $x=0$. In particular, the radius of the half-turn does not matter, a property we use in section~\ref{sec:reflection-stokes}.

\paragraph{Curvature perturbation of heavy channel.} We now evaluate~\eqref{eq:reflection} at late times, using the late-time behaviour~\eqref{eq:late-coefficients}. Under the half-turn, the constant is unchanged, and the two oscillating powers are multiplied by the constants~\eqref{eq:half-turn-powers}, with $\mu=\mueff$. The constant terms of~\eqref{eq:reflection} give, for $a=-1$,
\begin{equation}
\label{eq:heavy-constant}
\pi_{c,-}(0)=-\,\ee^{-\pi\rhohat/2}\,\pi_{c,+}^{*}(0)\,,
\end{equation}
which is the exact relation~\eqref{eq:asymmetry}, here with its phase. This relation holds with no correction in powers of $1/\rhohat$, and for any mass. The decaying solution of section~\ref{sec:waves}, which is odd in $x$, behaves in the same way. Its coefficient in the heavy channel is $\ee^{-\pi\rhohat/2}$ times the complex conjugate of its coefficient in the light channel. The non-oscillating part of the heavy channel is thus exponentially small as a whole.

\paragraph{Tails of the massive field.} The oscillating powers of~\eqref{eq:late-coefficients} obey~\eqref{eq:reflection-rule} with $\mu=\mueff$ and $\chg=a\rhohat/2$, the conjugate mode of the channel $a$ being the channel $-a$,
\begin{equation}
\label{eq:light-tail}
\bar T_a=-\,\ii\,\ee^{-\pi(\mueff-a\rhohat/2)}\;T_{-a}^{*}\,,\qquad a=\pm\,.
\end{equation}
The reflection thus expresses the tail $\bar T$ of each channel in terms of the tail $T$ of the other. For the light channel, it gives $\bar T_+=-\ii\,\ee^{-\pi(\mueff-\rhohat/2)}\,T_-^{*}$, with $T_-$ given in~\eqref{eq:T-minus}. For the heavy channel, on the other hand, $\bar T_-$ is expressed in terms of $T_+$, which the reflection leaves undetermined.

This last coefficient follows from the commutation relation~\eqref{eq:ccr} of the two fields, which has no counterpart for a single field. Using the late-time behaviour~\eqref{eq:late-coefficients}, the coefficient of $x^{3/2+\ii\mueff}$ in this relation reads
\begin{equation}
\label{eq:ccr-tails}
\sum_{a=\pm}\Big[\pi_{c,a}(0)\,\bar T_a^{\,*}-\pi_{c,a}^{*}(0)\,T_a\Big]=0\,.
\end{equation}
With~\eqref{eq:heavy-constant} for $\pi_{c,-}(0)$ and~\eqref{eq:light-tail} for $\bar T_+$ and $\bar T_-$, it gives
\begin{equation}
\label{eq:light-tail-plus}
T_+=\frac{\pi_{c,+}(0)}{\pi_{c,+}^{*}(0)}\;\ee^{-\pi\rhohat/2}\,\frac{1+\ii\,\ee^{-\pi(\mueff-\rhohat)}}{1+\ii\,\ee^{-\pi(\mueff+\rhohat)}}\,T_-\,.
\end{equation}
The denominator, which comes from the heavy channel, differs from one by a term of order $\ee^{-2\pi\rhohat}$.
Equations~\eqref{eq:heavy-constant}, \eqref{eq:light-tail} and~\eqref{eq:light-tail-plus} thus give the four exponentially small coefficients exactly, in terms of $\pi_{c,+}(0)$ and $T_-$. Since $\mueff-\rhohat=\mathcal O(1/\rhohat)$, $\pi_{c,-}(0)$ and $\bar T_+$ are smaller than $\pi_{c,+}(0)$ and $T_-$ by a factor $\ee^{-\pi\rhohat/2}$ at leading order, $T_+$ by $\sqrt2\,\ee^{-\pi\rhohat/2}$, and $\bar T_-$ by $\sqrt2\,\ee^{-2\pi\rhohat}$.

\paragraph{Reading in the field basis.} That the relation~\eqref{eq:heavy-constant} receives no correction in powers of $1/\rhohat$, and does not depend on the mass, is easily understood in the field basis. From~\eqref{eq:rotation-x}, the combination $\varphi_1+\ii a\,\varphi_2=x^{\ii a\rhohat/2}\,(\pic+\ii a\,\sigma)$ behaves at early times as the wave~\eqref{eq:earlytime} of the channel $a$ stripped of its logarithmic phase, as for a free field. The exponentials therefore come from late times only, where a term that oscillates as $x^{\ii\nu}$ in the field basis is multiplied by $\ee^{-\pi\nu}$ under the half-turn. The late-time constant of $\pic$ in the light channel is the term $x^{\ii\rhohat/2}$ of $\varphi_1+\ii\,\varphi_2$, which is the rotating vector of section~\ref{sec:field-basis}, and its angular velocity is exactly that of the basis, $\nu=\rhohat/2$, whatever the mass. One thus recovers~\eqref{eq:heavy-constant}. In the same way, the tail $x^{3/2-\ii\mueff}$ of the light channel becomes $x^{3/2-\ii(\mueff-\rhohat/2)}$ in the field basis, which gives the exponential of~\eqref{eq:light-tail}.

\subsection{Solution of the connection problem}
\label{sec:reflection-stokes}

In section~\ref{sec:reflection-argument}, the Bunch-Davies mode function of a free field was written at late times as the sum~\eqref{eq:free-late} of a positive-frequency and of a negative-frequency wave, and the Bogoliubov coefficients were the coefficients of this sum. The analogue of this sum for the mixed system is the decomposition~\eqref{eq:connection-problem} of the two channels on the light and heavy waves of section~\ref{sec:waves}. We now obtain its five constants from the late-time coefficients of section~\ref{sec:reflection-mixed} and the commutation relations. The constants of the light channel are ratios of late-time coefficients, $\mathcal C_+=\bar T_+/T_-^{*}$ and $\mathcal D_+=T_+/T_-$, which~\eqref{eq:light-tail} and~\eqref{eq:light-tail-plus} give exactly. In the same way, the tail $\bar T_-$ of the heavy channel, expressed by~\eqref{eq:light-tail} in terms of $T_+$, yields the constant $\mathcal E=\bar T_-/T_-^{*}=-\ii\,\ee^{-\pi(\mueff+\rhohat/2)}\,\mathcal D_+^{*}$, of order $\ee^{-2\pi\rhohat}$. Their values are displayed in the summary~\eqref{eq:bogoliubov-waves} below.

\paragraph{Waves and branches.} Since $\mathcal C_+$ and $\mathcal D_+$ are exponentially small, the light wave has the same expansion in powers of $1/\rhohat$ as the light channel. In particular, in the adiabatic regime, it is the positive-frequency wave of the light branch and, for $|x|\ll\rhohat$, it is given at leading order by the effective solution~\eqref{eq:eft-solution} with the constants~\eqref{eq:Apm}. The latter statement also holds for complex $x$, since the effective solution is the leading order of the expansion of appendix~\ref{app:series} at fixed $x^2/\rhohat$, which does not require $x$ to be real. Similarly, the heavy channel is the heavy wave plus the three other solutions of section~\ref{sec:waves}, with the coefficients $\pi_{c,-}(0)$, $\bar T_-$ and that of the decaying solution, which are all exponentially small. The heavy wave therefore has the same expansion in powers of $1/\rhohat$ as the heavy channel, given by~\eqref{eq:heavy-uniform-bis}. This wave is thus a positive-frequency wave of the heavy branch at all times, as expected for a branch that is adiabatic throughout (section~\ref{sec:adiabaticity}).

\paragraph{Heavy channel.} The two remaining constants, $\mathcal C_-$ and $\mathcal D_-$, are those of the non-oscillating part of the heavy channel. They follow from the commutation relations~\eqref{eq:exact-wronskian}, applied to the light and heavy waves, once we note that the light wave has a vanishing Wronskian with the heavy wave and with its complex conjugate. The light wave is a combination of the constant and decaying solutions, which behave as $x^{0}$ and $x^{3}$ times series in powers of $x^{2}$, whereas the heavy wave behaves as $x^{3/2+\ii\mueff}$ times such a series (see~\eqref{eq:exponents}). Their Wronskian~\eqref{eq:wronskian} is thus $x^{\ii\mueff}$ times a sum of real powers of $x$. Since it does not depend on $x$ and $\mueff\neq0$, it vanishes. The same holds with the complex conjugate of the heavy wave, so that
\begin{equation}
\label{eq:waves-orthogonal}
W[f_\brl,f_\brh]=W[f_\brl,f_\brh^{*}]=0\,,
\end{equation}
and the only non-vanishing Wronskians of the four waves are $W_\brl\equiv W[f_\brl,f_\brl^{*}]$ and $W_\brh\equiv W[f_\brh,f_\brh^{*}]$. Inserting the decomposition~\eqref{eq:connection-problem} in the relations $W[f_+,f_-]=W[f_+,f_-^{*}]=0$ gives, since the Wronskian is antisymmetric,
\begin{equation}
\label{eq:waves-ccr}
\mathcal C_-\,W_\brl=\big(\mathcal C_+-\mathcal D_+\mathcal E\big)\,W_\brh\,,\qquad
\mathcal D_-^{*}\,W_\brl=-\big(\mathcal D_+-\mathcal C_+\mathcal E^{*}\big)\,W_\brh\,,
\end{equation}
and the relations $W[f_a,f_a^{*}]=-4\ii\,|N_a|^{2}$, with $N_+=N_-$, then give $W_\brl=(1-|\mathcal E|^{2})\,W_\brh$. Hence
\begin{equation}
\label{eq:heavy-constants}
\mathcal C_-=\frac{\mathcal C_+-\mathcal D_+\mathcal E}{1-|\mathcal E|^{2}}\,,\qquad
\mathcal D_-=-\,\frac{\mathcal D_+^{*}-\mathcal C_+^{*}\mathcal E}{1-|\mathcal E|^{2}}\,.
\end{equation}
Up to terms of relative order $\ee^{-2\pi\rhohat}$, these relations reduce to $\mathcal C_-=\mathcal C_+$, which is~\eqref{eq:connection-equal}, now obtained for the exact waves, and to $\mathcal D_-=-\mathcal D_+^{*}$.

\paragraph{Solution of the connection problem.} Collecting these constants and those of~\eqref{eq:heavy-constants}, the decomposition~\eqref{eq:connection-problem} of the two channels on the light and heavy waves reads
\begin{equation}
\label{eq:bogoliubov-waves}
\boxed{\;\begin{aligned}
&f_+=f_\brl+\mathcal C_+\,f_\brh^{*}+\mathcal D_+\,f_\brh\,,\qquad
f_-=f_\brh+\mathcal C_-\,f_\brl^{*}+\mathcal D_-\,f_\brl+\mathcal E\,f_\brh^{*}\,,\\[4pt]
&\mathcal C_+=-\,\ii\,\ee^{-\pi(\mueff-\rhohat/2)}\,,\qquad
\mathcal D_+=\frac{\pi_{c,+}(0)}{\pi_{c,+}^{*}(0)}\;\ee^{-\pi\rhohat/2}\,\frac{1+\ii\,\ee^{-\pi(\mueff-\rhohat)}}{1+\ii\,\ee^{-\pi(\mueff+\rhohat)}}\,,\\[4pt]
&\mathcal E=-\,\ii\,\ee^{-\pi(\mueff+\rhohat/2)}\,\mathcal D_+^{*}\,,\qquad
\mathcal C_-=\frac{\mathcal C_+-\mathcal D_+\mathcal E}{1-|\mathcal E|^{2}}\,,\qquad
\mathcal D_-=-\,\frac{\mathcal D_+^{*}-\mathcal C_+^{*}\mathcal E}{1-|\mathcal E|^{2}}\,,
\end{aligned}\;}
\end{equation}
exactly, for any mass and mixing. The five constants depend on the mass, the mixing and the phase of $\pi_{c,+}(0)$ only. Once the waves are normalised like the channels, this decomposition is exactly a Bogoliubov transformation (see the insert below).

\begin{techinsert}{Exact Bogoliubov transformation}The waves are not exactly normalised like the channels. Inserting~\eqref{eq:bogoliubov-waves} in the relations $W[f_a,f_a^{*}]=-4\ii\,|N_a|^{2}$, with $N_+=N_-$, gives $W_\brh=-4\ii\,|N_+|^{2}\,\nu_\brh^{2}$ and $W_\brl=-4\ii\,|N_+|^{2}\,\nu_\brl^{2}$, with
\begin{equation*}
\nu_\brh^{-2}=1-|\mathcal E|^{2}+|\mathcal D_+|^{2}-|\mathcal C_+|^{2}\,,\qquad
\nu_\brl^{2}=\big(1-|\mathcal E|^{2}\big)\,\nu_\brh^{2}\,.
\end{equation*}
The waves $f_\brl/\nu_\brl$ and $f_\brh/\nu_\brh$ are thus normalised like the channels, and~\eqref{eq:bogoliubov-waves} is the transformation~\eqref{eq:excited-modes} from them to the channels, with
\begin{equation*}
\alpha=\begin{pmatrix}\nu_\brl&\mathcal D_+\,\nu_\brh\\\mathcal D_-\,\nu_\brl&\nu_\brh\end{pmatrix}\,,\qquad
\beta=\begin{pmatrix}0&\mathcal C_+\,\nu_\brh\\\mathcal C_-\,\nu_\brl&\mathcal E\,\nu_\brh\end{pmatrix}\,.
\end{equation*}
These matrices obey the Bogoliubov conditions of section~\ref{sec:excited} exactly, since the latter are the commutation relations~\eqref{eq:exact-wronskian} written for the normalised waves. The factors $\nu_\brl$ and $\nu_\brh$ differ from one by terms of order $\ee^{-\pi\rhohat}$, like $|\mathcal C_+|^{2}$ and $|\mathcal D_+|^{2}$.
\end{techinsert}

\subsection{Particle production and conversion}
\label{sec:production}

For the physical reading of the solution~\eqref{eq:bogoliubov-waves}, its leading order in the exponentially small constants suffices, and we restrict ourselves to it from now on. Up to terms of relative order $\ee^{-2\pi\rhohat}$, the constant $\mathcal E$ is negligible, $\mathcal C_-=\mathcal C_+$ and $\mathcal D_-=-\mathcal D_+^{*}$. We then drop the indices, $\mathcal C\equiv\mathcal C_+$ and $\mathcal D\equiv\mathcal D_+$, and the two channels read
\begin{equation}
\label{eq:bogoliubov-simple}
f_+=f_\brl+\mathcal C\,f_\brh^{*}+\mathcal D\,f_\brh\,,\qquad
f_-=f_\brh+\mathcal C\,f_\brl^{*}-\mathcal D^{*}\,f_\brl\,.
\end{equation}
At this order the waves are normalised like the channels, and the Bogoliubov transformation of the insert above reduces to
\begin{equation}
\label{eq:bogoliubov-matrices}
\alpha=\begin{pmatrix}1&\mathcal D\\-\mathcal D^{*}&1\end{pmatrix}\,,\qquad
\beta=\begin{pmatrix}0&\mathcal C\\\mathcal C&0\end{pmatrix}\,.
\end{equation}
In section~\ref{sec:excited}, such a transformation acted on the channels and described excited initial states. Here, by contrast, the state is the Bunch-Davies vacuum, and the transformation relates the in-modes of the problem (the channels) to its out-modes (the waves).

The two constants describe two different effects. The constant $\mathcal C$, which appears in $\beta$, relates the light channel to the complex conjugate of the heavy wave, and the heavy channel to the complex conjugate of the light wave. Like the coefficient $\beta$ of a free field in section~\ref{sec:reflection-argument}, it mixes positive and negative frequencies, and we refer to this effect as \emph{particle production}, or simply the production. As for the constant $\mathcal D$, which appears in $\alpha$, it relates the light channel to the heavy wave itself and the heavy channel to the light wave itself, without complex conjugation. A wave of one type is partly converted into a wave of the other, and we refer to this effect as the \emph{conversion}. This effect has no analogue for a single field.

The late-time curvature perturbation of the heavy channel is the sum of these two effects. At late times, where the heavy wave decays,~\eqref{eq:bogoliubov-simple} gives
\begin{equation}
\label{eq:heavy-constant-sum}
\begin{aligned}
\pi_{c,-}(0)&=\mathcal C\,\pi_{c,+}^{*}(0)-\mathcal D^{*}\,\pi_{c,+}(0)\\[4pt]
&=\Big[-\,\ii\,\ee^{-\pi(\mueff-\rhohat)}+\Big(-1+\ii\,\ee^{-\pi(\mueff-\rhohat)}\Big)\Big]\,\ee^{-\pi\rhohat/2}\,\pi_{c,+}^{*}(0)\\[4pt]
&=-\,\ee^{-\pi\rhohat/2}\,\pi_{c,+}^{*}(0)\,.
\end{aligned}
\end{equation}
In the second line, the first term is the production and the parenthesis is the conversion. Each of them depends on the mass, through the term in $\ee^{-\pi(\mueff-\rhohat)}$, which appears in the two with opposite signs and cancels in the sum. The exact relation~\eqref{eq:heavy-constant} owes its independence of the mass to this cancellation, which the exact constants of~\eqref{eq:bogoliubov-waves} preserve.

\paragraph{Origin of the conversion.} The relations~\eqref{eq:heavy-constants} are exact, but they do not show where the conversion comes from. The reflection~\eqref{eq:reflection} helps to answer this question, once combined with the effective solution. This relation expresses the heavy channel in terms of the light channel continued along a half-turn, and the non-oscillating part of the heavy channel, which contains $\mathcal C_-$ and $\mathcal D_-$, is the light wave continued in this way. At leading order in $1/\rhohat$ and for $|x|\ll\rhohat$, the light wave is the effective solution, namely a Hankel function of argument $x^2/2\rhohat$, and this argument makes a full turn when $x$ makes a half-turn. After a full turn, a Hankel function $H^{(1)}$ does not return to itself but acquires a term in $H^{(2)}$, which gives the complex conjugate of the light wave. As we show in the insert below, the heavy channel then reads
\begin{equation}
\label{eq:heavy-decomposition-leading}
f_-=f_\brh-\ii\,\ee^{-\pi\rhohat/2}\,f_\brl^{*}+\sqrt2\;\ee^{-2\ii\gamma}\,\ee^{-\pi\rhohat/2}\,f_\brl\,,
\end{equation}
up to corrections of relative order $1/\rhohat$ in its last two terms. This agrees with the exact constants~\eqref{eq:heavy-constants}, which reduce at leading order to $\mathcal C_-=-\ii\,\ee^{-\pi\rhohat/2}$ and, with the phase of $\pi_{c,+}(0)$ given in~\eqref{eq:pi-late}, to $\mathcal D_-=\sqrt2\;\ee^{-2\ii\gamma}\,\ee^{-\pi\rhohat/2}$.

The effective solution thus shows where each factor of the conversion comes from. The exponential $\ee^{-\pi\rhohat/2}$ is the prefactor of the reflection, as for the production. The factor $\sqrt2$ comes from the point $x=0$, where the positive-frequency and negative-frequency waves of the light branch coalesce and this branch crosses the horizon (see section~\ref{sec:beyond-why}). Indeed, the effective solution is not single-valued around this point, and the conversion comes from the term in $H^{(2)}$ that the Hankel function acquires after a full turn of its argument, whereas the production comes from the term in $H^{(1)}$. The phase, finally, is twice the phase $\gamma$ accumulated by the light branch.

\begin{techinsert}{Conversion from the effective solution}For $a=+1$, the reflection~\eqref{eq:reflection} reads $f_+(\ee^{\ii\pi}x)=-\ee^{\pi\rhohat/2}f_-^{*}(x)$, that is,
\begin{equation}
\label{eq:heavy-reflected}
f_-(x)=-\,\ee^{-\pi\rhohat/2}\,\Big[f_+\big(\ee^{\ii\pi}x\big)\Big]^{*}\,.
\end{equation}
The two oscillating solutions of the light channel are powers of $x$ times series in powers of $x^{2}$. The half-turn leaves the series unchanged and multiplies $f_\brh$ by $-\ii\,\ee^{-\pi\mueff}$ and $f_\brh^{*}$ by $-\ii\,\ee^{\pi\mueff}$ (see~\eqref{eq:half-turn-powers}). Using~\eqref{eq:bogoliubov-waves} in~\eqref{eq:heavy-reflected}, the term $\mathcal C_+f_\brh^{*}$ of the light channel thus gives $-\ee^{-\pi\rhohat/2}\,\big(\ii\,\ee^{\pi\mueff}\big)\,\mathcal C_+^{*}f_\brh=f_\brh$, and the term $\mathcal D_+f_\brh$ gives $\mathcal E\,f_\brh^{*}$. Hence, one obtains exactly
\begin{equation}
\label{eq:heavy-from-light}
f_-(x)=f_\brh(x)+\mathcal E\,f_\brh^{*}(x)-\ee^{-\pi\rhohat/2}\,\Big[f_\brl\big(\ee^{\ii\pi}x\big)\Big]^{*}\,.
\end{equation}
The half-turned light wave is a combination of the constant and decaying solutions, $f_\brl(\ee^{\ii\pi}x)=c_1\,f_\brl(x)+c_2\,f_\brl^{*}(x)$, and~\eqref{eq:heavy-from-light} gives $\mathcal C_-=-\ee^{-\pi\rhohat/2}\,c_1^{*}$ and $\mathcal D_-=-\ee^{-\pi\rhohat/2}\,c_2^{*}$. For real positive $x\ll\rhohat$, the light wave reads, at leading order,
\begin{equation}
\label{eq:light-wave-eft}
\pi_{c,\brl}(x)=A_+\,x^{5/2}\,H^{(1)}_{5/4}(z)\,,\qquad \sigma_\brl(x)=-A_+\,x^{5/2}\,H^{(1)}_{1/4}(z)\,,\qquad z\equiv\frac{x^{2}}{2\rhohat}\,.
\end{equation}
When $x$ makes a half-turn, $z$ makes a full turn, and~\cite{NIST:DLMF}
\begin{equation}
\label{eq:hankel-turn}
H^{(1)}_{\nu}\big(\ee^{2\ii\pi}z\big)=-\,H^{(1)}_{\nu}(z)-2\cos(\nu\pi)\,\ee^{-\ii\nu\pi}\,H^{(2)}_{\nu}(z)\,,
\end{equation}
with $2\cos(\nu\pi)=-\sqrt2$ for $\nu=5/4$ and $+\sqrt2$ for $\nu=1/4$. For the curvature perturbation, this gives
\begin{equation}
\label{eq:light-half-turn-pi}
\begin{aligned}
\pi_{c,\brl}\big(\ee^{\ii\pi}x\big)&=\ii\,A_+\,x^{5/2}\Big[-H^{(1)}_{5/4}(z)+\sqrt2\;\ee^{-5\ii\pi/4}\,H^{(2)}_{5/4}(z)\Big]\\[4pt]
&=-\,\ii\,\pi_{c,\brl}(x)+\sqrt2\;\ee^{-3\ii\pi/4}\,\frac{A_+}{A_+^{*}}\;\pi_{c,\brl}^{*}(x)\,,
\end{aligned}
\end{equation}
where we used that $H^{(2)}_\nu(z)=\big[H^{(1)}_\nu(z)\big]^{*}$ for real $\nu$ and real positive $z$, so that $A_+^{*}\,x^{5/2}H^{(2)}_{5/4}(z)=\pi_{c,\brl}^{*}(x)$. From~\eqref{eq:Apm} and~\eqref{eq:gamma}, $A_+/A_+^{*}=\ee^{2\ii(\gamma+7\pi/8)}$, and the last coefficient is $-\sqrt2\,\ee^{2\ii\gamma}$. The massive field, with the order $1/4$, gives the same two coefficients. Hence $c_1=-\ii$ and $c_2=-\sqrt2\,\ee^{2\ii\gamma}$ at leading order, which gives~\eqref{eq:heavy-decomposition-leading}.
\end{techinsert}

\paragraph{Relation to Stokes phenomenon.} In the adiabatic regime, the light and heavy waves are the adiabatic waves of the light and heavy branches. The terms $\mathcal C\,f_\brh^{*}$ of the light channel and $\mathcal C\,f_\brl^{*}$ of the heavy channel are therefore the terms~\eqref{eq:connection} acquired across the Stokes line. In other words, the production is the Stokes phenomenon of the branch points~\eqref{eq:branch-points}, and $\mathcal C$ is its constant, whose modulus is indeed $\ee^{-\pi\rhohat/2}$ at leading order. Equation~\eqref{eq:bogoliubov-waves} gives this constant exactly, at any mass and mixing. Its phase, $-\ii$, coincides with that of the Bogoliubov coefficient~\eqref{eq:free-bogoliubov} of a free field.

The conversion, however, has no counterpart in section~\ref{sec:stokes}. It relates the light wave and the heavy wave, both of positive frequency, which never coalesce since $\homh-\homl=\rhohat$ for all $x$, so that no branch point exchanges them. As we have just seen, its constant involves both places where the adiabatic description of the system fails, i.e.~the branch points, through the exponential that it shares with the production, and the point $x=0$, through its factor $\sqrt2$. This structure is reminiscent of the new Stokes lines of equations of order higher than two, which are attached to no branch point and arise where ordinary Stokes lines cross~\cite{Berk:1982,Honda:2015}.

\paragraph{Time of the conversion.} Equation~\eqref{eq:heavy-decomposition-leading} holds for all $x\ll\rhohat$, and for $x\gg\sqrt{\rhohat}$ its two light waves are the adiabatic waves of the light branch, of positive and of negative frequency. The converted wave is therefore already present in the modified dispersion regime, whereas each channel is a single adiabatic wave at early times. Like the production, the conversion thus takes place around $x\sim\rhohat$, well before the light branch crosses its horizon. This conclusion, however, requires the heavy wave, which is defined at late times, to be a single adiabatic wave in the modified dispersion regime up to terms smaller than $\ee^{-\pi\rhohat/2}$. We have checked that this is the case by following the heavy wave along the imaginary axis, where an admixture of the light wave of this size would be of order one. We have not studied the profile of this switching, i.e.~the analogue of~\eqref{eq:Stokes-multiplier} for the conversion.

\paragraph{Quanta.} The two effects are most transparent in terms of quanta. The waves define their own quanta, with the annihilation operators
\begin{equation}
\label{eq:operators-waves}
\hat b_{\brl,\bm k}=\hat b_{+,\bm k}-\mathcal D^{*}\,\hat b_{-,\bm k}+\mathcal C^{*}\,\hat b^{\dagger}_{-,-\bm k}\,,\qquad
\hat b_{\brh,\bm k}=\hat b_{-,\bm k}+\mathcal D\,\hat b_{+,\bm k}+\mathcal C^{*}\,\hat b^{\dagger}_{+,-\bm k}\,,
\end{equation}
that is, $\hat b_{w,\bm k}=\sum_a(\alpha_{aw}\,\hat b_{a,\bm k}+\beta^{*}_{aw}\,\hat b^{\dagger}_{a,-\bm k})$ with the matrices~\eqref{eq:bogoliubov-matrices}. Since these matrices obey the Bogoliubov conditions of section~\ref{sec:excited} at this order (and exactly for those of the insert of section~\ref{sec:reflection-stokes}), the operators $\hat b_{w,\bm k}$ obey the same commutation relations as the $\hat b_{a,\bm k}$ (see also~\eqref{eq:bogoliubov-ccr}). Using~\eqref{eq:bogoliubov-simple}, the expansion~\eqref{eq:quanta} of the fields on the quanta of the channels becomes an expansion on the quanta of the waves,
\begin{equation}
\label{eq:quanta-waves}
X_{\bm k}=\sum_{w=\brl,\brh}\Big[\,X_w(x)\,\hat b_{w,\bm k}+X^{*}_w(x)\,\hat b^{\dagger}_{w,-\bm k}\Big]\,,\qquad X=\pic,\sigma\,,
\end{equation}
of which the waves are the mode functions. These relations hold at all times. The operators $\hat b_{w,\bm k}$ describe particles, however, only where the waves are adiabatic waves of positive frequency, once the production and the conversion have taken place around $x\sim\rhohat$, until the horizon crossing of the light branch for the light wave and at all later times for the heavy wave. Earlier, the relevant notion of particles is that of the channels.

In~\eqref{eq:operators-waves}, the terms in $\mathcal C^{*}$ mix annihilation and creation operators, and describe the production. The Bunch-Davies vacuum, annihilated by the $\hat b_{a,\bm k}$, therefore contains quanta of the two waves, with the occupation number $|\mathcal C|^2$ for each. They are produced in pairs of one light and one heavy quantum of opposite momenta, $\langle\hat b^{\dagger}_{\brl,\bm k}\,\hat b^{\dagger}_{\brh,\bm k'}\rangle=(2\pi)^3\delta^{(3)}(\bm k+\bm k')\,\mathcal C$. The terms in $\mathcal D$, for their part, relate annihilation operators only, and describe the conversion. This effect conserves the number of quanta and changes their species, as in the conversion of neutrino flavours or of photons into axions. A quantum of the light channel is thus partly a quantum of the heavy wave, with the probability $|\mathcal D|^2$. The conversion acts on the quanta of an excited initial state (section~\ref{sec:excited}), whereas the vacuum contains no quanta of the channels. In the vacuum, it acts only on the produced quanta, at second order in the exponentially small constants. It turns, for instance, the heavy quantum of a pair into a light one, $\langle\hat b^{\dagger}_{\brl,\bm k}\,\hat b^{\dagger}_{\brl,\bm k'}\rangle=-(2\pi)^3\delta^{(3)}(\bm k+\bm k')\,\mathcal C\mathcal D$.

\paragraph{Field basis.} The two effects of production and conversion are also distinguished in the field basis of section~\ref{sec:field-basis}. In this basis, the light and heavy waves are the two circular polarisations, $a=+1$ and $a=-1$ in~\eqref{eq:branches-rotation}, and complex conjugation exchanges the two polarisations, since it turns $\varphi_2=-\ii a\,\varphi_1$ into $\varphi_2^{*}=\ii a\,\varphi_1^{*}$. The complex conjugate of the heavy wave thus has the polarisation of the light wave. The production relates two waves of the same polarisation and of opposite frequencies. It is thus the production of particles of the free fields of mass $\rho/2$, as found in section~\ref{sec:stokes}. The conversion, by contrast, relates the two polarisations. It would vanish if the Lagrangian were invariant under rotations, since the two polarisations would then evolve independently, and it therefore originates from the last two terms of~\eqref{eq:lagrangian-fixed}. These terms are negligible around $x\sim\rhohat$, where the converted wave appears (section~\ref{sec:rotating}). There is no contradiction, however. Just as the constant of the production is set by branch points that lie away from the real axis, the constant of the conversion is not determined at the time at which the wave appears, but rather involves the point $x=0$, where these terms are important.

\section{Cosmological collider signal}
\label{sec:collider}

The relations of the previous section have an observable consequence, the cosmological collider signal. In the squeezed limit of the bispectrum of the curvature perturbation, where one momentum is much smaller than the other two, the massive field imprints oscillations in the logarithm of the ratio of the momenta. In this section we compute this signal in the regime~\eqref{eq:regime}. We write the bispectrum as a time integral of two-point functions, take the squeezed limit, in which the signal comes from the exponentially small oscillation of the long mode obtained above, and evaluate the remaining integral over the short modes.

\subsection{Bispectrum and its squeezed limit}
\label{sec:collider-bispectrum}

\paragraph{Bispectrum.} At leading order in derivatives, the cubic interactions of the EFT of inflationary fluctuations (section~\ref{sec:theory}) read~\cite{Pinol:2023oux}
\begin{equation}
\label{eq:cubic}
\mathcal L_{\rm int}/a^{3}=-g\,\sigma^{3}-\frac{\alpha}{2}\,\dot{\pi}_c\,\sigma^{2}-\frac{\dot{\pi}_c^{2}\sigma}{2\Lambda_2}-\lambda_2\,\dot{\pi}_c^{3}-\frac{(\partial_i\pic)^{2}\sigma}{2\Lambda_1 a^{2}}\,.
\end{equation}
The coupling of the last one is dictated by the symmetries, $\Lambda_1^{-1}=\rho/f_\pi^2$, and the operator that generates the mixing also contributes to $\dot\pi_c^2\sigma$, with $\Lambda_2^{-1}=-\rho/f_\pi^2$ in the absence of other operators~\cite{Pinol:2023oux}. We consider these five interactions for definiteness. Interactions with more derivatives would be treated in the same way. The mixing is kept in the mode functions, so that at first order in these interactions the bispectrum is a contact diagram. With $H_{\rm int}=-\int\dd^3x\,\mathcal L_{\rm int}$, the in-in formula gives the late-time correlator as
\begin{equation}
\label{eq:in-in}
\big\langle\pi_{c,\bm k_1}\pi_{c,\bm k_2}\pi_{c,\bm k_3}\big\rangle=2\,\mathrm{Im}\int_{-\infty}^{0}\dd\tau\;a\;\big\langle\pi_{c,\bm k_1}(0)\,\pi_{c,\bm k_2}(0)\,\pi_{c,\bm k_3}(0)\;H_{\rm int}(\tau)\big\rangle\,,
\end{equation}
where the fields on the right-hand side are the interaction-picture fields~\eqref{eq:quanta}. By Wick's theorem, each of the three external fields is contracted with one of the fields of the vertex. Each leg of the diagram is therefore a two-point function between the late-time curvature perturbation and a field $X$ at the time of the vertex, given by the sum over the channels of section~\ref{sec:channels},
\begin{equation}
\label{eq:leg}
\langle\pic(0)\,X(x)\rangle'_{k}=\sum_{a=\pm}\pi_{c,a}(0)\,X_a^{*}(x)\,,\qquad x=-k\tau\,,
\end{equation}
where the prime indicates that the momentum-conserving delta function is removed, and the subscript denotes the momentum of the leg. For the interaction $\sigma^3$, the $3!$ contractions are equal, and
\begin{equation}
\label{eq:bispectrum-sigma3}
\langle\pi_c^3\rangle'\equiv\big\langle\pi_{c,\bm k_1}\pi_{c,\bm k_2}\pi_{c,\bm k_3}\big\rangle'=12\,g\;\mathrm{Im}\int_{-\infty}^{0}\dd\tau\;a^{4}\;\prod_{i=1}^{3}\,\langle\pic(0)\,\sigma(x_i)\rangle'_{k_i}\,,
\end{equation}
with $x_i=-k_i\tau$. The other interactions have the same structure, with $\langle\pic(0)\,\dot\pi_c(x)\rangle'$ on a leg with a time derivative, a factor $\ii\bm k_i$ for a spatial gradient on the leg of momentum $\bm k_i$, and a sum over the contractions.

\paragraph{Squeezed limit and long mode.} We now take $k_1=k_2=k\gg k_3$, and use the time variable of the short modes, $x=-k\tau$. The long mode is then at $x_3=(k_3/k)\,x\ll x$, so that it crosses its horizon while the short modes are still deep inside theirs. We assume for the moment that it has done so at the times that dominate the integral. The corresponding condition,~\eqref{eq:squeezed-condition}, is derived below, once these times are identified. The leg of the long mode can then be replaced by its late-time expression. Using the late-time behaviour~\eqref{eq:late-coefficients} of the massive field, the oscillating part of~\eqref{eq:leg} for $X=\sigma$ reads
\begin{equation}
\label{eq:leg-tails}
\langle\pic(0)\,\sigma(x)\rangle'_{\rm osc}=\Big[\sum_{a}\pi_{c,a}(0)\,\bar T_a^{\,*}\Big]\,x^{3/2+\ii\mueff}+\Big[\sum_{a}\pi_{c,a}(0)\,T_a^{*}\Big]\,x^{3/2-\ii\mueff}\,.
\end{equation}
The commutation relation~\eqref{eq:ccr-tails} states that the two brackets are complex conjugates of each other, so that the correlator is real. Moreover, in the first bracket, the heavy channel is smaller than the light one by a factor of order $\ee^{-2\pi\rhohat}$ (section~\ref{sec:reflection-mixed}). With~\eqref{eq:light-tail} for the light channel, we obtain
\begin{equation}
\label{eq:mixed-correlator}
\langle\pic(0)\,\sigma(x)\rangle'_{\rm osc}=\mathcal T\,x^{3/2+\ii\mueff}+\mathcal T^{*}\,x^{3/2-\ii\mueff}\,,\qquad
\mathcal T=\ii\,\ee^{-\pi(\mueff-\rhohat/2)}\;\pi_{c,+}(0)\,T_-\,.
\end{equation}
The coefficient $\mathcal T$ involves both channels. It is the product of the late-time curvature perturbation of the light channel~\eqref{eq:pi-late} and of the tail of the heavy channel~\eqref{eq:T-minus}, times an exponentially small factor, which is precisely $\mathcal C^{*}$, the constant of the particle production~\eqref{eq:bogoliubov-waves}. The conversion, on the other hand, does not appear, since the tail $T_+$ of the light channel, which contains it, only enters the second bracket of~\eqref{eq:leg-tails}, the complex conjugate of the first. In other words, in this second bracket, the conversion carried by $T_+$ cancels against the late-time curvature perturbation of the heavy channel, as in~\eqref{eq:heavy-constant-sum}, so that only the production remains. The signal is thus governed by the production alone. Explicitly, at leading order in $1/\rhohat$,
\begin{equation}
\label{eq:mixed-correlator-explicit}
\begin{aligned}
\langle\pic(0)\,\sigma(x)\rangle'_{\rm osc}\simeq&-\,\frac{\Ga(1/4)}{\sqrt{2\pi}}\;\frac{H^2}{k^3}\;\frac{\ee^{-\pi(\mueff-\rhohat/2)}}{\rhohat^{1/4}}\;x^{3/2}\\[4pt]
&\times\cos\!\Big[\mueff\log\frac{x}{\rhohat}-\frac{\mueff-\rhohat}{2}\,\log\rhohat+\rhohat+\frac{3\pi}{8}\Big]\,.
\end{aligned}
\end{equation}
The leg with a time derivative follows from it. According to~\eqref{eq:late-coefficients}, the oscillating tails of $\pic$ are $\rhohat/(3/2\pm\ii\mueff)$ times those of $\sigma$, and since $\dot\pi_c=-Hx\,\dd\pic/\dd x$,
\begin{equation}
\label{eq:mixed-correlator-dot}
\langle\pic(0)\,\dot\pi_c(x)\rangle'_{\rm osc}=-H\rhohat\;\langle\pic(0)\,\sigma(x)\rangle'_{\rm osc}\,.
\end{equation}
Evaluated at $x_3=(k_3/k)\,x$, these oscillations depend on the momenta through $(k_3/k)^{3/2\pm\ii\mueff}$, and thus generate the oscillations of the bispectrum in $\log(k_3/k)$. The non-oscillating part of the leg contributes to the non-oscillating part of the bispectrum, which is that of the effective theory, and a gradient on the long mode is suppressed by $k_3/k$. We consider neither in what follows.

\subsection{Resonance of the short modes}
\label{sec:collider-resonance}

\paragraph{Short modes and time integral.} The legs of the short modes are needed at the same conformal time, when these modes still oscillate. In the sum~\eqref{eq:leg}, the heavy channel is suppressed by its late-time curvature perturbation~\eqref{eq:heavy-constant}, so that
\begin{equation}
\label{eq:short-leg}
\langle\pic(0)\,X(x)\rangle'_{k}\simeq\pi_{c,+}(0)\,X_+^{*}(x)\,,
\end{equation}
up to a relative correction of order $\ee^{-\pi\rhohat/2}$. Once the three legs are inserted into the time integral~\eqref{eq:in-in}, the oscillating part of the bispectrum reads, for all the interactions of~\eqref{eq:cubic},
\begin{equation}
\label{eq:bispectrum-squeezed}
\langle\pi_c^3\rangle'_{\rm osc}=\frac{2k^{3}}{H^{4}}\;\mathrm{Im}\,\bigg\{\pi_{c,+}(0)^{2}\,\bigg[\mathcal T_{3}\,\Big(\frac{k_3}{k}\Big)^{3/2+\ii\mueff}\mathcal J_{+}+\mathcal T_{3}^{*}\,\Big(\frac{k_3}{k}\Big)^{3/2-\ii\mueff}\mathcal J_{-}\bigg]\bigg\}\,,
\end{equation}
where $\pi_{c,+}(0)$ is that of the short modes, $\mathcal T_3$ is the coefficient~\eqref{eq:mixed-correlator} of the long mode, and $\mathcal J_\pm$ depends on the short modes only, through the integrals
\begin{equation}
\label{eq:hard-integral}
I^{XY}_{\pm}\equiv\int_0^\infty\dd x\;x^{-5/2\pm\ii\mueff}\;X_+^{*}(x)\,Y_+^{*}(x)\,,
\end{equation}
where $X_+$ and $Y_+$ are the mode functions of the light channel for the fields that the short modes carry at the vertex, that is, $\sigma_+$, $\dot\pi_{c,+}$, or $(k/a)\,\pi_{c,+}=Hx\,\pi_{c,+}$ for a gradient, which we label by $\partial\pi$. For the five interactions of~\eqref{eq:cubic}, we find
\begin{equation}
\label{eq:J-list}
\renewcommand{\arraystretch}{1.6}\setlength{\arraycolsep}{3pt}
\begin{array}{c|c|c|c|c|c}
&\sigma^{3}&\dot\pi_c\,\sigma^{2}&\dot\pi_c^{2}\,\sigma&\dot\pi_c^{3}&(\partial_i\pic)^{2}\,\sigma\\\hline
\mathcal J_\pm&6g\,I^{\sigma\sigma}_\pm&\alpha\big[2I^{\dot\pi\sigma}_\pm-H\rhohat I^{\sigma\sigma}_\pm\big]&\Lambda_2^{-1}\big[I^{\dot\pi\dot\pi}_\pm-2H\rhohat I^{\dot\pi\sigma}_\pm\big]&-6\lambda_2H\rhohat I^{\dot\pi\dot\pi}_\pm&\Lambda_1^{-1}I^{\partial\pi\partial\pi}_\pm
\end{array}
\end{equation}
In each of them, the terms proportional to $H\rhohat$ are those in which the long mode carries a time derivative (see~\eqref{eq:mixed-correlator-dot}). For the gradient interaction, the long mode carries $\sigma$, and the two gradients bring $-\bm k_1\cdot\bm k_2\simeq k^2$.

\paragraph{Resonance.} It remains to evaluate the integrals~\eqref{eq:hard-integral}. Their integrand oscillates, and they are dominated by the times, if any, at which its phase is stationary, as is standard in the computation of cosmological collider signals (see, e.g.,~\cite{Chen:2015lza,Chen:2022vzh,Qin:2025xct}). To find them we use the adiabatic mode functions~\eqref{eq:wkb-leading} of the light channel, valid for $x\gg\sqrt{\rhohat}$. In this regime, the fields at the vertex are multiples of $\pi_{c,+}$, with $\sigma_+=-\ii\,\pi_{c,+}$ and $\dot\pi_{c,+}=-\ii H\homl\,\pi_{c,+}$. Up to these factors, which vary slowly, the integrand reads
\begin{equation}
\label{eq:hPhi}
\begin{aligned}
x^{-5/2\pm\ii\mueff}\,\big[\pi_{c,+}^{*}(x)\big]^{2}&=N_+^{2}\,\eta(x)\,\ee^{\ii\Phi_\pm(x)}\,,\\[4pt]
\eta(x)=\frac{x^{-1/2}}{\left(1+\rhohat^{2}/4x^{2}\right)^{1/2}}\,,&\qquad \Phi_\pm(x)=\pm\mueff\log x-2S_+(x)\,,
\end{aligned}
\end{equation}
and its phase varies as
\begin{equation}
\label{eq:phase-rate}
x\,\frac{\dd\Phi_\pm}{\dd x}=\pm\mueff-2\,\homl(x)\,.
\end{equation}
For the lower sign, the phase is nowhere stationary. Since the integrand is analytic and decays at both ends of the range of $\log x$, at early times with the usual $\ii\epsilon$ prescription and at late times as a power of $x$, $\mathcal J_-$ is exponentially smaller than $\mathcal J_+$, and we discard it in what follows. For the upper sign, however, the phase $\Phi\equiv\Phi_+$ is stationary when
\begin{equation}
\label{eq:resonance}
2\,\homl(x_*)=\mueff\,,\qquad\text{that is,}\qquad x_*=\frac{\rhohat}{2}\sqrt{\Big(1+\frac{\mueff}{\rhohat}\Big)^{2}-1}\;\simeq\;\frac{\sqrt3}{2}\,\rhohat\,.
\end{equation}
The stationary point is a resonance. Indeed, the frequency of the light branch decreases with time, and when it reaches $\mueff H/2$, the two short quanta together resonate with the late-time oscillation of the long mode, at the frequency $\mueff H$. The resonance takes place when $k/a\simeq(\sqrt3/2)\,\rho$, at the scale of the heavy mass. The short modes are then deep inside their horizon, where the adiabatic solutions~\eqref{eq:wkb-leading} apply, as assumed, but outside the domain of validity of the EFT, $x\ll\rhohat$.

Around the stationary point, $\Phi(x)\simeq\Phi(x_*)+\tfrac12\Phi''(x_*)\,(x-x_*)^2$ with $\Phi''(x_*)=-4/(\rhohat+\mueff)$, so that the resonance lasts $\Delta x\sim\sqrt{\rhohat}\ll x_*$, and the Gaussian integral gives
\begin{equation}
\label{eq:stationary-phase}
\int_0^\infty\dd x\;\eta(x)\,\ee^{\ii\Phi(x)}\simeq \eta(x_*)\,\sqrt{\frac{\pi\,(\rhohat+\mueff)}{2}}\;\ee^{\ii\left[\Phi(x_*)-\pi/4\right]}\,,
\end{equation}
whose modulus, $\sqrt\pi\,3^{1/4}/\sqrt2$ at leading order, does not depend on the mixing. The slowly varying factors are evaluated at the stationary point, where $\sigma_+^{*}=\ii\,\pi_{c,+}^{*}$, $\dot\pi_{c,+}^{*}=\ii\,(\mueff H/2)\,\pi_{c,+}^{*}$, and where a gradient brings $k/a=Hx_*$. Each combination of~\eqref{eq:J-list} therefore becomes
\begin{equation}
\label{eq:J-resonance}
\mathcal J_+\simeq N_+^{2}\;\mathcal V\;\eta(x_*)\,\sqrt{\frac{\pi\,(\rhohat+\mueff)}{2}}\;\ee^{\ii\left[\Phi(x_*)-\pi/4\right]}\,,
\end{equation}
where the factor $\mathcal V$ collects the coupling and the fields at the vertex. At leading order in $1/\rhohat$, where $\mueff\simeq\rhohat$ and $x_*\simeq(\sqrt3/2)\,\rhohat$, it reads
\begin{equation}
\label{eq:vertex-factors}
\renewcommand{\arraystretch}{1.6}\setlength{\arraycolsep}{8pt}
\begin{array}{c|c|c|c|c|c}
&\sigma^{3}&\dot\pi_c\,\sigma^{2}&\dot\pi_c^{2}\,\sigma&\dot\pi_c^{3}&(\partial_i\pic)^{2}\,\sigma\\\hline
\mathcal V&-6\,g&\frac{3\ii}{2}\,\alpha H&\frac34\,\rhohat^{2}H^{2}/\Lambda_2&\frac32\,\rhohat^{3}\lambda_2H^{3}&\frac34\,\rhohat^{2}H^{2}/\Lambda_1
\end{array}
\end{equation}
The interaction $\dot\pi_c\sigma^2$ is an exception. The two terms of~\eqref{eq:J-list}, in which the derivative is carried respectively by a short mode and by the long one, add up to $\alpha H\,[\rhohat-2\,\homl(x_*)]$, which vanishes at leading order owing to the resonance condition~\eqref{eq:resonance}. The value given in~\eqref{eq:vertex-factors} therefore comes from the first corrections, computed in appendix~\ref{app:cancellations}.

\paragraph{Condition on the long mode.} We can now state when the long mode is described by its late-time oscillation, as we have assumed. The condition compares two times. The long mode must have crossed its own horizon~\eqref{eq:xstar}, $k_3^2/(a^2\rho)\ll H$, by the time the short modes resonate, at $k/a\simeq(\sqrt3/2)\,\rho$. Since the long mode is at $x_3=(k_3/k)\,x_*$ at the resonance, this requires
\begin{equation}
\label{eq:squeezed-condition}
\frac{k_3}{k}\,x_*\ll\sqrt{\rhohat}\,,\qquad\text{that is,}\qquad\frac{k_3}{k}\ll\rhohat^{-1/2}\,.
\end{equation}
Otherwise, the oscillation of the long mode still contains the phase $x^2/2\rhohat$ of~\eqref{eq:wkb-heavy-late}, which here equals $(3/8)\,\rhohat\,(k_3/k)^2$. The condition~\eqref{eq:squeezed-condition} is the counterpart, for the quadratic dispersion relation, of the usual requirement that the long mode be outside its horizon when the short modes resonate with the massive field.

\subsection{Signal in closed form}
\label{sec:collider-signal}

\paragraph{Resulting signal.} Finally, we express the result in terms of the shape $S$ of the bispectrum of the curvature perturbation, defined by $\langle\zeta_{\bm k_1}\zeta_{\bm k_2}\zeta_{\bm k_3}\rangle'=(2\pi)^{4}\Delta_\zeta^{4}\,S/(k_1k_2k_3)^{2}$. With $\zeta=-H\pic/f_\pi^2$ and the power spectrum~\eqref{eq:R-def}, in which the heavy channel is negligible, the shape reads
\begin{equation}
\label{eq:shape-squeezed}
S(k,k,k_3)=-\frac{f_\pi^{2}}{4H}\,\Big(\frac{k_3}{k}\Big)^{2}\,\frac{\langle\pi_c^3\rangle'}{|\pi_{c,+}(0)|^{4}}\,,
\end{equation}
with $\pi_{c,+}(0)$ that of the short modes. We insert into it the bispectrum~\eqref{eq:bispectrum-squeezed}, with~\eqref{eq:J-resonance} for the short modes and~\eqref{eq:mixed-correlator-explicit} for the long one, each mode function being normalised with $N_a=H/(2k^{3/2})$ at its own momentum. The phases accumulated by the two branches appear through $\pi_{c,+}(0)^{2}$ and $\mathcal T_3$, namely three times the phase $\gamma$ of the light channel~\eqref{eq:pi-late} and once that of the heavy channel~\eqref{eq:T-minus}. They combine with the phase at the resonance, $\Phi(x_*)$, into $\rhohat\,\log(3\sqrt3/4)-\tfrac32\,(\mueff-\rhohat)\log\rhohat$, up to terms of order $1/\rhohat$. We obtain, for the oscillating part of the shape,
\begin{equation}
\label{eq:collider}
\boxed{\;\begin{aligned}
S_{\rm osc}(k,k,k_3)&\simeq\frac{\pi\,3^{1/4}}{8\,\Ga(1/4)}\;\frac{f_\pi^{2}}{H^{3}}\;\frac{\ee^{-\pi(\mueff-\rhohat/2)}}{\rhohat^{3/4}}\;\Big(\frac{k_3}{k}\Big)^{1/2}\,\mathrm{Re}\Big[\mathcal V\,\ee^{\ii\vartheta}\Big]\,,\\[4pt]
\vartheta&=\mueff\log\frac{k_3}{k}+\rhohat\,\log\frac{3\sqrt3}{4}-\frac32\,(\mueff-\rhohat)\log\rhohat+\frac{3\pi}{8}\,,
\end{aligned}\;}
\end{equation}
at leading order in $1/\rhohat$ and under the condition~\eqref{eq:squeezed-condition}, with $\mathcal V$ given in~\eqref{eq:vertex-factors}. The third term of $\vartheta$, of order $\rhohat^{-1}\log\rhohat$, comes from the logarithmic phase of~\eqref{eq:gamma}, and the corrections to~\eqref{eq:collider} are of order $1/\rhohat$.

\paragraph{Features of the signal.} Three features of the signal are worth highlighting, namely its exponential dependence on the mass and the mixing, its growth with the number of time derivatives at the vertex, and its value when the interactions are generated by the mixing alone.

\emph{Exponential.} The exponential of~\eqref{eq:collider} is the modulus of the production constant $\mathcal C$, given exactly by~\eqref{eq:bogoliubov-waves}, and we have kept it as such, although the rest of~\eqref{eq:collider} holds at leading order in $1/\rhohat$ only. It follows from the reflection symmetry at any mass and mixing. More precisely, the factor $\ee^{\pi\rhohat/2}$ comes from the half-turn of the logarithmic phase of the early-time wave~\eqref{eq:half-turn-wave}, and the Boltzmann factor $\ee^{-\pi\mueff}$ from the half-turn of the late-time powers~\eqref{eq:half-turn-powers}. The integral over the short modes, dominated by a resonance at real time, adds no exponential in the regime~\eqref{eq:regime}, as shown above, nor at small mixing. This exponential reduces to the usual $\ee^{-\pi\mueff}$ at small mixing and, in the regime~\eqref{eq:regime}, to $\ee^{-\pi\mueff/2}$, half of the usual Boltzmann suppression, as found in~\cite{Belrhali:2026uxn}. This half has a simple origin. The particles are produced in the modified dispersion regime, where the two fields of the field basis are free with the mass $\rho/2$ (section~\ref{sec:stokes}), and since $\mueff\simeq\rhohat$, $\ee^{-\pi\mueff/2}$ coincides with the Bogoliubov coefficient $\ee^{-\pi\rhohat/2}$ of these free fields. The same law was obtained over the whole plane of mass and mixing in~\cite{Pinol:2026hey}, from the exact bispectra, whereas here it follows from a symmetry of the equations of motion. At large mass and fixed mixing, it agrees with the exponential dependence of the closed form obtained for the interaction $\sigma^3$ in~\cite{Wang:2026lff}.

\emph{Time derivatives.} The powers of $\rhohat$ in~\eqref{eq:vertex-factors} are easily understood. On a short leg, a time derivative brings the frequency at the resonance, $\mueff H/2$. On the long leg, by contrast, it brings the mixing itself, $\rhohat H$ (see the late-time relation~\eqref{eq:mixed-correlator-dot}). The signal therefore grows with the number of time derivatives of the curvature perturbation at the vertex, as found and explained in~\cite{Belrhali:2026uxn}. A gradient on a short leg brings $k/a\simeq(\sqrt3/2)\,\rho$ at the resonance, so that the gradient interaction grows in the same way.

\emph{Signal of the mixing itself.} The cancellation noted above for $\dot\pi_c\sigma^2$ also occurs between $\dot\pi_c^2\sigma$ and the gradient interaction, which was left aside in~\cite{Belrhali:2026uxn}, when both are generated by the mixing alone, $\Lambda_2^{-1}=-\Lambda_1^{-1}=-\rho/f_\pi^2$. The two interactions then combine into $-(\rho/2f_\pi^2)\,[(\partial_i\pic)^2/a^2-\dot\pi_c^2]\,\sigma$, their contributions to~\eqref{eq:vertex-factors} cancel, again because of the resonance condition, and the first corrections give (see appendix~\ref{app:cancellations}) $\mathcal V=-(21\,\ii/4)\,\rhohat H^2/\Lambda_1$, that is, $\mathcal V f_\pi^2/H^3=-(21\,\ii/4)\,\rhohat^{2}$. This value gives the cosmological collider signal of the mixing itself, which has no free parameter.

\begin{figure}[t]
\centering
\includegraphics[width=\textwidth]{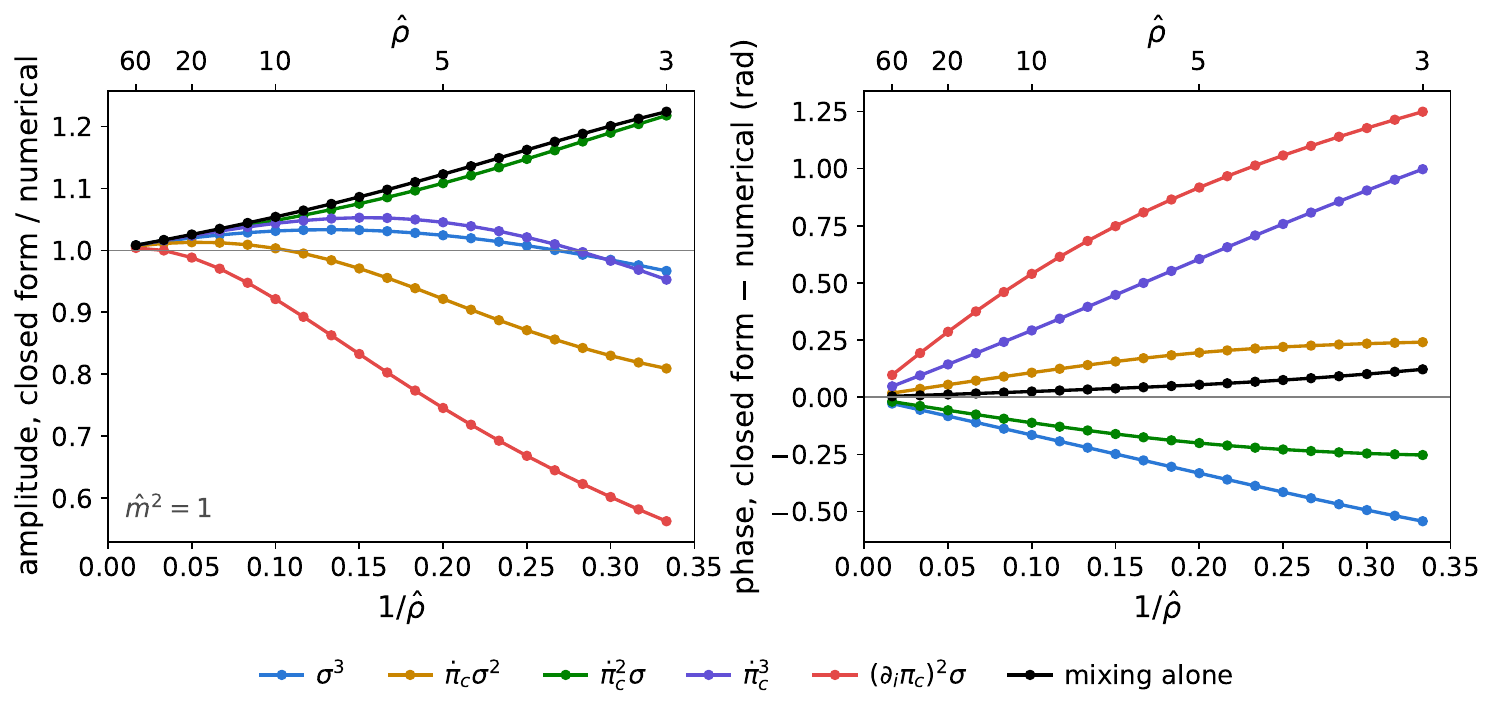}
\caption{The closed form~\eqref{eq:collider} of the cosmological collider signal against its numerical evaluation, as functions of $1/\rhohat$, for $\mhat^2=1$, for the five interactions of~\eqref{eq:cubic} and for the combination generated by the mixing alone. \emph{Left}: ratio of the amplitudes of the oscillation in $\mueff\log(k_3/k)$. \emph{Right}: difference of the phases, that is, the shift between the two oscillations in their argument $\mueff\log(k_3/k)$.}
\label{fig:collider}
\end{figure}

\paragraph{Accuracy.} To assess the accuracy of~\eqref{eq:collider}, we have computed the same signal numerically, with the exact mode functions in the legs~\eqref{eq:leg} and the time integrals evaluated by direct integration. Figure~\ref{fig:collider} shows the ratio of the amplitudes and the difference of the phases of the two results, for the five interactions and for the combination generated by the mixing alone. The differences decrease as $1/\rhohat$, the expansion parameter. They are below $10\%$ in amplitude and $0.55$ radian in phase at $\rhohat=10$, and below $1.5\%$ and $0.15$ radian at $\rhohat=40$.

\section{Conclusions}
\label{sec:conclusions}
Massive fields coupled to the curvature perturbation during inflation leave signatures in primordial correlators, which have most often been computed by treating the mixing between the two as a perturbation. The regime in which this mixing is strong has recently received a lot of attention, both theoretically~\cite{Werth:2023pfl,Pinol:2023oux,Jazayeri:2023xcj,Huenupi:2026abj,Huenupi:2026aqc,Wang:2026lff,Pinol:2026xnl,Belrhali:2026uxn,Pinol:2026hey,Chen:2026khu} and observationally~\cite{Philcox:2026rpn,Philcox:2026tjj}. In this paper we have considered the regime in which the mixing is not only strong but much larger than the Hubble rate, $\rhohat\equiv\rho/H\gg1$, and than $m^2/H$. We have approached it from the side opposite to perturbation theory, as an expansion in the inverse mixing that keeps the two fields throughout. The question that we have answered is a connection problem, namely how the solutions defined at early times by the Bunch-Davies vacuum decompose in terms of those defined by their behaviour at late times. Its coefficients give the late-time amplitude of the fluctuations and the Bogoliubov coefficients of particle production. Answering it has allowed us to derive the single-field effective theory that describes this regime at low energies, to compute the effects that lie beyond it, and to obtain from them the cosmological collider signal.

The expansion rests on the two branches of the local dispersion relation of the mixed system. A time-dependent rotation of the two fields removes the mixing, and in this rotated basis the system reduces at high frequencies to two free fields of mass $\rho/2$, whose two circular polarisations are the two branches. The heavy mode evolves adiabatically at all times, and the light mode until it crosses its own horizon, around which it obeys the equation of motion of the EFT. The EFT emerges as the leading order of the expansion, and matching the two descriptions selects its vacuum and sets a phase, accumulated before the EFT becomes valid, that it cannot determine on its own, and on which the phase of the cosmological collider signal depends. The heavy mode, which the EFT does not contain, is obtained in closed form at all times. The expansion in the inverse mixing thus provides the effective theory, supplemented by the heavy mode in its adiabatic vacuum, with corrections that can be computed systematically.

The production of particles at the origin of the cosmological collider signal is exponentially small in the mixing, and therefore lies beyond all orders of this expansion, which is asymptotic. We have shown that it arises from a Stokes phenomenon at the complex branch points of the dispersion relation, which lie at the scale of the heavy mass and which the EFT cannot see. In the rotated basis, this phenomenon reduces to the production of particles of free fields of mass $\rho/2$, with the Bogoliubov coefficient $\ee^{-\pi\rhohat/2}$. A symmetry of the equations of motion under the reflection of conformal time, $\tau\to-\tau$, combined with the canonical commutation relations, then gives all the exponentially small coefficients exactly, at any mass and mixing, in terms of those that the expansion determines, and with them the solution of the connection problem. It yields in particular the exponential dependence of the cosmological collider signal on the mass and the mixing found in~\cite{Pinol:2026hey}, which reduces at large mixing to this Bogoliubov coefficient~\cite{Belrhali:2026uxn}. Physically, the solution of the connection problem is a Bogoliubov transformation between the solutions defined at early and at late times, which contains two effects, particle production and conversion, both acquired at the scale of the heavy mass, well before the light mode crosses its horizon. Particle production creates pairs of one light and one heavy quantum, while conversion, which has no analogue for a single field, changes the species of the quanta without changing their number.

The exact relations given by the symmetry, combined with the adiabatic solutions of the expansion, give the cosmological collider signal in the squeezed limit of the bispectrum, with its amplitude and its phase in closed form, equation~\eqref{eq:collider}, for the cubic interactions of the EFT of inflationary fluctuations. The signal is generated by a resonance between the late-time oscillation of the long mode and the two short quanta, at the energy scale of the heavy mass, outside the domain of validity of the EFT. Each time derivative at the vertex brings in general a factor of the order of the mixing, except where the leading contributions cancel at the resonance, so that the signal grows with the number of time derivatives of the curvature perturbation, as found in~\cite{Belrhali:2026uxn}. These results are complementary to the analytical bispectra obtained there. The latter hold for any mass and mixing, and are given by series, while equation~\eqref{eq:collider} gives their cosmological collider signal at large mixing in closed form, and makes its physical origin transparent.

Beyond this signal, the rest of the tree-level bispectrum is within the reach of the expansion in the inverse mixing. The heavy mode contributes to the late-time curvature perturbation only through an exponentially small term, so that the external legs of a diagram involve the light channel only. A contact diagram therefore gives the bispectrum of the EFT at leading order, with corrections in powers of the inverse mixing that the expansion can compute. On internal lines, in exchange diagrams or in loops, the heavy mode is no longer exponentially suppressed, and its mode functions, obtained here in closed form, give access to its contributions. We leave these computations for future work.

Exponentially small effects of heavy fields have also been studied recently from the point of view of resurgence. For heavy fields whose mixing is treated perturbatively, the derivative expansion of the EFT is asymptotic~\cite{Jazayeri:2023xcj,Jazayeri:2023kji,DuasoPueyo:2025lmq}, and its resummation, supplemented by boundary conditions, was shown to recover the effects of heavy particle production~\cite{Li:2026xhw,Dorigoni:2026mbh}. The present analysis is complementary. It is carried out at strong mixing and at the level of the mode functions, whose adiabatic solutions we follow in the complex time plane, so that these effects are computed directly, from the Stokes phenomenon at the branch points of the dispersion relation and from the reflection symmetry, without resumming the expansion in the inverse mixing. It would be interesting to relate the two descriptions.

Finally, the approach followed in this paper does not rely on the existence of an exact solution, which has only served here as a check. The expansion in the inverse mixing only requires a separation of scales and the adiabaticity of the heavy mode, and the branch points at the origin of the exponentially small effects follow from the dispersion relation alone. The reflection symmetry that determines these effects exactly is not specific to this system either. It already gives the particle production of a free field, including in the presence of a chemical potential, without solving its equation of motion, and it holds for many systems with constant parameters in de Sitter space, possibly up to an exchange of their modes. We therefore expect the strategy followed here to apply more generally, for instance in the presence of more than two fields. When the mixing or the mass vary in time, the symmetry is lost in general. The production could still be computed as a series in the inverse mixing, by a local analysis at the branch points matched to the adiabatic waves on both sides, whereas the conversion, which no branch point produces on its own, would require a finer analysis of the Stokes phenomenon. Particle production in cosmology can always be formulated as a connection problem between the solutions adapted to early and to late times, and we hope that this strategy will help to solve it beyond the cases that admit an exact solution.

\acknowledgments
We are very grateful to Nathan Belrhali, Arthur Poisson, Zhiming Shuai and Tao Zhu for useful conversations. The authors used Claude (Anthropic) to assist with this work, and take full responsibility for its content. SGS was supported by the Guangdong Provincial Key Laboratory of Topological Matter (2026B1212010004) and by Provincial Grants (2023QN10X389, 2026B0303050001).
YL thanks SIMIS for support.

\appendix

\section{Decoupled fourth-order equations}
\label{app:fourth-order}
The two second-order equations~\eqref{eq:system} can be decoupled at the price of raising their order. Differentiating the first equation and eliminating $\sigma''$ with the second, then differentiating the second and eliminating $\pic''$ with the first, one obtains four relations that are linear in $\sigma$, $\sigma'$, $\sigma''$ and $\sigma'''$, and can be solved for them in terms of $\pic$ and its derivatives. The result for $\sigma$ itself is
\begin{equation}
\label{eq:app-sigma-of-pi}
\sigma=-\frac{x}{\rhohat\,(x^{2}+\mhat^{2})}\Big[x^{2}\pic'''+\big(x^{2}-2+\rhohat^{2}\big)\pic'+2x\,\pic\Big]\,.
\end{equation}
In the range~\eqref{eq:eft-range}, and for a $\pic$ that varies slowly in the sense of section~\ref{sec:eft-constraint}, each derivative bringing a factor of order $\max(x/\rhohat,1/x)$, the bracket is dominated by the term $\rhohat^2\pic'$ and~\eqref{eq:app-sigma-of-pi} reduces to the constraint~\eqref{eq:constraint}. Substituting~\eqref{eq:app-sigma-of-pi} back into the first equation of~\eqref{eq:system} gives the fourth-order equation obeyed by $\pic$ alone,
\begin{equation}
\label{eq:app-fourth-pi}
\begin{aligned}
&\pic''''-\frac{2x}{x^{2}+\mhat^{2}}\,\pic'''+\frac{\rhohat^{2}+\mhat^{2}+2x^{2}-2}{x^{2}}\,\pic''\\
&\qquad-\frac{2\big(x^{4}+2\rhohat^{2}x^{2}+\mhat^{2}x^{2}-4x^{2}+\rhohat^{2}\mhat^{2}+\mhat^{4}-2\mhat^{2}\big)}{x^{3}\,(x^{2}+\mhat^{2})}\,\pic'\\
&\qquad+\frac{x^{4}+2\mhat^{2}x^{2}-6x^{2}+\mhat^{4}-2\mhat^{2}}{x^{2}\,(x^{2}+\mhat^{2})}\,\pic=0\,,
\end{aligned}
\end{equation}
and the same procedure with the roles of the two fields exchanged gives that obeyed by $\sigma$,
\begin{equation}
\label{eq:app-fourth-sigma}
\begin{aligned}
&\sigma''''-\frac{2}{x}\,\sigma'''+\frac{2x^{2}+4+\rhohat^{2}+\mhat^{2}}{x^{2}}\,\sigma''\\
&\qquad-\frac{2}{x}\Big(1+\frac{2+2\rhohat^{2}+2\mhat^{2}}{x^{2}}\Big)\sigma'
+\Big(1+\frac{\mhat^{2}}{x^{2}}+\frac{6\rhohat^{2}+6\mhat^{2}}{x^{4}}\Big)\sigma=0\,.
\end{aligned}
\end{equation}
These equations are exact. They are the starting point of the series of appendix~\ref{app:series}.

\section{WKB expansion to second order}
\label{app:wkb}

With the ansatz~\eqref{eq:wkb-ansatz}, the system~\eqref{eq:canonical-system} becomes, after division by $\ucan$,
\begin{align}
\label{eq:app-E1}
&-\Psi'^2+\ii\varepsilon\Psi''+1-\frac{2\varepsilon^2}{\xi^2}-\frac{2\varepsilon}{\xi}Z'-\frac{2\ii}{\xi}Z\Psi'+\frac{4\varepsilon}{\xi^2}Z=0\,,\\
\label{eq:app-E2}
&\varepsilon^2Z''+2\ii\varepsilon Z'\Psi'+\ii\varepsilon Z\Psi''-Z\Psi'^2+Z+\frac{(\mhat^2-2)\varepsilon^2}{\xi^2}Z+\frac{2\ii}{\xi}\Psi'+\frac{2\varepsilon}{\xi^2}=0\,.
\end{align}
Inserting the expansions~\eqref{eq:wkb-ansatz} and collecting powers of $\varepsilon$ gives, at order $\varepsilon^0$,
\begin{equation}
\label{eq:app-order0}
1-p_0^{2}-\frac{2\ii}{\xi}\,Z_0p_0=0\,,\qquad (1-p_0^{2})\,Z_0+\frac{2\ii}{\xi}\,p_0=0\,.
\end{equation}
The second equation gives $Z_0=-2\ii p_0/[\xi(1-p_0^2)]$, and the first then reads $(1-p_0^2)^2=4p_0^2/\xi^2$. Its two positive roots, $1-p_0^2=\mp2p_0/\xi$, are $p_0=(r\pm1)/\xi$, with $Z_0=\pm\ii$. We thus recover the dispersion relation~\eqref{eq:omega-xi}, $\hat\omega_{\brh,\brl}=x\,p_0$, together with the relation between the two fields on each branch. At order $\varepsilon^1$, the equations for $p_1$ and $Z_1$ read
\begin{equation}
\label{eq:app-order1}
-\frac{2r}{\xi}\,p_1-\frac{2\ii p_0}{\xi}\,Z_1=-\ii p_0'-\frac{4}{\xi^{2}}\,Z_0\,,\qquad \mp\frac{2\ii r}{\xi}\,p_1\mp\frac{2p_0}{\xi}\,Z_1=-\ii Z_0p_0'-\frac{2}{\xi^{2}}\,,
\end{equation}
where we have used $Z_0'=0$ and $1-p_0^2=\mp2p_0/\xi$ on the left-hand side. At order $\varepsilon^2$, the same matrix multiplies $(p_2,Z_2)$, and the right-hand side involves the first-order solution and, for the first time, the mass,
\begin{equation}
\label{eq:app-order2}
\begin{aligned}
-\frac{2r}{\xi}\,p_2-\frac{2\ii p_0}{\xi}\,Z_2&=p_1^{2}+\frac{2\ii}{\xi}\,Z_1p_1-\ii p_1'+\frac{2}{\xi}\,Z_1'-\frac{4}{\xi^{2}}\,Z_1+\frac{2}{\xi^{2}}\,,\\[2pt]
\mp\frac{2\ii r}{\xi}\,p_2\mp\frac{2p_0}{\xi}\,Z_2&=Z_0p_1^{2}+2Z_1p_0p_1-2\ii Z_1'p_0-\ii Z_0p_1'-\ii Z_1p_0'-\frac{\mhat^{2}-2}{\xi^{2}}\,Z_0\,.
\end{aligned}
\end{equation}
The same structure holds at every order $n\geq1$. The coefficients of $\varepsilon^n$ in~\eqref{eq:app-E1} and~\eqref{eq:app-E2} are linear in $p_n$ and $Z_n$, with the matrix of~\eqref{eq:app-order1}, of determinant $8p_0q/\xi$ (up to sign), and a right-hand side built from the lower orders, so that $p_n$ and $Z_n$ are uniquely determined. The computation is conveniently done in the variable $t=\xi+\sqrt{1+\xi^2}=\ee^{\,\mathrm{arcsinh}\,\xi}$, in which $\xi=(t-t^{-1})/2$, $r=(t+t^{-1})/2$ and $\ee^{\mathrm{arcsinh}(1/\xi)}=(t+1)/(t-1)$ are rational, so that every coefficient is a rational function of $t$ and the integration of $p_2$ is elementary. One finds, in terms of $r=\sqrt{1+\xi^2}$ and with the upper sign for the heavy branch,
\begin{align}
\label{eq:app-results}
p_1&=\frac{\ii}{2}\frac{q'}{q}=-\frac{\ii}{2\xi\,r^2}\,,\qquad
Z_1=\pm\frac{3}{2(r\pm1)}\,,\\
\label{eq:app-W2}
Z_2&=\ii\,\frac{2\mhat^2r^2(r\pm1)-6r^3\mp21r^2+12r\pm6}{8\,r^2(r\pm1)^2}\,,\\
\label{eq:app-p2}
\mathrm{Re}\,p_2&=\frac{\dd\Psi_2^{(\brh,\brl)}}{\dd\xi}\,,\qquad \mathrm{Im}\,p_2=0\,,
\end{align}
with $\Psi_2^{(\brh,\brl)}$ given in~\eqref{eq:Psi2}. The mass appears only through the last term of~\eqref{eq:app-order2}, and hence only in $Z_2$ and $\Psi_2$. The limits of $\Rpm$ are
\begin{equation}
\label{eq:app-R-limits}
\Rp\to\frac{71}{96}\,,\quad \Rm\simeq\frac{21}{8\xi^2}\qquad(\xi\to0)\,;\qquad\qquad \Rpm\simeq\frac{19}{16\,\xi}\qquad(\xi\to\infty)\,.
\end{equation}
At early times, $\xi\to\infty$, both $\ash$ and $\Rpm$ vanish, so that $\Psi_2\to0$ on both branches. The second-order correction thus leaves the Bunch-Davies normalisation of~\eqref{eq:wkb-leading} unaffected. At late times, $\xi\to0$, $\ash\simeq\log(\rhohat/x)$, and on the heavy branch
\begin{equation}
\label{eq:app-Psi2-late}
\frac{2\Psi_2^{(\brh)}}{\rhohat}\;\to\;(\mueff-\rhohat)\log\frac{x}{\rhohat}+\frac{71}{48\,\rhohat}\,,
\end{equation}
that is, the shift of the exponent of~\eqref{eq:wkb-heavy-late} from $x^{\ii\rhohat}$ to $x^{\ii\mueff}$, together with a constant phase. On the light branch, on the other hand, $2\Psi_2^{(\brl)}/\rhohat\simeq21\rhohat/(16x^2)$ diverges, as noted in~\eqref{eq:Rminus-div}.

\section{Exact solution and numerical checks}
\label{app:numerics}

The exact solution of section~\ref{sec:exact} is the pair of channels, that is, the solutions of~\eqref{eq:system} with the Bunch-Davies behaviour~\eqref{eq:BD}. The approximations of this paper are compared with it in sections~\ref{sec:wkb-second}, \ref{sec:uniform} and~\ref{sec:collider-signal}. Its late-time coefficients are known in closed form (see~\eqref{eq:pi-exact} and~\eqref{eq:tails-exact} below). At finite times, however, we evaluate it by numerical integration of the equations of motion.

\paragraph{Exact late-time tails.} At large $\lambda$, the spectral density~\eqref{eq:sigmaray} decays as the sum of two powers~\cite{Belrhali:2026uxn},
\begin{equation}
\label{eq:tails-D}
\begin{aligned}
&B_a(\lambda)\simeq D^{(a)}_{+}\,\lambda^{-\frac12+\nueff}+D^{(a)}_{-}\,\lambda^{-\frac12-\nueff}\,,\\[4pt]
&D^{(a)}_{\pm}=\frac{\Ga(1-2s)\,\Ga(\pm2\nueff)\;2^{\frac12\mp\nueff}}{\Ga\big(\tfrac12\pm\nueff\big)\,\Ga\big(\tfrac12\pm\nueff-2s\big)}\,,
\end{aligned}
\end{equation}
and the power $\lambda^{-1/2\mp\nueff}$ is the dual image of the late-time power $x^{3/2\pm\nueff}$ of the massive field. With the normalisation $-a\,N_a\,\kappa_a$ of~\eqref{eq:sigmaray}, the tails of~\eqref{eq:late-coefficients} read
\begin{equation}
\label{eq:tails-exact}
\begin{aligned}
T_a&=-a\,N_a\,\kappa_a\;\ee^{\ii\pi/4}\,\ee^{-\ii\pi\nueff/2}\;\Ga\big(\tfrac12-\nueff\big)\,D^{(a)}_{-}\,,\\[4pt]
\bar T_a&=-a\,N_a\,\kappa_a\;\ee^{\ii\pi/4}\,\ee^{+\ii\pi\nueff/2}\;\Ga\big(\tfrac12+\nueff\big)\,D^{(a)}_{+}\,.
\end{aligned}
\end{equation}
Together with~\eqref{eq:pi-exact}, they give the six late-time coefficients exactly, for any mass and mixing. We use them for the long mode of the cosmological collider signal evaluated numerically in section~\ref{sec:collider-signal} (figure~\ref{fig:collider}). These expressions also provide a check of several results of this paper. They obey the relations~\eqref{eq:heavy-constant}, \eqref{eq:light-tail} and~\eqref{eq:light-tail-plus}, which section~\ref{sec:reflection} derives from the reflection symmetry and the commutation relations, and at large $\rhohat$ they reduce to~\eqref{eq:pi-late} and~\eqref{eq:T-minus}, with the accuracy stated there.

\paragraph{Numerical integration.} Each channel is integrated from $x_{\rm in}=1500\,\rhohat$, deep in the relativistic regime, where it is initialised with the second-order adiabatic solution of section~\ref{sec:wkb-second}. The truncation of these initial data seeds the other channel at a level of order $1/x_{\rm in}$. This matters for the heavy channel only. Indeed, the heavy channel decays at late times, as $x^{3/2}$, whereas the light channel freezes, so that a small admixture of the latter grows relative to it. The initial point is therefore taken far enough for this admixture to remain below the accuracy at which we test the heavy channel, of order $\rhohat^{-3}$ in phase. We have checked that the numerical solution agrees with the integral representation~\eqref{eq:sigmaray} to one part in $10^{7}$ at $\rhohat=20$, from the relativistic regime down to $x=0.2$.

\begin{figure}[t]
\centering
\includegraphics[width=\textwidth]{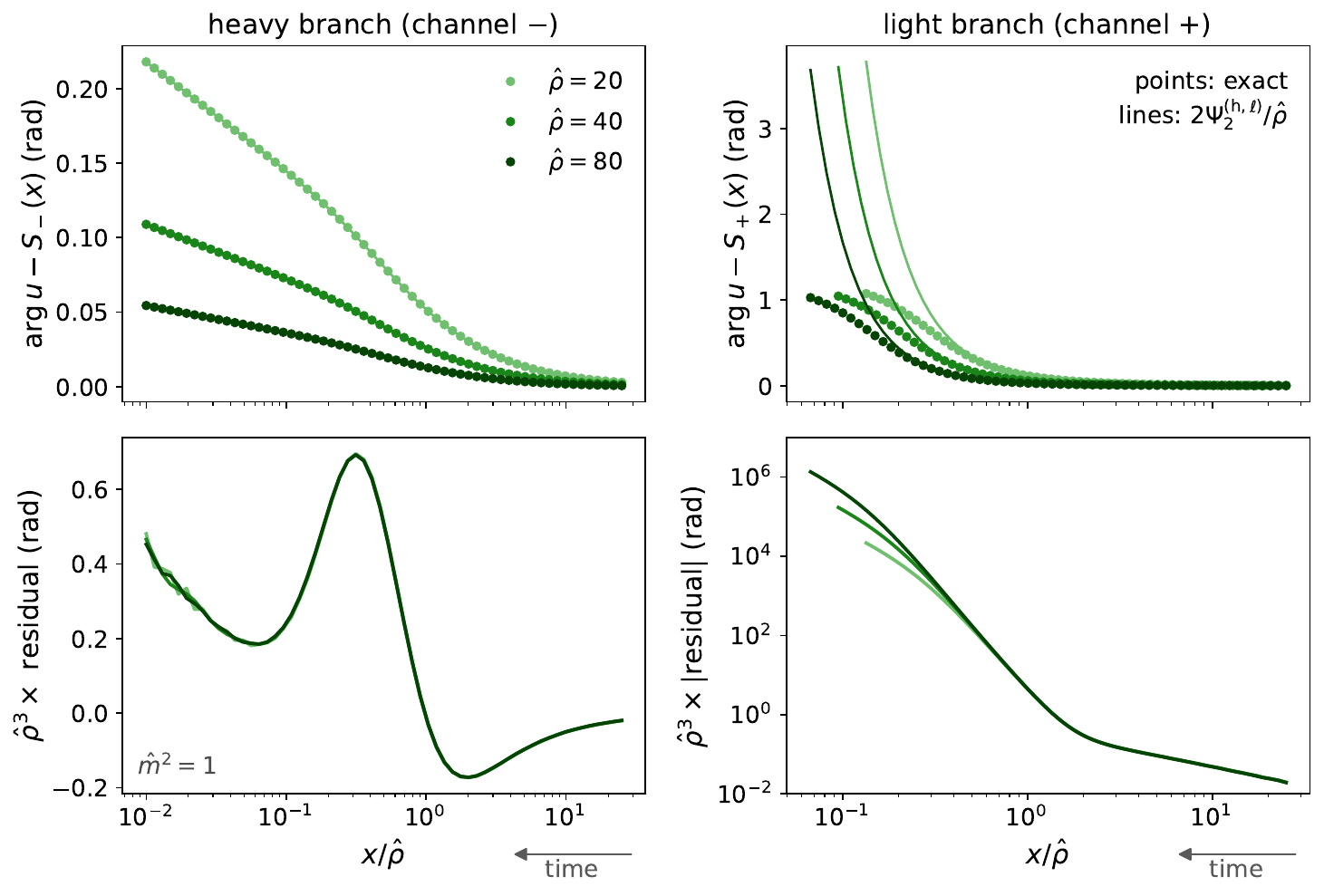}
\caption{Second-order WKB against the exact solution, for $\mhat^2=1$ and $\rhohat=20$, $40$, $80$. \emph{Top}: phase of the exact canonical mode function $\ucan=\pi_{c,a}/x$ once the leading phase $S_a(x)$ of~\eqref{eq:Sa} is subtracted (points), compared with the second-order prediction $2\Psi_2^{(\brh,\brl)}/\rhohat$ of~\eqref{eq:Psi2} (lines), for the heavy branch (left) and the light branch (right), as functions of $x/\rhohat$. \emph{Bottom}: the residual after subtraction of $2\Psi_2^{(\brh,\brl)}/\rhohat$, multiplied by $\rhohat^3$, in absolute value and on a logarithmic scale for the light branch. The three curves collapse for the heavy branch at all times, and for the light branch as long as $x\gg\sqrt{\rhohat}$, where they follow the next order of the expansion. They separate at $x\sim\sqrt{\rhohat}$, that is $x/\rhohat\sim\rhohat^{-1/2}$, where the light branch loses adiabaticity~\eqref{eq:Rminus-div}.}
\label{fig:wkb-residuals}
\end{figure}

\paragraph{WKB solutions.} Figure~\ref{fig:wkb-residuals} shows the phase of the numerical solution, once the leading phase of~\eqref{eq:wkb-leading} has been subtracted, together with the second-order prediction $2\Psi_2/\rhohat$, and the residual after subtraction of the latter, for the two branches. On the heavy branch, the residual scales as $\rhohat^{-3}$ over the full range, down to $x\ll1$, and the amplitude, not shown, agrees with $q^{-1/2}$ to order $\rhohat^{-2}$. On the light branch, however, the curves collapse only as long as $x\gg\sqrt{\rhohat}$, and they separate when $x/\sqrt{\rhohat}$ becomes of order one, as the divergence~\eqref{eq:Rminus-div} anticipates.

\begin{figure}[t]
\centering
\includegraphics[width=\textwidth]{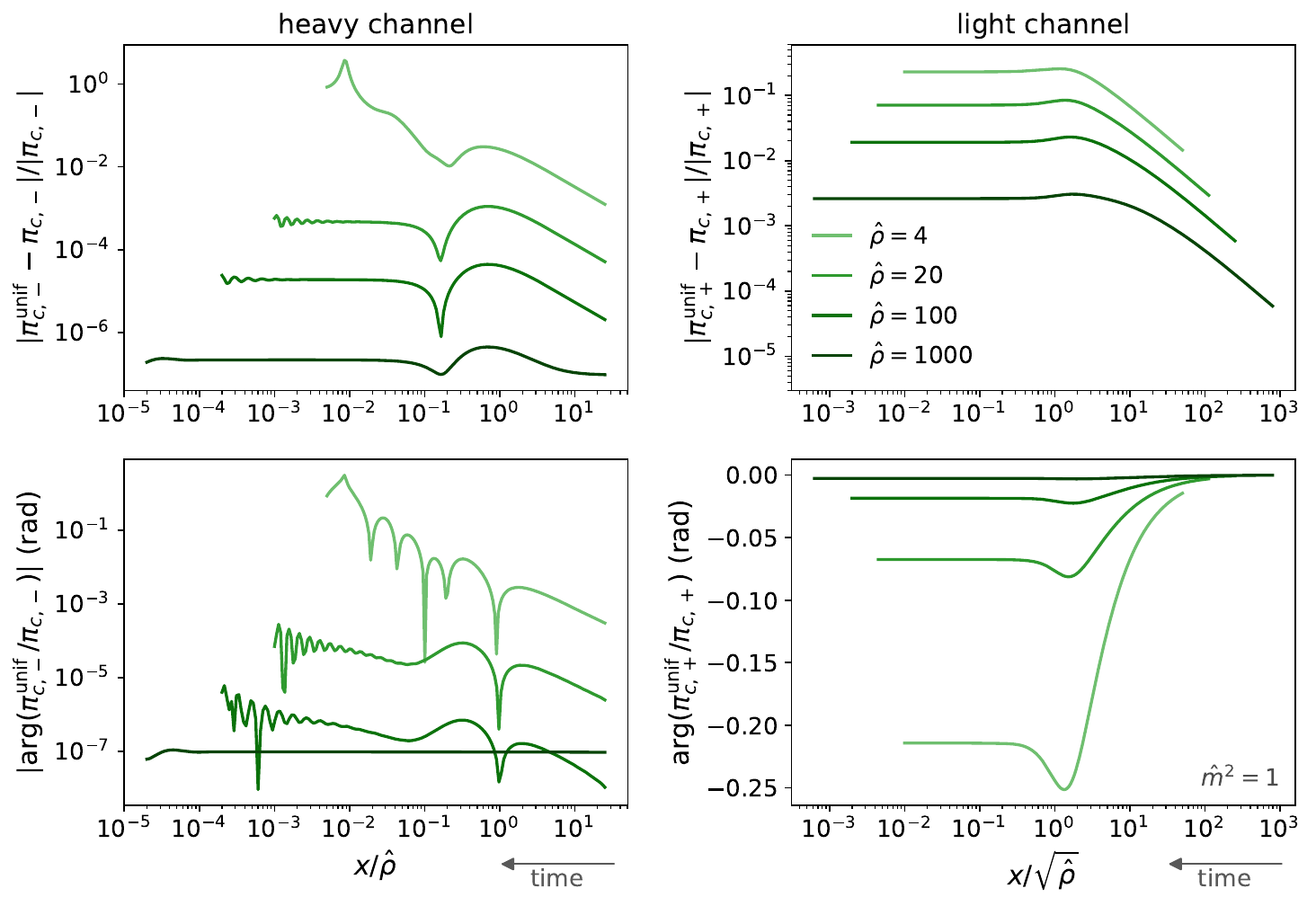}
\caption{Accuracy of the uniform approximations, for $\mhat^2=1$ and four values of the mixing. \emph{Left}: heavy channel, uniform approximation~\eqref{eq:heavy-uniform-bis} of $\pi_{c,-}$: relative error (top) and absolute value of the error on the phase, on a logarithmic scale (bottom), as functions of $x/\rhohat$. \emph{Right}: light channel, uniform approximation~\eqref{eq:light-uniform} of $\pi_{c,+}$: relative error (top) and error on the phase (bottom), as functions of $x/\sqrt{\rhohat}$. The late-time failure of the heavy channel at $\rhohat=4$ is the exponentially small constant discussed in section~\ref{sec:beyond}.}
\label{fig:uniform-x}
\end{figure}

\paragraph{Uniform approximations.} Figure~\ref{fig:uniform-x} shows the errors of~\eqref{eq:heavy-uniform-bis} and~\eqref{eq:light-uniform} as functions of time, in terms of both the relative error on the mode function of $\pic$ and the error on its phase. Their maxima over the range of $x$ shown are those of figure~\ref{fig:uniform-rho}. For the heavy channel at $\rhohat=4$, the approximation is still accurate to a few percent through the modified dispersion regime, and it fails at late times only, where the exact heavy channel freezes to the constant~\eqref{eq:asymmetry}. For the light channel, the error peaks at horizon crossing, $x\sim\sqrt{\rhohat}$, and settles at late times to the error on the late-time constant. The error on its phase also contains a logarithm, since the approximation~\eqref{eq:light-uniform} is built on the leading phase $S_+(x)$ and does not include the phase due to the mass, which becomes the logarithmic term of~\eqref{eq:gamma} at late times. The mode function of the massive field, not shown, has errors of the same size as that of $\pic$ in the light channel, and errors that decrease in the same way, as $\rhohat^{-2}$, in the heavy channel.

\begin{figure}[t]
\centering
\includegraphics[width=0.5\textwidth]{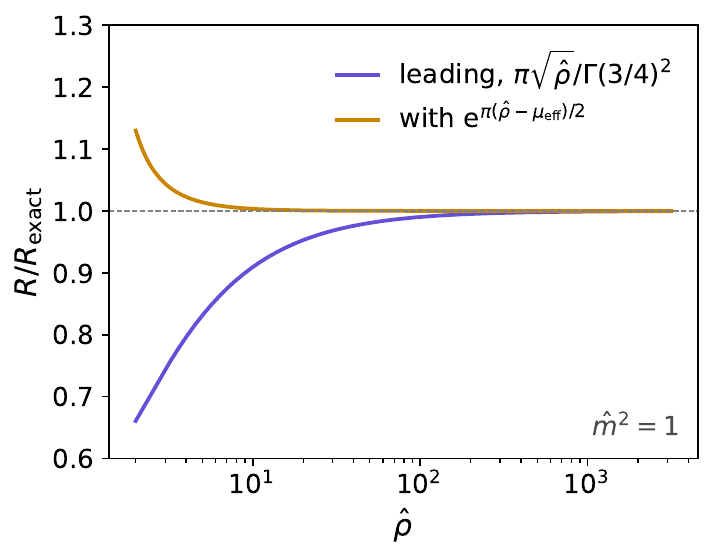}
\caption{Amplification $R$ of the power spectrum, normalised to the exact result~\eqref{eq:R}, for the leading formula~\eqref{eq:R-leading} and the exponential form~\eqref{eq:R-NLO}, as a function of the mixing, for $\mhat^2=1$.}
\label{fig:R}
\end{figure}

\paragraph{Power spectrum.} Figure~\ref{fig:R} compares the leading formula~\eqref{eq:R-leading} and the exponential form~\eqref{eq:R-NLO} with the exact amplification~\eqref{eq:R}. The former approaches it as $1/\rhohat$ and the latter as $1/\rhohat^{2}$, with the errors quoted in section~\ref{sec:uniform}.

\section{Series in the relativistic regime and its resummation}
\label{app:series}
This appendix gives an independent derivation of the results of sections~\ref{sec:wkb} and~\ref{sec:eft}. Its second part, the expansion at fixed $x$, organises order by order the corrections to the effective solution across horizon crossing, where the WKB expansion does not apply.

\paragraph{Series in the relativistic regime.} At large $x$, the fourth-order equation~\eqref{eq:app-fourth-pi} admits solutions of the form
\begin{equation}
\label{eq:app-subH-ansatz}
\pic=x^{1-\ii s_1\rhohat/2}\exp\Bigg[\ii s_2\,x\Big(1+\sum_{n\ge1}a_n\,\frac{\rhohat^{2n}}{(2x)^{2n}}\Big)+\sum_{n\ge1}b_n\,\frac{\rhohat^{2n}}{(2x)^{2n}}\Bigg]\,,\qquad s_1,s_2=\pm1\,,
\end{equation}
one for each of the four solutions, the two branches with positive and negative frequency, with coefficients that depend on $\rhohat$ and $\mhat$ but tend, as $\rhohat\to\infty$, to the universal values
\begin{equation}
\label{eq:app-subH-coeffs}
a_n=\frac{(-1)^{n}\,\Ga(n+\tfrac12)}{\sqrt\pi\,(2n-1)^{2}\,\Ga(n+1)}\,,\qquad b_n=\frac{(-1)^{n}}{4n}\,,
\end{equation}
independent of the branch. These are the Taylor coefficients of $\sqrt{1+u^2}-u\,\mathrm{arcsinh}\,u$ and of $-\tfrac14\log(1+u^2)$ in $u=\rhohat/2x$, so that the series resum to the adiabatic solution~\eqref{eq:wkb-leading}, with $s_1=a$ the channel and $s_2=+1$ the positive frequency. A truncation of~\eqref{eq:app-subH-ansatz} is only valid for $x\gg\rhohat$. Its resummation, however, being the adiabatic solution, is valid down to $x\gg\sqrt{\rhohat}$ on the light branch and at all times on the heavy one, as sections~\ref{sec:wkb} and~\ref{sec:uniform-accuracy} show.

\paragraph{Series at fixed $x$.} If one expands instead in powers of $1/\rhohat$ at fixed $x$, the system~\eqref{eq:system} is solved order by order by two quadratures, and the solution takes the form
\begin{equation}
\label{eq:app-fixedx}
\begin{aligned}
\pic&=\pi_0\sum_{n\ge0}\frac{\mathcal P_n(x)}{\rhohat^{2n}}-\frac{\sigma_0\,x^{3}}{\rhohat}\sum_{n\ge0}\frac{\mathcal Q_n(x)}{\rhohat^{2n}}\,,\\
\sigma&=\sigma_0\,x^{3}\sum_{n\ge0}\frac{\mathcal R_n(x)}{\rhohat^{2n}}-\frac{\pi_0\,x^{2}}{\rhohat}\sum_{n\ge0}\frac{\mathcal S_n(x)}{\rhohat^{2n}}\,,
\end{aligned}
\end{equation}
where $\pi_0$ and $\sigma_0$ are two integration constants, and $\mathcal P_n$, $\mathcal R_n$, $\mathcal S_n$ and $\mathcal Q_n$ are even polynomials in $x$, of degree $4n$ for the first three and $4n+2$ for the last, with $\mathcal P_0=\mathcal R_0=\mathcal S_0=1$ and $\mathcal Q_0=x^2/5+\mhat^2/3$. Their coefficients depend on the mass, with the exception of those of their highest monomials, which we denote by $\hat{\mathcal P}_n$, $\hat{\mathcal Q}_n$, $\hat{\mathcal R}_n$ and $\hat{\mathcal S}_n$. These obey two-term recurrence relations, whose solutions are
\begin{equation}
\label{eq:app-fixedx-coeffs}
\begin{aligned}
\hat{\mathcal P}_n&=\frac{(-1)^n}{2^{4n}\,n!}\,\frac{\Ga(-\tfrac14)}{\Ga(n-\tfrac14)}\,,\qquad &
\hat{\mathcal Q}_n&=\frac{(-1)^n}{2^{4n+2}\,n!}\,\frac{\Ga(\tfrac54)}{\Ga(n+\tfrac94)}\,,\\
\hat{\mathcal R}_n&=\frac{(-1)^n}{2^{4n}\,n!}\,\frac{\Ga(\tfrac54)}{\Ga(n+\tfrac54)}\,,\qquad &
\hat{\mathcal S}_n&=-\frac{(-1)^n}{2^{4n+2}\,n!}\,\frac{\Ga(-\tfrac14)}{\Ga(n+\tfrac34)}\,.
\end{aligned}
\end{equation}
The series formed by the highest monomials are those of Bessel functions of argument $x^2/2\rhohat$, and they resum to
\begin{equation}
\label{eq:app-fixedx-bessel}
\begin{aligned}
\pic&=x^{5/2}\Big[\mathcal A\,J_{-5/4}\Big(\frac{x^{2}}{2\rhohat}\Big)-\mathcal B\,J_{5/4}\Big(\frac{x^{2}}{2\rhohat}\Big)\Big]\,,\\
\sigma&=x^{5/2}\Big[\mathcal B\,J_{1/4}\Big(\frac{x^{2}}{2\rhohat}\Big)+\mathcal A\,J_{-1/4}\Big(\frac{x^{2}}{2\rhohat}\Big)\Big]\,,
\end{aligned}
\end{equation}
with $\mathcal A=\Ga(-\tfrac14)\,(4\rhohat)^{-5/4}\,\pi_0$ and $\mathcal B=\Ga(\tfrac54)\,(4\rhohat)^{1/4}\,\sigma_0$. One recognises the effective solution~\eqref{eq:eft-solution}, written in the basis of Bessel functions of the first kind instead of Hankel functions. The lower monomials of each polynomial are suppressed, at fixed $x^2/\rhohat$, by powers of $1/x^2$, and hence contribute at relative order $1/\rhohat$ at horizon crossing and at relative order $\rhohat^{-2}$ at late times. The effective solution is thus the leading order of this expansion, resummed at fixed $x^2/\rhohat$, uniformly down to $x\to0$, as used in section~\ref{sec:eft-solution}. The lower monomials, for their part, are where the mass first appears in the light branch.

\section{Excited initial states}
\label{app:excited}
\paragraph{Late-time power spectrum.} The conditions of section~\ref{sec:excited} state that the operators of the new state are canonical. Expressing the Bunch-Davies operators in terms of them, $\hat b_b=\sum_a(\alpha_{ab}\,\tilde b_a+\beta^*_{ab}\,\tilde b^\dagger_a)$, and requiring that these be canonical as well gives the equivalent form
\begin{equation}
\label{eq:bogoliubov-ccr}
\alpha^\dagger\alpha-\beta^T\beta^*=\mathds{1}\,,\qquad \alpha^\dagger\beta=\beta^T\alpha^*\,.
\end{equation}
At late times, the heavy channel and its complex conjugate decay at every order in $1/\rhohat$, so that $\tilde\pi_{c,a}(0)=\alpha_{a+}\pi_{c,+}(0)+\beta_{a+}\pi^{*}_{c,+}(0)$. With $\pi_{c,+}(0)=|\pi_{c,+}(0)|\,\ee^{\ii\theta_0}$, the sum over channels in~\eqref{eq:R-def} becomes $|\pi_{c,+}(0)|^2\sum_a|\alpha_{a+}+\beta_{a+}\ee^{-2\ii\theta_0}|^2$, and the $(+,+)$ element of the first relation in~\eqref{eq:bogoliubov-ccr}, $\sum_a|\alpha_{a+}|^2=1+\sum_a|\beta_{a+}|^2$, gives~\eqref{eq:excited-late}. In these terms, $n_+$ is the mean number of Bunch-Davies quanta of the light channel in the new state.

\paragraph{States invariant under rotations.} Deep inside the horizon, $x\gg\rhohat$, the terms in $\rhohat/x^2$ and $\mhat^2/x^2$ of~\eqref{eq:canonical-system} are negligible, and the remaining mixing terms, in $\rhohat/x$, couple the two fields antisymmetrically. The dynamics is then invariant under rotations of $(\pic,\sigma)$ in field space, and the channels are the eigenstates of charge $a=\pm1$ under these rotations, namely the combinations $\pic+\ii a\,\sigma$ of~\eqref{eq:earlytime}, with opposite logarithmic phases. The conjugate of the channel $a$ has the charge $-a$. In a state invariant under rotations, $\alpha$ can only connect channels of equal charge and $\beta$ channels of opposite charges. Hence $\alpha$ is diagonal, $\beta$ is off-diagonal, and $\mathcal I=0$. Applying the same single-field transformation $(\alpha,\beta)$ to $\pic$ and to $\sigma$ gives such a state, $\alpha_{ab}=\alpha\,\delta_{ab}$ and $\beta_{ab}=\beta\,(1-\delta_{ab})$, with $n_+=|\beta|^2$.

\paragraph{Excitation of the curvature perturbation.} We consider an excitation acting on $\pic$ alone at a time $x_f\gg\rhohat$.\footnote{A step in the inflaton potential produces such an excitation at leading order. In the adiabatic and entropic decomposition of the perturbations of a two-field model, the adiabatic fluctuation obeys the Mukhanov-Sasaki equation with the term $z''/z$, where $z\propto a\sqrt{\epsilon}$ is sensitive to a step in the slow-roll parameter $\epsilon$, whereas the entropic one involves $a''/a$ only~\cite{Langlois:2008mn}. At $x_f\gg\rhohat$, the mixing, of relative size $\rhohat/x_f$, does not transmit the excitation from one field to the other at leading order.} From~\eqref{eq:BD}, the channel mode functions are then $X_a\propto x_f^{-\ii a\rhohat/2}\,(1,-\ii a)$ up to a common factor, so that the solutions whose $\sigma$ component, respectively $\pic$ component, vanishes at that time are
\begin{equation}
\label{eq:app-single-field-modes}
X_\pi=\frac{x_f^{\ii\rhohat/2}X_++x_f^{-\ii\rhohat/2}X_-}{\sqrt2}\,,\qquad
X_\sigma=\ii\,\frac{x_f^{\ii\rhohat/2}X_+-x_f^{-\ii\rhohat/2}X_-}{\sqrt2}\,.
\end{equation}
They are the mode functions of the two fields taken separately at $x_f$. A Bogoliubov transformation $(\alpha,\beta)$ acting on $\pic$ alone at $x_f$ replaces $X_\pi$ by $\alpha X_\pi+\beta X_\pi^*$ and leaves $X_\sigma$ unchanged. Written again in terms of the channels, it gives $\alpha_{++}=(\alpha+1)/2$, $\alpha_{-+}=(\alpha-1)\,x_f^{\ii\rhohat}/2$, $\beta_{++}=\beta\,x_f^{-\ii\rhohat}/2$ and $\beta_{-+}=\beta/2$, which satisfy~\eqref{eq:bogoliubov-ccr}. Hence $n_+=|\beta|^2/2$ and $\mathcal I=\tfrac12\alpha\beta^*\,x_f^{\ii\rhohat}$. The phase of the interference term is readily interpreted. The coefficients of a transformation acting at $x_f$ are defined with respect to the mode functions~\eqref{eq:app-single-field-modes} at that time, whose phase is $x_f$. If we define the phase $\delta$ intrinsic to the transformation by $\alpha\beta^*=|\alpha\beta|\,\ee^{\ii(\delta-2x_f)}$, the phase of the interference term in~\eqref{eq:excited-late} is $\delta+2[\theta_0-S_+(x_f)]$, where $S_+(x_f)\simeq x_f-(\rhohat/2)\log x_f$ is the phase~\eqref{eq:Sa} of the light channel at $x_f$. The second term is twice the phase accumulated by the light channel between the transformation and its freezing. It contains $\gamma$ because this phase is accumulated across the transition at $x\sim\rhohat$, after the transformation.

\section{Cancellations at the resonance}
\label{app:cancellations}

In this appendix we derive the factor $\mathcal V$ of~\eqref{eq:vertex-factors} in the two cases in which the leading contributions cancel, namely the interaction $\dot\pi_c\sigma^2$, and the combination of $\dot\pi_c^2\sigma$ and $(\partial_i\pic)^2\sigma$ generated by the mixing alone. In both cases, the sum over the legs produces in the integrand of~\eqref{eq:hard-integral} a factor that vanishes at the stationary point~\eqref{eq:resonance}, so that the result comes from the first corrections.

We write the fields that the short modes carry at the vertex as multiples of $\pi_{c,+}$, with $\sigma_+=Z\pi_{c,+}$ and $\dot\pi_{c,+}=-H\,d\;\pi_{c,+}$, where $d\equiv x\,\dd\log\pi_{c,+}/\dd x$. To the order needed,
\begin{equation}
\label{eq:app-W-d}
Z=-\ii-\frac{3}{\rhohat\,(r-1)}\,,\qquad d=\ii\,\homl+1+\frac{1}{2r^2}\,,
\end{equation}
the first being the relation between the two fields at first order, given below~\eqref{eq:wkb-order1}, and the second following from the phase and the amplitude of~\eqref{eq:wkb-leading}. The integrals to evaluate are then of the form $\int\dd x\,\eta\,\ee^{\ii\Phi}\,\mathcal V(x)$, with $\eta$ and $\Phi$ given in~\eqref{eq:hPhi} and $\mathcal V(x)$ a combination of $Z^*$, $d^*$ and $x$. Since $x\,\Phi'=\mueff-2\homl$, a factor that vanishes at the stationary point can be integrated by parts. For a slowly varying function $G$, one has
\begin{equation}
\label{eq:app-by-parts}
\int\dd x\;\eta\,\ee^{\ii\Phi}\,\big(\mueff-2\homl\big)\,G=\ii\int\dd x\;\ee^{\ii\Phi}\,\frac{\dd(x\,\eta\,G)}{\dd x}\,,
\end{equation}
so that, under the integral, $(\mueff-2\homl)\,G$ is equivalent to $\ii\,[G\,(\tfrac12+r^{-2})+x\,G']$, where we used $x\,\eta'/\eta=-\tfrac12+r^{-2}$. The other corrections of relative order $1/\rhohat$, from the second-order phase~\eqref{eq:Psi2} and from the stationary phase approximation itself, multiply a factor that vanishes at the stationary point, and therefore do not contribute at this order.

\paragraph{Interaction $\dot\pi_c\sigma^2$.} The long mode carries either the time derivative, which brings $-H\rhohat$ and leaves two short legs with $\sigma$, or $\sigma$, which leaves one short leg with $\dot\pi_c$ and one with $\sigma$, with twice as many contractions. With the coupling $\alpha/2$, this gives $\mathcal V(x)=-\alpha H\,Z^*\,[\rhohat\,Z^*+2\,d^*]$, and~\eqref{eq:app-W-d} yields
\begin{equation}
\label{eq:app-case1}
\rhohat\,Z^*+2\,d^*=\ii\,\big(\mueff-2\homl\big)+\ii\,\big(\rhohat-\mueff\big)-\frac{3}{r-1}+2+\frac{1}{r^2}\,.
\end{equation}
The first term is the factor that vanishes at the resonance. Using~\eqref{eq:app-by-parts} with $G=1$, it is equivalent to $-\tfrac12-r^{-2}$. The second, of order $1/\rhohat$, is negligible. The sum is therefore $\tfrac32-3/(r-1)$, which equals $-\tfrac32$ at $r_*=2$, and with $Z^*=\ii$ we obtain $\mathcal V(x_*)=(3\ii/2)\,\alpha H$. We thus recover the factor $\mathcal V$ of~\eqref{eq:vertex-factors}, which, for all the interactions, is the value of $\mathcal V(x)$ at the stationary point.

\paragraph{Interactions generated by the mixing.} With $\Lambda_2^{-1}=-\Lambda_1^{-1}$, the two interactions combine into the Hamiltonian density $(2\Lambda_1)^{-1}\,[(\partial_i\pic)^2/a^2-\dot\pi_c^2]\,\sigma$. When the long mode carries $\sigma$, the two short legs carry either the two gradients, which bring $(k/a)^2=H^2x^2$, or the two time derivatives. When it carries one of the time derivatives, the short legs carry the other one and $\sigma$. This gives $\mathcal V(x)=(H^2/\Lambda_1)\,[x^2-d^{*2}-2\rhohat\,d^*Z^*]$. With~\eqref{eq:app-W-d} and the dispersion relation of the light branch~\eqref{eq:omega-pm}, $x^2=\homl(\homl+\rhohat)$, we find
\begin{equation}
\label{eq:app-case2}
x^2-d^{*2}-2\rhohat\,d^*Z^*=-\homl\big(\mueff-2\homl\big)+2\ii\,\big(\homl-\rhohat\big)\Big(1+\frac{1}{2r^2}\Big)-\frac{6\,\ii\,\homl}{r-1}+\mathcal O(\rhohat^{0})\,.
\end{equation}
The first term, of order $\rhohat^2$ away from the resonance, vanishes there. Using~\eqref{eq:app-by-parts} with $G=-\homl$ and $x\,\homl'=\rhohat\,(r^2-1)/2r$, this term is equivalent to $-\ii\,[\homl(\tfrac12+r^{-2})+x\,\homl']$. At the stationary point, where $r=2$ and $\homl=\rhohat/2$, the three terms of~\eqref{eq:app-case2} are $-9\ii\rhohat/8$, $-9\ii\rhohat/8$ and $-3\ii\rhohat$, so that
\begin{equation}
\label{eq:app-V-mixing}
\mathcal V=-\frac{21\,\ii}{4}\;\frac{\rhohat H^2}{\Lambda_1}\,,
\end{equation}
one power of $\rhohat$ below the contribution of each interaction taken separately. These two results are compared with the numerical evaluation of the signal in figure~\ref{fig:collider}.

\bibliographystyle{JHEP}
\bibliography{references}

\end{document}